\documentclass[12pt]{article}
\usepackage{amsmath}
\usepackage{graphicx}
\usepackage{natbib}
\usepackage{url} % not crucial - just used below for the URL 

\usepackage{algorithm,algpseudocode}
\usepackage{amssymb}
\usepackage{amsthm}
\usepackage{bbm}
\usepackage{centernot}
\usepackage{changebar}
\usepackage{changepage} 
\usepackage{color, colortbl}
\usepackage{import}
\usepackage{mathrsfs}
\usepackage{mathtools}
\usepackage{setspace}
\usepackage{titling}
\usepackage{bm}
\usepackage{caption}
\usepackage{hyperref}
\usepackage{float}
\usepackage{rotating}
\usepackage{comment}
\usepackage{xr}

\newcommand{\blind}{0}

\newtheorem{myexample}{Example}
\newtheorem{myproposition}{Proposition} 
\newtheorem{mydef}{Definition}

\begin{document}

\def\spacingset#1{\renewcommand{\baselinestretch}%
{#1}\small\normalsize} \spacingset{1}

%%%%%%%%%%%%%%%%%%%%%%%%%%%%%%%%%%%%%%%%%%%%%%%%%%%%%%%%%%%%%%%%%%%%%%%%%%%%%%

\if0\blind
{
  \title{\bf Exploiting disagreement between high-dimensional variable selectors for uncertainty visualization}
  \author{Christine Yuen
  \hspace{.2cm}\\
    Department of Statistics, London School of Economics and Political Science\\
    and \\
    Piotr Fryzlewicz \\
    Department of Statistics, London School of Economics and Political Science}
  \maketitle
} \fi

\if1\blind
{
  \bigskip
  \bigskip
  \bigskip
  \begin{center}
    {\LARGE\bf Exploiting disagreement between high-dimensional variable selectors for uncertainty visualization}
\end{center}
  \medskip
} \fi

\bigskip

\begin{abstract}
We propose Combined Selection and Uncertainty Visualizer  (CSUV), which estimates the set of true covariates in high-dimensional linear regression and visualizes selection uncertainties by exploiting the (dis)agreement among different base selectors. Our proposed method selects covariates that get selected the most frequently by the different variable selection methods on subsampled data. The method is generic and can be used with different existing variable selection methods. We demonstrate its variable selection performance using real and simulated data. The variable selection method and its uncertainty illustration tool are publicly available as R package \verb+CSUV+ (\url{https://github.com/christineyuen/CSUV}). The graphical tool is also available online via \url{https://csuv.shinyapps.io/csuv}.
\end{abstract}

\noindent%
{\it Keywords:} high-dimensional data, variable selection, uncertainty visualization
\vfill

\newpage
\spacingset{1.5} % DON'T change the spacing!
\section{Introduction} \label{intro}
	Model and variable selection in high-dimensional regression settings have been widely discussed in the past decades. In the context of the linear model, the best subset selection (dated back to at least \citeauthor{beale1967discarding}, \citeyear{beale1967discarding}) is computationally infeasible when the number of covariates $p$ is large. Regularization methods with convex penalties, such as the Lasso \citep{tibshirani1996regression}, are capable of performing variable selection in large-$p$ settings and yet they are computationally efficient. Elastic Net \citep{zou2005regularization} is believed to be particularly suitable for designs with a high degree of correlation between the covariates. Group Lasso \citep{yuan2006model} is designed for situations in which the covariates are best considered in groups.
Regularized regression methods with non-convex penalties such as the smoothly clipped absolute deviation (SCAD, \citeauthor{fan2001variable}, \citeyear{fan2001variable}) and minimax concave penalty (MCP, \citeauthor{zhang2010nearly}, \citeyear{zhang2010nearly})
methods are designed to reduce estimation bias. The theoretical evaluation of the properties of these and many others variable selection methods has been the subject of intense research effort.
For example, the irrepresentable condition \citep{zhao2006model} is sufficient and almost necessary for the Lasso to be sign consistent. \cite{fan2010selective} provide a detailed review of different variable selection methods in high-dimensional settings. \par
	There has also been a growing focus on post-selection inference. \cite{van2014asymptotically}, \cite{zhang2014confidence} and \cite {javanmard2018debiasing} advocate the de-biasing approach, which constructs confidence intervals for covariates by de-sparsifying the Lasso estimators. \cite{lee2016exact}, \cite{tibshirani2016exact} and \cite{tibshirani2018uniform} propose a conditional approach which provides confidence intervals for the selected covariates using the distribution of a post-selection estimator conditioning on the selection event. \cite{chatterjee2011bootstrapping} and \cite{liu2013asymptotic} suggest using bootstrapping on some existing variable selection methods. \par
%------------------------------------------------
	In this paper we focus on identifying the true set of covariates and illustrating the selection uncertainty in the linear model. We assume that the observed data are the realization 
of:
    \begin{equation} \label{model}
        Y_i = \beta_0 + \sum_{j=1}^{p} \beta_j X_i^{j}+\epsilon_i,\quad i = 1,...,n,
    \end{equation}
	where $p$ is the number of covariates, $n$ is the number of observations, and we potentially have $p > n$. $X_i^j$ is the $j^{th}$ covariate of the $i^{th}$ observation of $\bm{X}$ and $\bm{X}$ is a fixed $n \times p$ design matrix. $\bm{X}$ is standardized with each covariate $X^{j}$ has $\sum_{i=1}^n X_i^{j}/n = 0$ and $\sum_{i=1}^n (X_i^{j})^2/n = 1$. $\epsilon$ is i.i.d. noise with mean zero and variance $\sigma^2$. Furthermore, the model is assumed to be sparse with the set of true covariates $S= \{ j \in \{1,...,p\}: \beta_j \ne 0\}$, $s = |S| \ll p$. \par
\begin{figure}[!h]
		\begin{center}
 			\includegraphics[width=0.8\linewidth, scale=0.5]{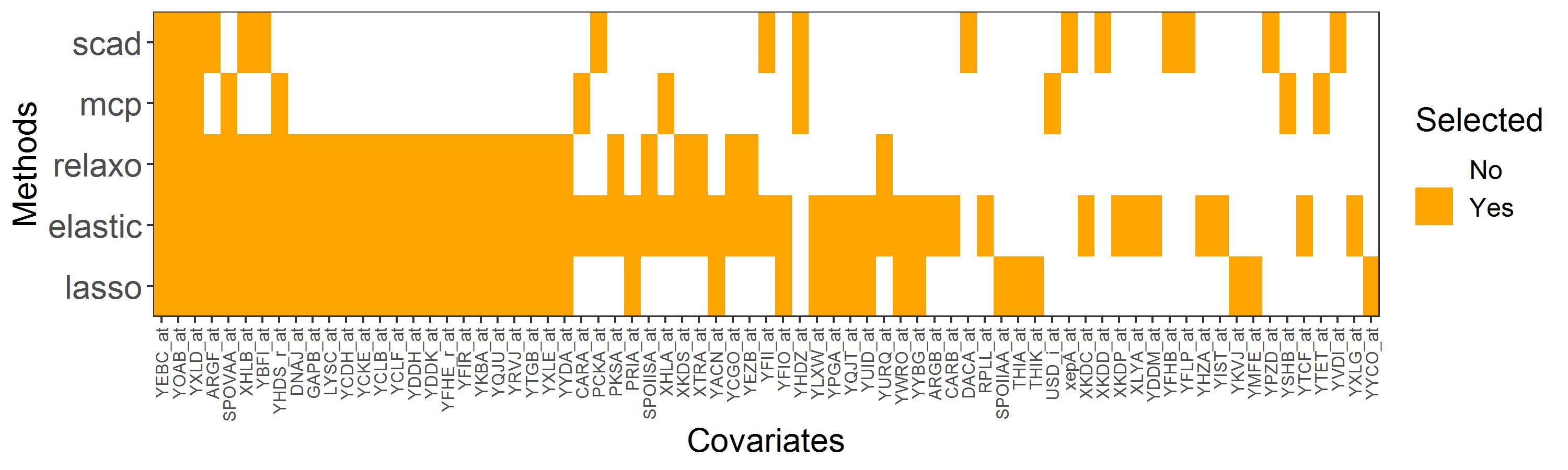}
 		\end{center}
 		\caption{Graphical illustration of selections by different variable selection methods (Lasso, Elastic Net, relaxed Lasso, MCP and SCAD) with default tuning using the riboflavin dataset from Example \ref{ribo1}. Covariates that are not selected by any methods are not shown in the graph for readability.}
 		\label{plot:selection:ribo}
	\end{figure}

\begin{figure}[!h]
		\begin{center}
 			\includegraphics[width=0.8\linewidth, scale=0.5]{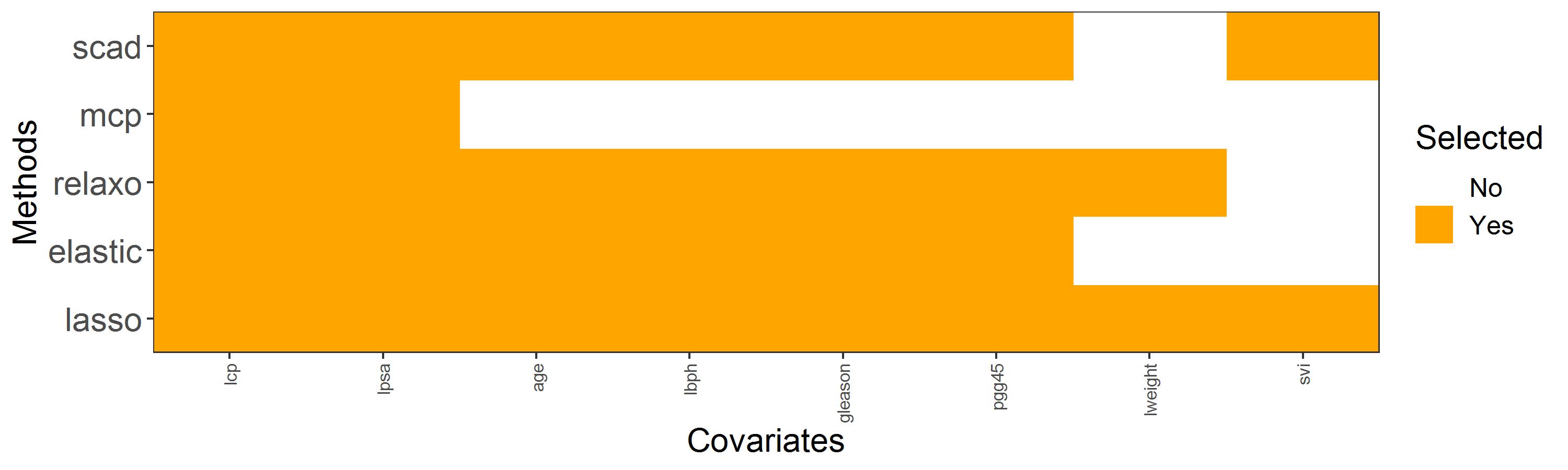}
 		\end{center}
 		\caption{Graphical illustration of selections by different variable selection methods (Lasso, Elastic Net, relaxed Lasso, MCP and SCAD) with default tuning using the prostate dataset from Example \ref{pros}.}
 		\label{plot:selection:pros}
	\end{figure}

	
%------------------------------------------------
	Less effort has been devoted in the literature to {\em selecting the best variable selection method} for the data at hand. Various theoretical performance guarantees are available for a range of methods, but
many of them are not testable in practice; for instance, checking the irrepresentable condition usually requires knowing the true set of covariates. Therefore, this type of theory can be of limited use in method selection. 
How to select a method remains an open and yet very important question to ask, as it affects our selection of the set of relevant variables. To illustrate this impact, let us consider two real-life datasets in Examples \ref{ribo1} and \ref{pros}.
	\begin{myexample}[Riboflavin data] \label{ribo1}
		{\normalfont The riboflavin dataset concerns the riboflavin (vitamin B2) production by bacillus subtilis. The response is the logarithm of the riboflavin production rate by bacillus subtilis and the $p = 4088$ covariates are the logarithms of the expression levels of $4088$ genes. The number of samples is $n = 71 \ll p$. The dataset is available in the R package \verb+hdi+.} \par % {\bf hdi}
	\end{myexample}
	\begin{myexample}[Prostate cancer data, \citeauthor{stamey1989prostate}, \citeyear{stamey1989prostate}]\label{pros}
		{\normalfont The prostate cancer dataset comes from a study that examined the relationship between the level of prostate-specific antigen and $p=8$ clinical measures (logarithm of weight, age, Gleason score, among others) in men who were about to receive a radical prostatectomy. The sample size is $n = 97$. The dataset is available in the R package \verb+lasso2+.}
	\end{myexample}
	We process the datasets using five different variable selection methods: the Lasso, Elastic Net, relaxed Lasso \citep{meinshausen2007relaxed}, MCP and SCAD in R with default tuning in the corresponding R packages (see Section \ref{implement_R} for more details). We justify the choice of these particular methods in Section \ref{methods}. Working with default parameters would be a commonly used starting point for the non-expert applied user. The selection results are shown in Figures \ref{plot:selection:ribo} and \ref{plot:selection:pros}. \par
	Figure \ref{plot:selection:ribo} shows that for the riboflavin dataset the sets of covariates selected vary significantly among the methods, which makes it difficult to justify the validity of the set of covariates selected using any one method. For the prostate cancer dataset, even though there are only eight covariates to choose from, there is still selection disagreement among the methods (Figure \ref{plot:selection:pros}). \par
%------------------------------------------------	
	Such disagreement among methods as shown in Figures \ref{plot:selection:ribo} and \ref{plot:selection:pros} is not an exception but a common observation. The distance heat maps in Appendix \ref{app:heatmap} show that selection disagreement manifests itself across different simulation settings (see Section \ref{sim} for more details on the simulation settings). Having observed disagreement, one possible way to proceed would be to rank the different models considered (e.g. using cross-validation or an information criterion) and select the highest-ranked one.
In this paper, we consider eBIC \citep{chen2008extended} and delete-$n/2$ cross-validation \citep{zhang2015cross} as they are suitable for high-dimensional settings. Further details of these two methods are discussed in Section \ref{modsel}. Our simulation results show that in general eBIC performs better than the delete-$n/2$ cross-validation in terms of variable selection (see Tables \ref{1_1}-\ref{7_1} in Appendix \ref{app:table}). In fact, eBIC in many simulation settings performs very similarly to the best performing individual variable selection method. \par
	Although eBIC seems to be able to select a single good model fit, can more be said regarding the uncertainty of variable selection, based on the disagreement between the methods tested? The similarities and disagreements among the different variable selectors, which is a piece of information not typically used by any one of them, may provide us with some useful insight. For example, in Figure \ref{plot:selection:ribo} all of the methods select the first three covariates whereas the remaining covariates are selected by some of the methods only. Does it mean that the first three covariates are more likely to be the true covariates? This question is central to this paper, and motivates our main development, described next.
 %------------------------------------------------	
	In this paper, we propose a new tool for variable selection with uncertainty visualization, termed Combined Selection and Uncertainty Visualizer (CSUV). CSUV combines, in a particular way, a number of different base variable selection methods into a new variable selector, and illustrates the output of this new selector together with a graphical representation of its uncertainty. It makes use of sets of covariates selected on different subsamples of the data with different variable selection methods. A full description of the proposed method is in Section \ref{method} and \ref{graph:illustration}. The variable selection part of the proposed procedure can be summarized as follows: first, split the data into the training and test sets and fit different variable selection methods on the training set over a grid of tuning parameter values. Estimate the performance of the fitted models on the test set, and retain only the $k$ best-performing models. Repeat the process a number of times and select the covariates that appear the most frequently in the collection of the retained fitted models. \par
\begin{figure}[!h]
		\begin{center}
 			\includegraphics[width=\linewidth, scale=0.8]{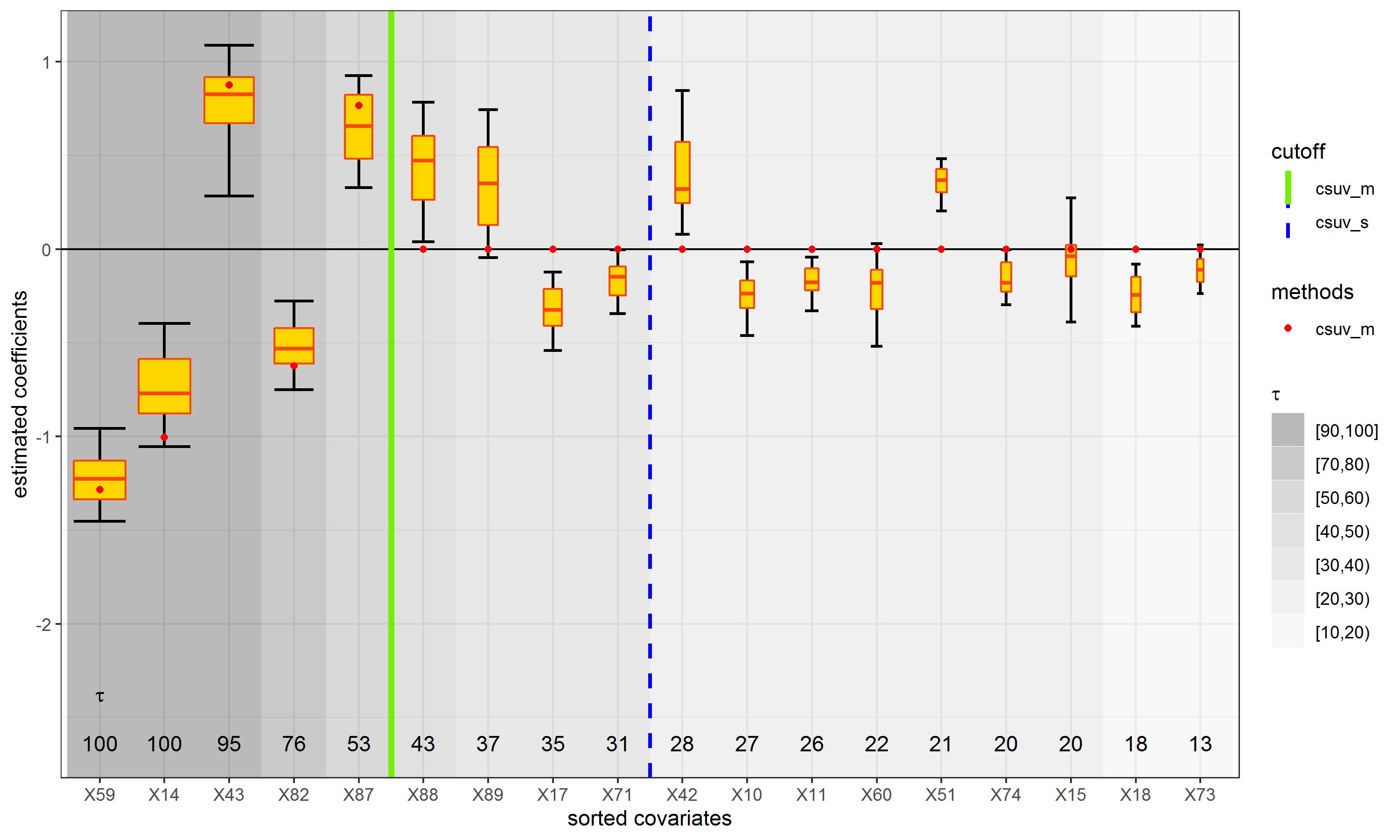}
 		\end{center}
 		\caption{Example of the CSUV graphical tool with simulated data from model 2 parameter setting 5 (see Section \ref{syn} for more details on the simulation setting). Box plots illustrate the empirical distributions of the estimated coefficients conditional on them being non-zero, and the whiskers represent their $5\%$ and $95\%$ percentiles. The ordering of the covariates is according to the CSUV solution path (see Definition \ref{sol:path:def} in Section \ref{sol:path}) and the width of each box plot along the x-axis is proportional to the level of the relative same sign frequency $\tau_j$ (see Definition \ref{rf} in Section \ref{new:method}; heuristically, the higher the value of $\tau_j$, the higher the frequency with which the corresponding variable has been selected with the same positive or negative sign). The numbers at the bottom of the graph show the actual values of $\tau_j$ times $100$ and the shade in the background corresponds to the level of $\tau_j$ with ranges as shown in the legend. Dots (red in the color version) are the estimated coefficients by CSUV-m (see Definition \ref{method:1} in Section \ref{def2}). The solid vertical line (green in the color version) represents the cut-off of CSUV-m, and the dotted vertical line (blue in the color version) represents the cut-off of CSUV-s (see Definition \ref{method:2} in Section \ref{sol:path}). Covariates with $\tau_j < 0.1$ are not shown for readability.}
		\label{boxplot:intro}
	\end{figure}

	The other component of CSUV is a graphical tool designed to visualize the selection uncertainty by using disagreement among the different model fits. See Figure \ref{boxplot:intro} as an example of a graphical output of CSUV. The plot shows the frequency with which each covariate is selected and the variability of the non-zero estimated coefficients. As we will see in Section \ref{compare:ci}, the graphical tool can be used to assist variable selection. \par 
\begin{figure}[!h]
		\begin{center}
			\includegraphics[width=\linewidth]{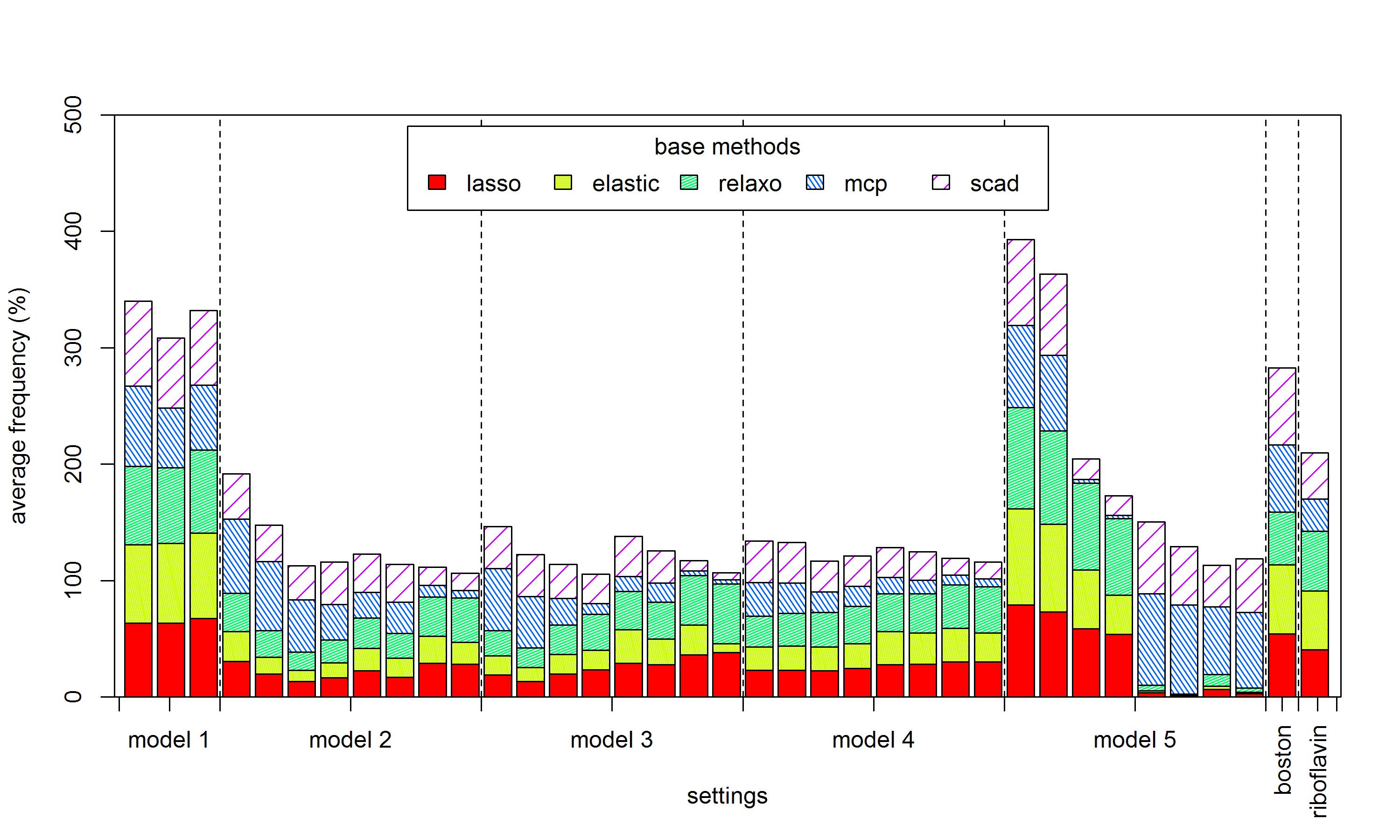}
		\end{center}
		\caption{Average relative frequency of the constituent methods selecting the same set of covariates as the fitted models retained by CSUV when the Lasso, Elastic Net, relaxed Lasso, MCP and SCAD are used as the constituent variable selection methods for CSUV in our simulations (see Section \ref{syn} for more details on the simulation settings). The sum of the average frequency of methods can be more than 100\% as multiple methods can select the same set of covariates.}
		\label{mod:freq}
	\end{figure}

	Our numerical experience (see Figure \ref{mod:freq} for a summary) suggests that the fitted models selected by CSUV tend to be distributed fairly uniformly over the entire range of the base variable selection methods used. This shows CSUV generally makes use of most of the base variable selection methods to get the final fitted model.\par
	The paper is organized as follows. In Section \ref{sec:rw}, we describe some related work. In Section \ref{method}, we discuss the main ideas behind CSUV, and we present the variable selection and coefficient estimation part of CSUV. In Section \ref{graph:illustration}, we introduce the graphical tool of CSUV to illustrate the disagreement in variable selection and the variability in coefficient estimation, and demonstrate its capability in assisting variable selection. In Section \ref{sim}, we present the simulation results. We conclude the paper with a discussion in Section \ref{conclusion}.
%------------------------------------------------
\section{Related work} \label{sec:rw}
	\subsection{Model selection procedures} \label{modsel}
		One possibility open to analysts when faced with competing fitted models is to select one of them. For example, \cite{chen2008extended} propose eBIC, an extension of BIC to high-dimensional data which takes into account both the number of unknown parameters and the complexity of the model space. \cite{zhang2015cross} advocate the use of the delete-$n/2$ cross-validation to select a method among all the candidate methods. For each iteration, delete-$n/2$ cross-validation uses half of the data for fitting and half for evaluation. The authors argue that in order to consistently identify the best variable selection procedure by cross-validation, the evaluation part has to be sufficiently large so that there are (1) more observations in the testing part to provide better evaluation and (2) fewer observations in the training part to magnify the difference in performance between methods. \par
	\subsection{Model combination with a single method}
		Model combination with subsampling has been used to improve variable selection performance of a single variable selection method. For example, Bolasso \citep{bach2008bolasso} fits the Lasso on each bootstrap sample and takes the intersection of all the selections. \cite{wang2014median} propose the median selection subset aggregation estimation (MESSAGE) algorithm which aims to perform variable selection on large-$n$ datasets. It runs a variable selection method (e.g. the Lasso) in parallel on each subset of the data and selects the set of covariates whose median is non-zero. The ranking-based variable selection \citep{baranowski2018ranking} algorithm uses subsampling to identify the set of consistently highly-ranked covariates. Stability selection (\citeauthor{meinshausen2010stability}, \citeyear{meinshausen2010stability} and \citeauthor{shah2013variable}, \citeyear{shah2013variable}) provides control over the finite sample familywise type I errors via subsampling. The stability selection procedure repeatedly samples observations and fits the sampling data using a variable selection method (e.g. the Lasso). It then keeps the covariates with selection frequency higher than a certain threshold. \par
		Similarly to the methods above, CSUV, our proposal, fits variable selection methods on subsampled data and selects the covariates that appear the most frequently. Unlike these other approaches, however, CSUV makes use of different variable selection methods as we observe that no one method outperforms all other methods in all settings. This brings various advantages, including obtaining access to good model fits from different variable selection methods, and being able to exploit disagreement between the selectors to evaluate selection uncertainty. We elaborate on these points later.\par
	\subsection{Model combination with multiple methods}
		Adaptive regression by mixing (ARM, \citeauthor{yang2001adaptive}, \citeyear{yang2001adaptive}) and its variation, adaptive regression by mixing with screening (ARMS, \citeauthor{yuan2005combining}, \citeyear{yuan2005combining}), aggregate fits from different methods by estimating weights through subsampling. ARM uses half of the data to fit some candidate models/procedures (e.g. smoothing splines with cross-validation tuning) and estimate $\sigma$. The remaining data is used to evaluate the prediction loss. The weight for each candidate model/procedure is calculated using $\hat{\sigma}$ and prediction loss. ARM gets the final weights by averaging the weights from different iterations. Finally it fits the full set of data using all the candidate models/procedures and obtains the final model by averaging the fits using the estimated weights. ARMS is similar to ARM except it uses half of the data to calculate AIC or BIC and retains only the models that have low AIC or BIC. The final fitted model from ARM or ARMS is not necessarily sparse as it is a weighted average of a number of models. Variable selection deviation measures (VSD, \citeauthor{nan2014variable}, \citeyear{nan2014variable}) aim to provide a sense of how trustworthy a set of selected covariates is. The VSD of a target model $m$ is the weighted cardinality of the symmetric difference between $m$ and each candidate model. \cite{nan2014variable} suggests using the sets of fitted models on the solution paths from the Lasso, SCAD and MCP as candidate models and the weight of each candidate model is calculated based on information criteria or ARM. The simulation results in \cite{nan2014variable} show that a large VSD compared to the size of the target model means that the target model is not trustworthy, but a small VSD does not necessarily mean that the target model is close to the true model. \cite{yang2017toward} propose to select a set of covariates that minimizes the total Hamming distance with all the candidate models in terms of VSD (we refer to this method as VSD-minimizing in the remainder of the paper). The authors also propose using different thresholds, where the threshold of $0.5$ is equivalent to minimizing the standard Hamming distance. For variable selection method combinations that do not involve subsampling, \cite{tsai2010combining}, \cite{mares2016combining} and \cite{pohjalainen2015feature} provide empirical results on combining sets of selected covariates from different variable selection methods by intersection, union and/or some other set operations. \par
		Both our method and VSD use resampling and different variable selection methods to provide an assessment of how good the final set of covariate selection is. VSD focuses on the whole model fit. Our method focuses on the uncertainty of individual covariates and a graphical tool is designed to illustrate these uncertainties. In terms of methodology detail, our method combines the sets of covariates selected in resampling fits whereas VSD combines sets of covariates selected on the solution path when fitting using all the data. Resampling data is only used in VSD for calculating the weight of each set of covariates. In our simulation study we compare the variable selection performance of our method to the VSD-minimizing method proposed by \cite{yang2017toward}, as it is the method the most similar to CSUV. The simulation results in Section \ref{compare:cv} show that in general our method outperforms the VSD-minimizing model.
%------------------------------------------------	
\section{CSUV variable selection methodology} \label{method}
	\subsection{Simple aggregation} \label{sa}
		The first goal of this paper is to use the similarity of fits from different methods to obtain the final set of covariates. One naive way to do so would be as follows.
		\begin{itemize}
			\item Step 1: fit the data using different variable selection methods.
			\item Step 2: record the percentage of times a covariate $X_j$ is selected among the different methods. Denote it by $\theta_j$.
			\item Step 3: get the final set of covariates by selecting covariates with high $\theta_j$'s. For example, select the set of covariates $\{X_j: \theta_j \geq 0.5\}$.
		\end{itemize}
		Different variable selection methods optimize different objective functions. In the case of regularized regression, the difference among methods is usually in terms of the penalty. If a covariate is selected by the majority of methods, it means the covariate is chosen to minimize many different objective functions. We expect that a true covariate $j$ should have a high $\theta_j$, i.e. 
it should frequently be chosen regardless of the objective function used.
		This simple procedure, however, suffers from the following drawbacks.
		\begin{itemize}
			\item Some variable selection methods can be similar in terms of selection regardless of the data as their objective functions are similar. Taking an extreme example, if we include two equivalent variable selection methods, such as the constrained and the penalized forms of the Lasso with equivalent regularization parameters, the sets selected by both methods will be the same. Such set is selected twice not because it maximizes two different object functions but merely because two equivalent methods are considered. This issue can cause an uneven ``sampling'' of methods and the corresponding fitted models.
			\item The above procedure assigns the same weight to all the base methods. When the performance across methods is very different (for example one method is substantially better than the others), such equal weight assignment is not ideal.
			\item Several methods can be wrong at the same time. For example, a false covariate can be wrongly selected by most methods if it has a spuriously high sample correlation with the response. When all methods are not performing well, a false covariate may have a high $\theta_j$. 
		\end{itemize}
		In the next section, we discuss how to overcome these drawbacks.\par 
%------------------------------------------------	
	\subsection{CSUV variable selection}	\label{new:method}
		Motivated by the above discussion, the variable selection in CSUV uses the general simple aggregation principles introduces in the previous section, but is also supplemented with the additional principles below:
		\begin{itemize}
			\item Only include the fitted models that exhibit good performance, in the sense specified in Section \ref{perf}.
			\item Repeat the fitting on subsampled data, to incorporate the variability in selection caused by the variability in data.
		\end{itemize}
		The variable selection procedure of CSUV can be summarized as follows. First, randomly split the data into training and test sets, and fit different variable selection methods on the training set over a grid of regularization parameters without tuning (see Section \ref{methods} and \ref{implement_R} for more the details on the grid of regularization parameters considered). Then, use the test set to calculate the performance of the fitted models and retain only the first $k$ fitted models that have the best performance (see Section \ref{perf} for more details on performance measure). Repeat the process many times to record a list of retained fitted models. Finally, select the covariates that appear the most frequently with the same positive or negative sign in the retained fitted models. The pseudo-code in Algorithm \ref{code:vs} provides a more detailed description of the variable selection part of CSUV. Coefficient estimation on the selected set is discussed in Section \ref{coef:est}.\par
		Before we present Algorithm \ref{code:vs}, we define the relative same sign frequency $\tau_j$, which measures the percentage of times that the $j$th covariate is selected with the same sign.
		\begin{mydef}[Relative same sign frequency $\tau_j$] \label{rf}
			Assume we have a set of fitted models $\mathcal{M}$. The relative same sign frequency of covariate $X_j$ is defined as:
				$$\tau_j = \frac{1}{|\mathcal{M}|}\max\left(\sum_{M_k \in \mathcal{M}} \mathbbm{1}_{\hat{\beta}_j^{M_k}>0}, \sum_{M_k \in \mathcal{M}} \mathbbm{1}_{\hat{\beta}_j^{M_k}<0}\right)$$
		 	where $\hat{\beta}_j^{M_k}$ is the estimated coefficient of the $j^{th}$ covariate on the fitted model $M_k \in \mathcal{M}$, and $\mathbbm{1}_{x}$ is the indicator function.
		 \end{mydef}	
		 \scalebox{0.9}{
			\begin{minipage}{1.1\linewidth}
				\begin{algorithm}[H]
					\caption{Select a set of covariates in CSUV} \label{code:vs}
 						\begin{algorithmic}[1]
						\Require
							variable selection methods $\mathcal{A}_1,...,\mathcal{A}_R$ with the corresponding generation of the grid of regularization parameters; $n$ observations with $p$ covariates $\bm{X}$ and response $Y$; number of repetitions $B$, percentile parameter $q$; frequency threshold $t$; percentage of data used in training set $w\%$; performance measure. 
						\Ensure
							set of selected covariates $\hat{S}$.
						\For{$b$ in \{$1,..., B$\}}
							\State randomly assign $w\%$ of the observations as training data with labels $I_{train}^b$ and the rest as test data with label $I_{test}^b$. Fit data with label $I_{train}^b$ using $\mathcal{A}_1,...,\mathcal{A}_R$ over grids of $K_r'$ different values of the corresponding regularization parameters, $r \in \{1,...,R\}$. For each method $\mathcal{A}_r$, denote the fitted models as $\tilde{M}_{r,1}^b,...,\tilde{M}_{r,{K'_r}}^b$ and the set of covariates selected by each fitted model as $S^{\tilde{M}_{r,k}^b} = \{j : \tilde{\beta_j}^{\tilde{M}_{r,k}^b} \ne 0\}, k \in \{1,...,K'_r\}$.
								\label{code:vs:fit}
							\State remove any duplication \textit{within each method} in terms of variable selection to get $S^{\tilde{M}_{r,1}^b},...,S^{\tilde{M}_{r,K_r}^b} \text{ such that for each } r, S^{\tilde{M}_{r,k}^b} \ne S^{\tilde{M}_{r,{k'}}^b} \ \forall k \ne k' \in \{1,..., K_r\}$. Record the sets of covariates selected by each fitted model $S^{\tilde{M}_{1,1}^b},...,S^{\tilde{M}_{1,K_1}^b},...,S^{\tilde{M}_{R,1}^b},...,S^{\tilde{M}_{R,K_R}^b}$ and re-index as $S^{\tilde{M}_{1}^b},...,S^{\tilde{M}_{K^b}^b}$, where $K^b$ is the number of fitted models recorded.
								\label{code:vs:unique}
							\State if the number of selected covariates $|S^{\tilde{M}_k^b}| < |I_{train}^b|$, refit the selected set of covariates $S^{\tilde{M}_k^b}$ using ordinary least squares (OLS), to get the fitted models $\hat{M}_1^b,...,\hat{M}_{K^b}^b$ with the estimated coefficients $\hat{\beta_j}^{\hat{M}_{k}^b}$. Otherwise, set $\hat{\beta_j}^{\hat{M}_{k}^b} =  \tilde{\beta_j}^{\tilde{M}_{k}^b}$.
								\label{code:vs:refit}
							\State use data with label $I_{test}^b$ to estimate the performance of each fitted model $\hat{M}_k^b$ from Step (\ref{code:vs:refit}). Order the models $\hat{M}_1^b,...,\hat{M}_{K^b}^b$ by the performance measure calculated from the best to the worst to obtain $\hat{M}_{(1)}^b,...,\hat{M}_{(K^b)}^b$.
								\label{code:vs:mse}
 							\State retain the first $q\%$ of the fitted models $\hat{M}_{(1)}^b,...,\hat{M}_{(K_q^b)}^b$, where $K_q^b= \mbox{round}(K^b \times q/100)$.
								\label{code:vs:retain}
						\EndFor						
 							\label{code:vs:loop}
						\State denote the set of retained fitted models by $\mathcal{M} = \{\hat{M}_{(1)}^1,...,\hat{M}_{(K_q^1)}^1,...,\hat{M}_{(1)}^B,...,\hat{M}_{(K_q^B)}^B \}$.
							\label{code:vs:M}
						\State calculate the relative same sign frequency $\tau_j$ for each variable $j$ according to Definition \ref{rf} and select the covariates such that:
			 				$$\hat{S} = \{j : \tau_j \ge t \}$$
							\label{code:vs:freq}
						\State \Return $\hat{S}$.
					\end{algorithmic}
				\end{algorithm}
			\end{minipage}
		}
		\\ \par
%------------------------------------------------	
        In Step (\ref{code:vs:unique}) of Algorithm \ref{code:vs}, duplicated sets of covariates selected within each method from Step (\ref{code:vs:fit}) are removed. This is because while multiple selections of a set of covariates in Step (\ref{code:vs:fit}) may suggest that the covariates in the set are likely to be the true covariates, it can also just be because many similar regularization parameter values have been used in Step (\ref{code:vs:fit}). In order to reduce the dependency of the frequency of the appearance of a set of covariates on the choice of the grid of regularization parameters, duplicated sets of covariates selected within each method from Step (\ref{code:vs:fit}) are removed. Note that duplicated sets of covariates selected across methods are not removed. \par
		Algorithm \ref{code:vs} involves repeated fits on subsamples of data, and this can be computationally expensive. Fortunately, the algorithm can easily be parallelized by running iterations on different cores/machines. This makes the algorithm feasible for high-dimensional data analysis. For example, on a 3-core machine, Applying CSUV on the riboflavin dataset of Example \ref{ribo1} takes less than $2$ minutes when following the specifications recommended in Section \ref{param} ($B = 100$, the constituent methods = \{Lasso, MCP, SCAD\}, etc.). Applying CSUV with the recommended specification on each simulated dataset in Section \ref{syn} with $p = 300$ takes less than $30$ seconds. \par
%------------------------------------------------		
	\subsection{Specifications for CSUV variable selection} \label{param}
		 Algorithm \ref{code:vs} provides a general framework for the CSUV variable selection approach. Here we discuss how the various parameters should or may be set for practical use. 	 
 		\subsubsection{Coefficient estimation} \label{coef:est}
			Algorithm \ref{code:vs} only selects a set of covariates without estimating the $\beta$ coefficients. In our implementation we use ordinary least squares (OLS) to estimate the $\beta$ coefficients on the selected set $\hat{S}$ using the full set of data to form the final fitted model. If the number of covariates selected is larger than the number of observations, we use ridge regression to estimate the coefficients by cross-validation (in this case, we use the default cross-validation setting from the \verb+glmnet+ R package).
		 \subsubsection{Performance measure} \label{perf}
		 	Step (\ref{code:vs:mse}) of Algorithm \ref{code:vs} aims to rank the fitted models based on their variable selection performance. As we do not know the true covariates, we are not able to measure variable selection performance directly. In general, in attempting to select fitted models or methods with good variable selection performance, it is common to use prediction measures such as MSE or information criteria such as BIC or eBIC. Theoretically, BIC is consistent in model identification when $p$ is fixed and eBIC is consistent in high-dimensional settings \citep{chen2008extended}. Our empirical experiments, however, show that when using BIC or eBIC as performance measures in Algorithm \ref{code:vs}, the resulting fitted models tend to select too few covariates so the final selection by CSUV omits too many true covariates. By contrast, using MSE as the performance measure in Algorithm \ref{code:vs} in our simulation settings provides good variable selection performance. Although MSE measures prediction rather than variable selection performance, MSE is often used for variable selection methods such as in selecting tuning parameter $\lambda$ for SCAD (\cite{fan2001variable}). \par
		\subsubsection{Percentage of data used in training set $w\%$}
			Following \cite{yang2001adaptive}, \cite{yuan2005combining} and \cite{zhang2015cross}, we use $50\%$ of the data for fitting and the remaining $50\%$ for testing; this splitting ratio attempts to ensure a sufficiently large sample size for both. Stability selection (\citeauthor{meinshausen2010stability}, \citeyear{meinshausen2010stability} and \citeauthor{shah2013variable}, \citeyear{shah2013variable}) also uses the same splitting ratio although their rationale is that subsampling with such a ratio behaves similarly to bootstrapping. \par
			Empirically, our simulations show that using a smaller training set (25\% of the data) results in the selecting fitted models with too few covariates. When using a large training set (75\% of the data), the selected fitted models are too similar to each other, which causes CSUV to select too many false covariates.
		\subsubsection{Frequency threshold $t$}\label{def2}
			 The frequency threshold $t$ features in Step (\ref{code:vs:freq}) of Algorithm \ref{code:vs}. In this paper, we set $t = 1/2$, which means that covariates with $\tau_j \geq 1/2$ are selected. We have the following definition,
in which the ``m" stands for median, because selecting covariates with $\tau_j \geq 1/2$ is equivalent to selecting covariates with a non-zero median in $\mathcal{M}$.
		 	\begin{mydef}[CSUV-m] \label{method:1}
				The CSUV method described by Algorithm \ref{code:vs} and using $t = 1/2$ is denoted by \textit{CSUV-m}.
			\end{mydef}	
\noindent The following results hold.
		 	\begin{myproposition}[] \label{propo:jaccard}
				If the signs for the non-zero $\hat{\beta}_j^k$'s for all $k$ for which $M_k \in \mathcal{M}$ are the same, i.e.
					\[
					\tau_j = \frac{\sum_{\{k | M_k \in \mathcal{M}\}} \mathbbm{1}_{\beta_j^k \ne 0}}{|\mathcal{M}|},
					\]
				then selecting a covariate $j$ when $\tau_j \geq 1/2$ is equivalent to minimizing the average Hamming distance between the final selected sets of covariates $\hat{S}$ and all the fitted models $M_k \in \mathcal{M}$.
			\end{myproposition}
			\begin{myproposition}[] \label{propo:sign}
				Let $\tau_j^+ = \frac{1}{|\mathcal{M}|}\sum_{\{k|M_k \in \mathcal{M}\}} \mathbbm{1}_{\hat{\beta}_j^k>0}$ and $\tau_j^- =  \frac{1}{|\mathcal{M}|}\sum_{\{k|M_k \in \mathcal{M}\}} \mathbbm{1}_{\hat{\beta}_j^k<0}$ (note $\tau_j = max(\tau_j^+, \tau_j^-)$). Consider the following distance function between a model $M$ and all the fitted models $M_k \in \mathcal{M}$:
					$$ dist(M, \mathcal{M}) = \sum_{j=1}^p \sum_{\{k|M_k \in \mathcal{M}\}} |s_j^M-sign(\hat{\beta}_j^k)| $$
				where $s_j^M$ is the sign of the coefficient of the covariate $j$ in model $M$ which can take the value $-1$, $0$ or $1$. Selecting a covariate $j$ when $\tau_j \geq 1/2$ and setting $s_j^M=+1$ when $\tau_j^+\geq 1/2$ and $-1$ when $\tau_j^- \geq 1/2$ minimizes $dist(M, \mathcal{M})$.
			\end{myproposition}
			The proofs are in Appendix \ref{prop:prove} and \ref{prop:prove:sign}. Selecting covariates via  thresholding $\tau_j$ in CSUV-m (Definition \ref{method:1}) is not the only option. In Section \ref{sol:path} we introduce CSUV-s, which uses information provided by the sizes of the retained models.	
		\subsubsection{Constituent variable selection methods $\mathcal{A}_1, ..., \mathcal{A}_R$} \label{methods}
		 	CSUV is designed to be generic so that any variable selection methods can be used as the constituent methods $\mathcal{A}_1,...,\mathcal{A}_R$ in CSUV. Ideally, all the methods $\mathcal{A}_r$ should have good variable selection performance, and there should be some variability among the methods in terms of false selection. The constituent methods should also be computationally efficient as Algorithm \ref{code:vs} fits the constituent methods on subsampled data multiple times. In this paper, we choose the Lasso, MCP and SCAD to be the default constituent methods as they are optimizing different objective functions. Methods like Elastic Net or relaxed Lasso are not selected as the default constituent methods as they are relatively similar to Lasso. The default constituent methods we choose are also computationally feasible in high-dimensional settings with efficient fitting algorithms available, and there is also a default way to compute the grid of regularization parameters to consider. For example, the R package \verb+ncvreg+ for MCP and SCAD by default computes a sequence of parameters $\lambda$ with equal spacing on the log scale and of length $100$, starting from the smallest value $0.001$. See Section \ref{implement_R} for more details on the R packages used. We do not consider some two-stage methods such as the adaptive Lasso \citep{zou2006adaptive} as they are relatively slow. We also do not consider methods without default parameter tuning in R (e.g. the Dantzig selector, \citeauthor{candes2007dantzig}, \citeyear{candes2007dantzig}) as it makes the comparison with other methods like delete-$n/2$ cross-validation more complicated. \par			 	
		 	CSUV can also tolerate duplicated or very similar methods, although it is not recommended due to the computational time. Including duplicated or very similar methods, though not preferable as it extends the computation time, in our experience it does not affect the variable selection performance much when the percentage parameter $q$ is small. In our simulation, when methods that usually select similar sets as the Lasso (such as the Elastic Net or relaxed Lasso) are included, the performance of CSUV is close to when these similar methods are not included. \par		
		 \subsubsection{Percentile parameter $q$}
		 	In our simulation $q = 0$ and $q = 5$ are used with MSE as the performance measure recommended in Section \ref{perf}, with $q=0$ corresponds to selecting one single fitted model with the lowest MSE. The performance is similar with $q=0$ and $q=5$, although $q=0$ provides slightly better results. When the larger percentile $q = 20$ is used, again the performance is still close to that of $q=0$. With $q=50$, CSUV performs poorly as it includes too many fitted models.\par		 	
		\subsubsection{Number of repetitions $B$}
		 	The number of repetitions $B$ should be large enough to stabilize the value of $\tau_j$ and at the same time it should not be too large so that Algorithm \ref{code:vs} can be run within a reasonable time. $B=100$ is used in our simulation when $n = 100$ and $p = {100, 300}$ and it provides a good compromise between stability and computational time. \par	 		
%------------------------------------------------
	\subsection{Solution path and selection with other thresholds} \label{sol:path}
		The CSUV-m uses $t=1/2$, which is equivalent to selecting the covariates for which $ \tau_j \geq 1/2$. Empirically, based on our simulation results, CSUV-m provides good variable selection results by striking a good balance between false inclusion and false omission. Comparing to other variable selection methods, CSUV-m usually includes many fewer false covariates, with the trade off being that it occasionally omits some true covariates. When the analyst's focus is on performance criteria other than variable selection, for example on prediction, they may want to select more covariates. This can be done by considering other thresholds $t$ on the sign frequency $\tau_j$, or a threshold on the model size as described in Algorithm \ref{code:thr}.\par
		 Algorithm \ref{code:thr} generates a solution path (see Definition \ref{sol:path:def}) by ordering covariates from the highest to the lowest relative same sign frequency $\tau_j$. This solution path can be regarded as a series of nested sets of covariates with increasing model sizes. Given a fixed model size $s$ as the size threshold, Algorithm \ref{code:thr} selects the first $s$ covariates on the solution path and returns them as the final selection set. 
		\begin{mydef}[CSUV solution path] \label{sol:path:def}
			The CSUV solution path orders covariates so that
				$$R_j < R_{j'} \text{ if } \tau_j > \tau_{j'} \text{ or } (\tau_j = \tau_{j'} \text{ and } |\bar{\hat{\beta}}_j| >|\bar{\hat{\beta}}_{j'}|)$$
			where $R_j$ is the position of covariate $j$ on the solution path, $\tau_j$ is the relative same sign frequency calculated in Step (\ref{code:vs:freq}) of Algorithm \ref{code:vs} and $\bar{\hat{\beta}}_j$ is the average of the estimated coefficients in $\mathcal{M}$ in Step (\ref{code:vs:M}) of Algorithm \ref{code:vs}. 
			\end{mydef}
			\begin{algorithm}[H]
				\caption{CSUV with a given model size} \label{code:thr}
				\begin{algorithmic}[1]
					\Require
						relative same sign frequency $\tau_j$ calculated in Step (\ref{code:vs:freq}) and $\mathcal{M}$ in Step (\ref{code:vs:M}) of Algorithm \ref{code:vs}; size threshold $s$.
					\Ensure
 						set of selected covariates.
					\State obtain the solution path (Definition \ref{sol:path:def}) using $\tau_j$ and $\mathcal{M}$.
						\label{code:thr:path}
					\State \Return the first $s$ covariates ordered in Step (\ref{code:thr:path}), i.e.
						$$\{j | R_j \leq s\}$$
				\end{algorithmic}
			\end{algorithm}
			The standardization of the design matrix in Equation (\ref{model}) ensures the comparison of the size of the estimated coefficients is meaningful. In the particular implementation of CSUV described in this paper, we set the size threshold $s$ equals to the median size of the selected sets in $\mathcal{M}$ in Step (\ref{code:vs:M}) of Algorithm \ref{code:vs} and we define CSUV with this threshold as CSUV-s.
			 \begin{mydef}[CSUV-s] \label{method:2}
			 	The CSUV method in Algorithm \ref{code:thr} with size threshold $s = median(|S^{\tilde{M}_{(1)}^1}|,$ $...,|S^{\tilde{M}_{(K)}^1}|,...,|S^{\tilde{M}_{(1)}^B}|,...,|S^{\tilde{M}_{(K)}^B}|)$, i.e. $s$ equal to the median size of the selected sets in $\mathcal{M}$ in Step (\ref{code:vs:M}) of Algorithm \ref{code:vs} is denoted by \textit{CSUV-s}, where s stands for size.
			 \end{mydef}
%------------------------------------------------
\section{CSUV visualization of uncertainty} \label{graph:illustration}
	\subsection{Graphical component of CSUV}
	 	In this section, we introduce the graphical component of CSUV, which is a tool designed to illustrate the variable selection and estimation uncertainty. An example of a plot is shown in Figure \ref{boxplot:intro} and the graphical tool is available interactively via a Shiny app at \url{https://csuv.shinyapps.io/csuv} and in the R package \verb+CSUV+. It has the following ingredients.\par
\begin{figure}[!h]
			\includegraphics[width=\linewidth]{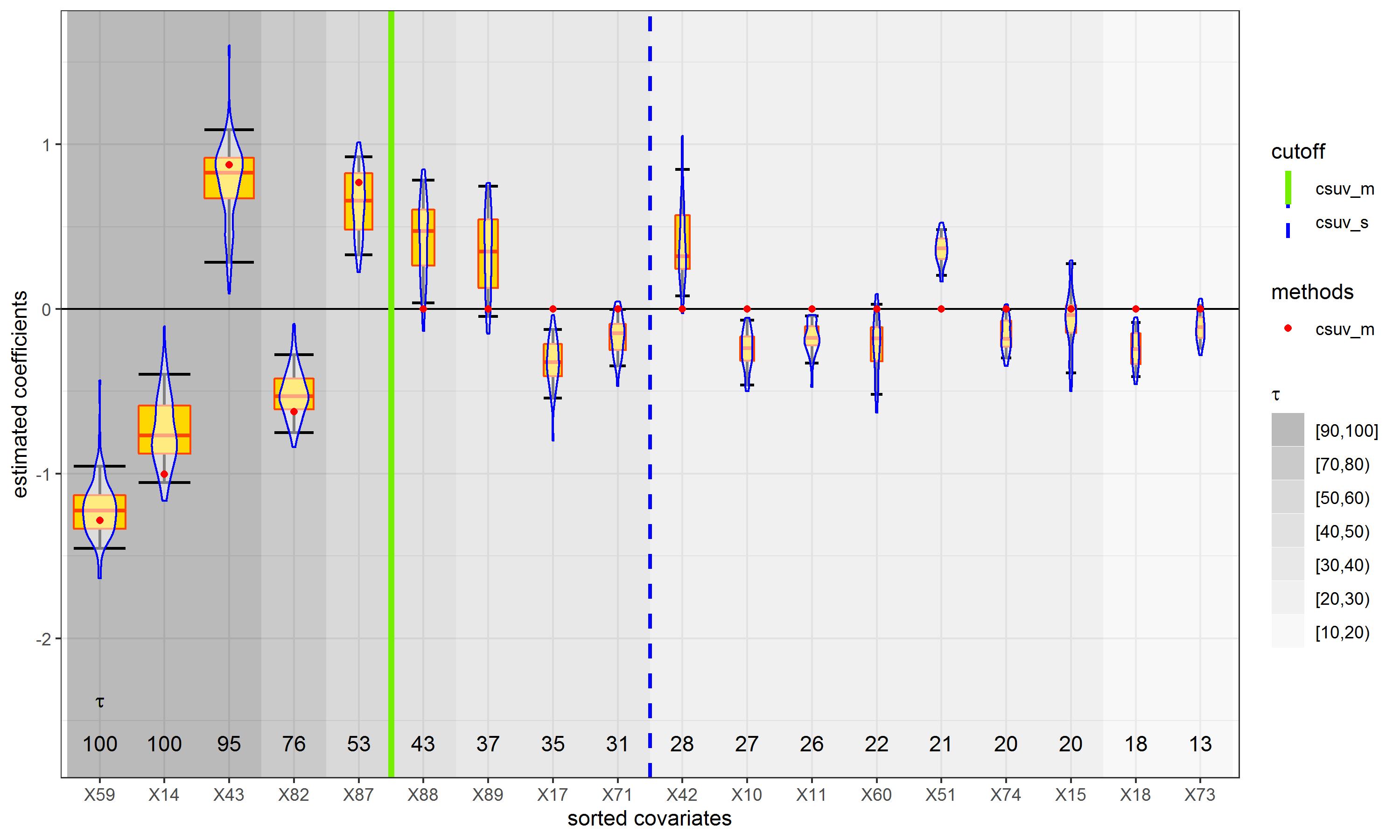}
			\caption{Same as Figure \ref{boxplot:intro} but with violin plots superimposed to show the conditional kernel density.}
			\label{bp_violin}
		\end{figure}

		\begin{itemize}
		 	\item Box plots that visualize the estimated coefficient uncertainty: each box plot correspond to a covariate $X_j$ and it shows the lower and the upper quartiles of the empirical distributions of the estimated coefficients conditional on them being non-zero, i.e. only take into account of the non-zero coefficients $\{\hat{\beta}_j^{M_k}|\hat{\beta}_j^{M_k}\ne0, \ M_k \in \mathcal{M}\}$ from Step (\ref{code:vs:M}) of Algorithm \ref{code:vs}. Its whiskers corresponding to the $5\%$ and $95\%$ percentile of the non-zero estimated coefficients (default, level can be changed in the \verb+CSUV+ R package). The width of each box is proportional to the relative same sign frequency $\tau_j$ (Definition \ref{rf}). The median value of the non-zero estimated coefficients is shown as a horizontal line in each box (red in the color version). The box plots are ordered according to the solution path (Definition \ref{sol:path:def}). Together, the width and the vertical aspect of each box plot visually describe the variability of the corresponding estimated coefficient over the different data subsamples drawn.
		 	\item Shaded background representing $\tau_j$: the background behind each box plot is shaded according to the relative same sign frequency $\tau_j$ of the corresponding covariate. The darker the color, the higher the value of $\lfloor 100\% \tau_j / 10 \rfloor$. The actual value of $\tau_j$ is displayed in black underneath the box plots.
		 	\item Lines showing the cut-off points for variable selection by the various versions of CSUV: CSUV-m (Definition \ref{method:1}) selects all covariates to the left of the solid vertical line. CSUV-s (Definition \ref{method:2}) selects all those to the left of the dotted vertical line.
		\end{itemize}
		Covariates with $\tau_j < 0.1$ are not included in the plot for readability. Users wishing to have a more detailed look into the empirical distribution of the non-zero estimated coefficients can superimpose the corresponding violin plots on the box plots in the \verb+CSUV+ package. See Figure \ref{bp_violin} as an example of such a plot. \par	
\begin{figure}[!h]
			\includegraphics[width=\linewidth]{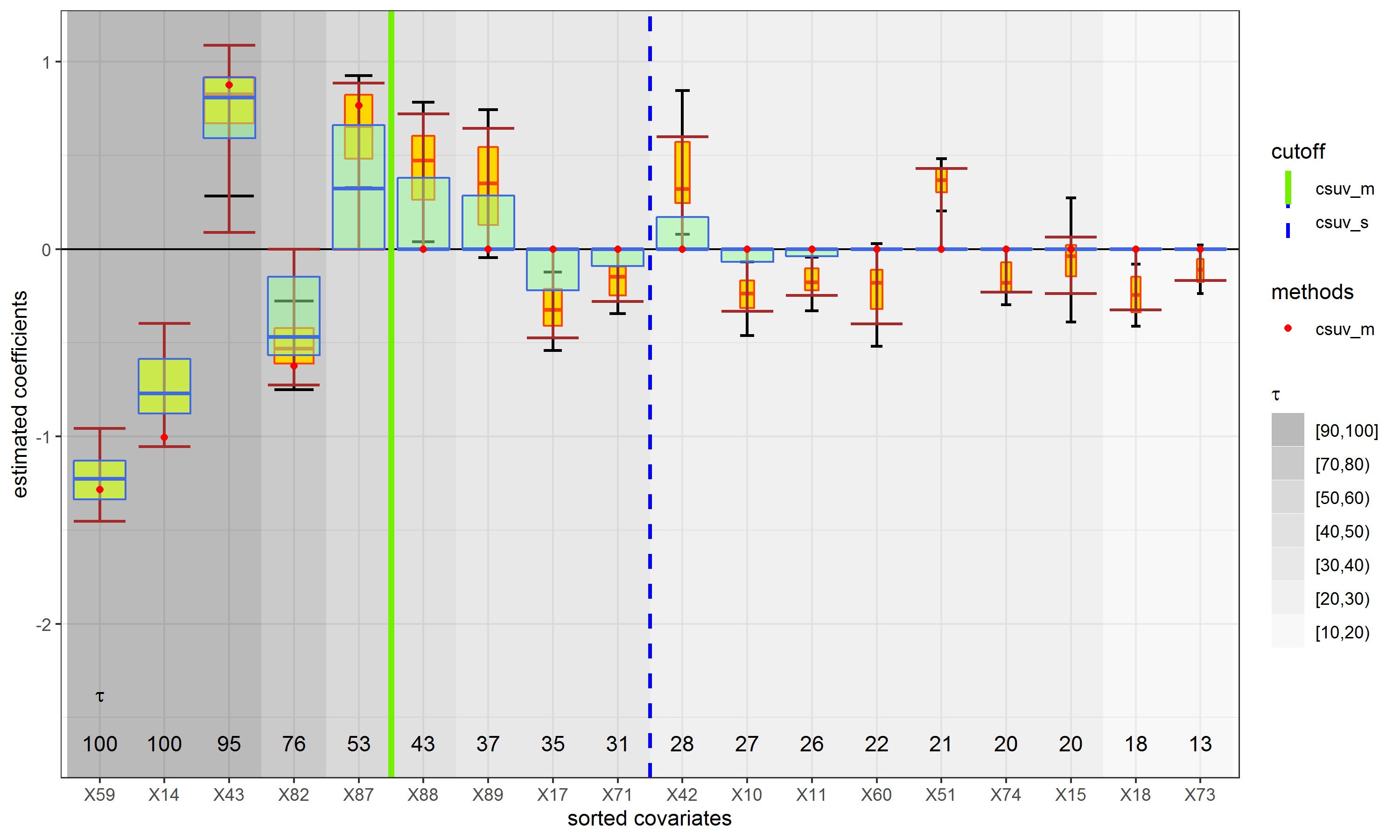}
			\caption{Same as the box plot in Figure \ref{boxplot:intro} but with the semi-transparent boxes (green in the color version, usually they are wider than the conditional boxes underneath them) which represent all the estimated coefficients in $\mathcal{M}$ in Step (\ref{code:vs:M}) superimposed on top of it.}
		\label{bp_unconditional}
		\end{figure}

		The default plot such as the one shown in Figures \ref{boxplot:intro} and \ref{bp_violin} only considers the empirical distributions of the estimated coefficients conditional on them being non-zero (we refer to them as ``conditional box plots''). This is because box plots that use all the estimated coefficients in $\mathcal{M}$ in Step (\ref{code:vs:M}) of Algorithm \ref{code:vs} that are both the zero and non-zero ones (``unconditional box plots'', see Figure \ref{bp_unconditional} for an example) hardly provide useful information beyond that already provided in the value of $\tau_j$, the latter also being reflected in the width of the conditional boxes. Nevertheless, the \verb+CSUV+ package allows users to create the unconditional box plots as well. \par
		The \verb+CSUV+ package users wishing to compare the results returned by CSUV with any individual variable selection procedures of their choice (as long as their outputs are in a compatible format stated in the R package documentation) are also able to produce an enhanced CSUV plot, showing all of the above, and with addition of the items below.
		 \begin{itemize}
		 	\item Graphical representation of the selection by a group of user-provided variable selection methods: the number (blue in the color version) in the bottom part of the graph shows the percentage of user-provided methods that have selected the corresponding covariate when fitting with all the observations. 
		 	\item Graphical representation of the selection by any single user-provided method: the coefficient estimates by the given method are shown as empty circles (white circles with a blue outline in the color version).
		 \end{itemize}
		\begin{figure}[!h]
			\includegraphics[width=\linewidth]{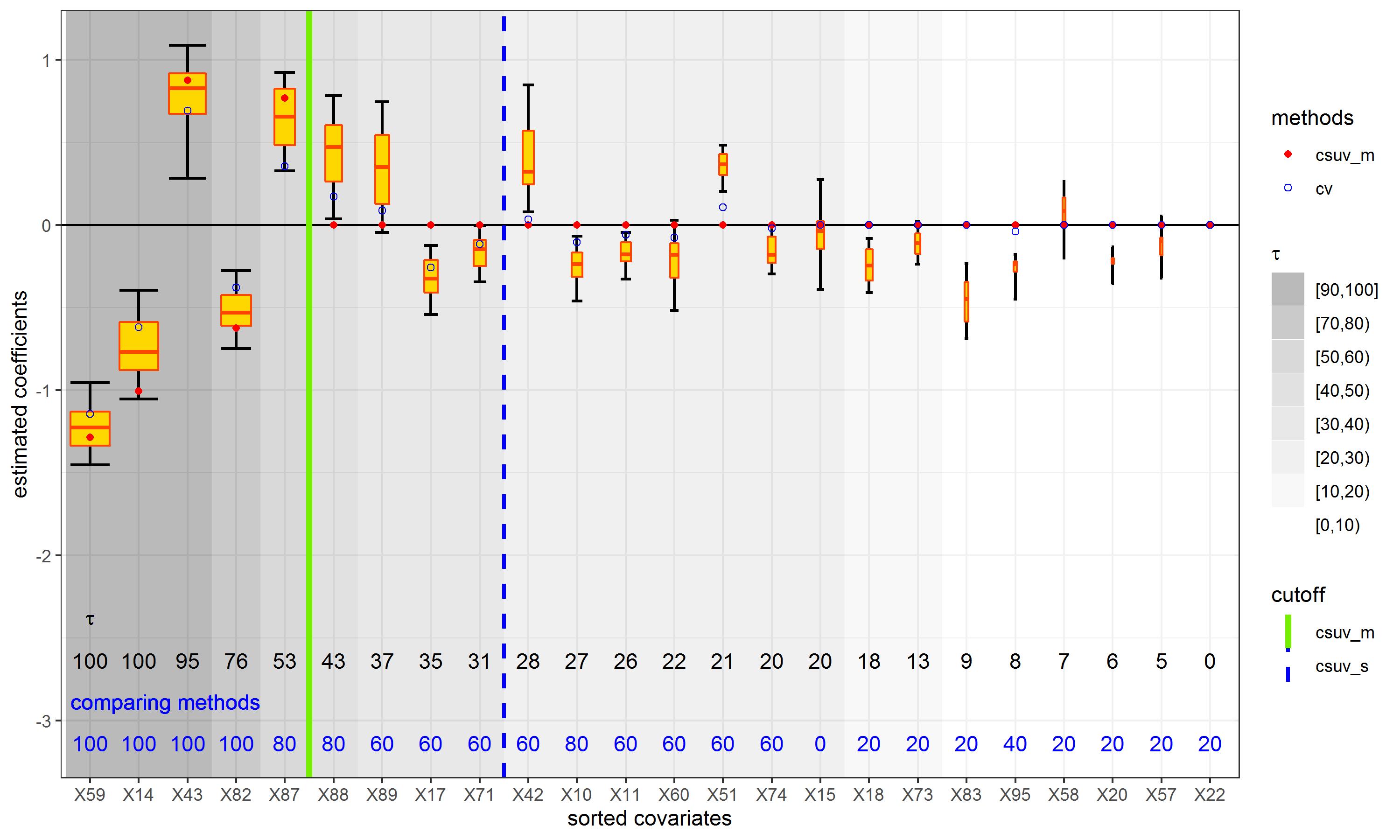}
			\caption{Example of the CSUV graphical tool with additional information of the fitting results from five individual variable selection methods (Lasso, Elastic Net, relaxed Lasso, MCP and SCAD) and delete-$n/2$ cross-validation, using simulated data from model 2 parameter setting 5 (see Section \ref{syn} for more details on the simulation setting). The plot is the same as Figure \ref{boxplot:intro} with the following extra information: Empty circles (white circles with blue outline in the color version) represent the coefficients estimated by a single method (here is delete-$n/2$ cross-validation). Numbers at the bottom (blue in the color version) represent the relative percentage proportion of the group of the individual methods that select the corresponding covariates. Covariates that are not selected by any methods and $\tau_j < 0.1$ are not shown for readability.}
			\label{bp1}
		\end{figure}
		See Figure \ref{bp1} for an example for such a plot. The user can decide if a covariate should be selected by considering if the corresponding CSUV box plot, the coefficient estimated by a single method and the percentage of selection by a group of comparing methods agree to some extent. \par
		Note that it is common that CSUV and other model selection procedures agree to some extent. For example, in Figure \ref{bp1}, CSUV, cross-validation and all the individual variable selection methods select the first four covariates. The methods, however, have some disagreements over the other covariates. For example, the fifth covariate is selected by both versions of CSUV, cross-validation and $80\%$ of the individual variable selection methods, but one of the individual variable selection methods does not select the covariate. The next ten covariates are selected by the majority of the individual methods and cross-validation, but they are not chosen by CSUV-m. These non-selection decisions taken by CSUV-m are correct, as in this particular simulation setting only the first five covariates have non-zero coefficients.
%------------------------------------------------
	\subsection{CSUV assessment of uncertainty} \label{compare:ci}
	   The CSUV plot provides a graphical tool to illustrate both the selection and the estimation uncertainty in the coefficients. The uncertainty illustrated by the CSUV plot should be interpreted to originate from the randomness of $\epsilon$. This is similar to the classical confidence intervals in fixed-$p$, fixed-design regression. \par
	   In this section, our focus is on the uncertainty illustration by the default conditional boxes and whiskers, and on whether and how the information they carry can be used to assess the uncertainty in selection and estimation. Therefore our mentions of ``boxes'' or ``whiskers'' in this section refer to the conditional boxes and whiskers. Roughly speaking, the selection uncertainty is represented by the width of the boxes along the x-axis, and the estimation uncertainty is represented by the range of the boxes and whiskers along the y-axis. The plot provides a graphical aid to help users to decide whether to select a covariate by considering both dimensions of the corresponding box. The following similarities between the CSUV boxes and confidence intervals can be identified.
		\begin{itemize}
			\item Both provide intervals that likely cover the value of the true coefficient.
			\item Both aid the users in deciding if a covariate should be selected.
		\end{itemize} 
		However, we also highlight the following differences between the two.
		\begin{itemize}
			\item {\em Information content.} Unlike the classical confidence interval, which is one-dimensional, the CSUV box is two-dimensional: both its width and its range should be used in deciding whether or not to include the corresponding covariate. This is because the ranges of CSUV boxes only contain information on non-zero estimated coefficients (i.e. any zero estimates for the coefficient are not reflected in the range of the box, but only in its width). For this reason, a covariate that is rarely chosen (and in particular, is not selected by CSUV-m) may have a box plot that does not cross $0$. Therefore, the width of the box plot, which is directly proportional to the same-sign frequency with which the corresponding coefficient is selected, should also be considered in deciding whether or not to include the corresponding covariate in the model. \par
			\item {\em Covering percentiles.} The boxes in the CSUV plot represent the upper and the lower quartiles (i.e. 25\% and 75\% percentile) of the non-zero estimated coefficients. By contrast, classical confidence intervals are often considered in the context of much larger coverage; frequently, 90 or 95\%. With this in mind, we set the whiskers in the box plots to describe the [5\%, 95\%] range
(of the non-zero estimated coefficients) by default. This default range for the whiskers can be changed by users in the R package \verb+CSUV+.
		\end{itemize}
		Moreover, the box plots are based on the individual empirical estimated coefficients, and do not take into account the effect of the selection uncertainty in other covariates. For example, if covariates $X_1$ and $X_2$ are highly correlated, whether $X_2$ is selected affects the estimated coefficients of $X_1$. While the conditional approach considered by \cite{loftus2014significance} and \cite{tibshirani2016exact}, and the debiased approached considered by \cite{zhang2014confidence} in principle can be used here, the generalization to CSUV is not straightforward and the conditional approach is computationally intensive. \par
		The intertwining of the selection uncertainty and the estimation uncertainty makes it difficult to propose one simple interval that covers the true covariate with a given confidence level without a complicated adjustment e.g. as in \cite{loftus2014significance} or \cite{tibshirani2016exact}. We instead restrict ourselves to investigating if the whiskers are useful in deciding if a covariate selected by CSUV-m should be chosen, without providing a confidence level guarantee. \par
		Our investigation is as follows: using the simulated data from model settings 2-5 in Section \ref{syn}, for covariates selected by CSUV-m, we want to find out if the covariates for which the whiskers cover zero are more likely to be the false covariates. For each realization of the simulated data, we separate the CSUV-m selected covariates into two sets: (1) whiskers covering zero, and (2) whiskers not covering zero. We then find out the frequency with which the covariates in the two sets are the true covariates. \par
		\begin{figure}[!h]
			\begin{center}
				\includegraphics[width=0.9\linewidth]{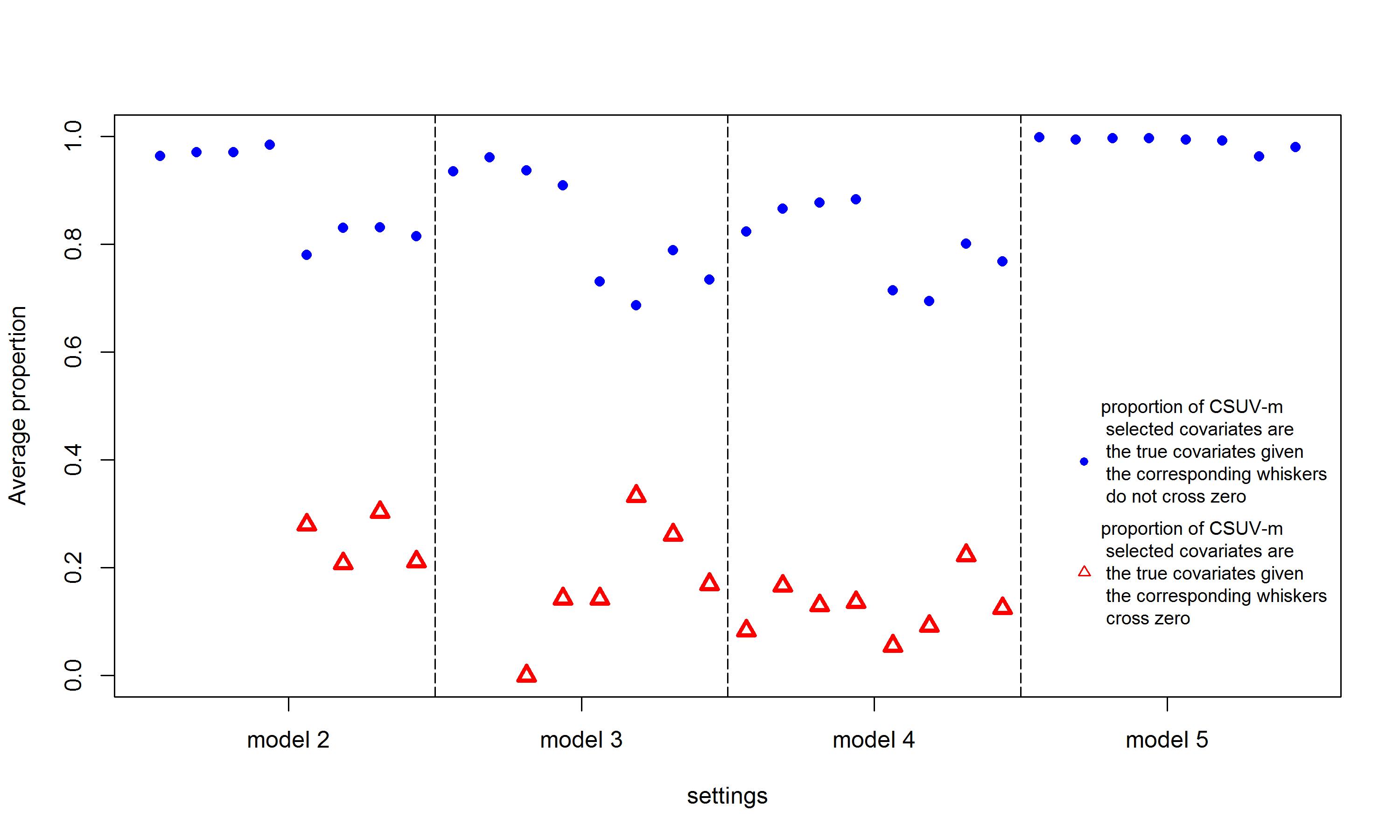}
				\caption{Average proportion of the CSUV-m selected covariates are the true covariates, using simulated data from simulation model 2-5 with eight different parameter settings under each model setting (see Section \ref{syn} for more details on the simulation settings). Circles (blue in the color version) show the average proportions of the CSUV-m selected covariates are the true covariates given the corresponding whiskers do not cross zero whereas the triangles (red in the color version) show the average proportions of the CSUV-m selected covariates are the true covariates given the corresponding whiskers cross zero. If there is no triangle for a particular setting, it means that none of the CSUV-m selected covariates have whiskers crossing zero.}
				\label{ci_cover}
			\end{center}
		\end{figure}
		 The simulation results show that a covariate with whiskers crossing zero is much more likely to be a false covariate than a covariate with whiskers not crossing zero (Figure \ref{ci_cover}). This indicates that observing if the whiskers of a covariate cross zero does provide useful information in deciding if the covariate is a true one. 
%------------------------------------------------
\section{Simulation study} \label{sim}
	In this section, we evaluate the performance of CSUV with numerical examples which consists of five simulated data settings and two real datasets. The main focus of our simulation is to compare the performance of CSUV with some model selection procedures including cross-validation and information criteria as they are popular approaches when there are different variable selection methods available. We also compare the performance of CSUV under different specifications (e.g. percentile parameter $q = 0$ vs $q = 5$, different constituent methods) to verify some claims we made in Section \ref{param}. 
%------------------------------------------------
	\subsection{Simulation settings}
		\subsubsection{R implementations} \label{implement_R}
			In the simulation, we consider CSUV with different sets of constituent methods:
			\begin{enumerate}
				\item Lasso, MCP and SCAD (default)
				\item Lasso, Elastic Net, relaxed Lasso, MCP and SCAD
				\item MCP
			\end{enumerate}
			The first set is our primary interest. When we mention CSUV without specifying the corresponding constituent methods, we implicitly assume that this set of methods is used. The second combination is used to verify the claim that adding some similar methods does not affect the performance too much. The third set is used to verify the claim that using more constituent methods in general provides better results. We use MCP here because in the majority of the simulation settings it has the best variable selection performance among the individual variable selection methods in terms of the F-measure and the number of false classifications. \par
			We use publicly available R packages for the implementation of the constituent methods (Lasso, Elastic Net, relaxed Lasso, MCP, SCAD) used in CSUV. See Table \ref{table:base_method} for the list of the corresponding R packages, functions and parameter settings used in the \verb+CSUV+ package and also in this simulation. The concavity values of SCAD and MCP are set to the value recommended by the original papers from \cite{fan2001variable} and \cite{zhang2010nearly} respectively, which are also the default values in the \verb+ncvreg+ R package. For Elastic Net, we use $\alpha = 0.5$. \par
			\begin{table}[ht]
				\begin{center}
					\resizebox{.95\textwidth}{!}{
						\begin{tabular}{ |l|l|l|l|l|c| } \hline
						Method 														& R package & R function		& Parameters		& $\lambda$ tuning\\ \hline
						Lasso \citep{tibshirani1996regression}							& glmnet 	& cv.glmnet		& 					& default 10-fold cross-validation \\ 
						Elastic Net \citep{zou2005regularization}						& glmnet 	& cv.glmnet 		& $\alpha$: 0.5		& default 10-fold cross-validation \\
						Relaxed Lasso \citep{meinshausen2007relaxed}				& relaxo 	& cvrelaxo 		& 					& default 5-fold cross-validation \\ 
						SCAD \citep{fan2001variable}									& ncvreg 	& cv.ncvreg 		& concavity: 3.7		& default 10-fold cross-validation\\ 
						MCP \citep{zhang2010nearly}									& ncvreg 	& cv.ncvreg 		& concavity: 3		& default 10-fold cross-validation\\ \hline
						\end{tabular}
						\label{method_table}
						}
					\end{center}
					\caption{Variable selection methods and the corresponding R packages and functions used in CSUV}
					\label{table:base_method}
				\end{table}
%------------------------------------------------
		\subsubsection{Methods to compare} \label{compare}
			 We use eBIC and delete-$n/2$ cross-validation as the major comparing methods to CSUV. We use eBIC instead of BIC as eBIC is designed for high-dimensional data. The details of the two methods are described in Section \ref{modsel}. We also include the simulation results of each constituent method (Lasso, Elastic Net, relaxed Lasso, MCP and SCAD), VSD-minimizing method \citep{yang2017toward} and BIC for readers' reference. \par
			 Both eBIC and delete-$n/2$ cross-validation uses the Lasso, MCP and SCAD (i.e. the methods used in the default case of CSUV) as the base methods. eBIC selects the fitted model that minimizes the corresponding information criterion value while delete-$n/2$ cross-validation selects the method that has the lowest estimated prediction error. The R packages and the parameter values used for the base methods are the same as what we use in CSUV for fair comparison. All the variable selection methods require tuning the regularization parameter $\lambda$. Default tuning in the R packages are used to simplify the analysis and the details of the tuning are shown in Table \ref{table:base_method}. eBIC and cross-validation have their own parameters and we set them as follow: For eBIC, we set $\gamma = 0.5$, which is one of the values considered in the simulations of the original paper \citep{chen2008extended} and the value used in \cite{lim2016estimation}. For the delete-$n/2$ cross-validation, we set the number of resampling $B = 100$, which is the same as the number of iterations we use in CSUV. \par
			For the VSD-minimizing method, we use the \verb+glmvsd+ R package to calculate the weight on each candidate model and then select the covariates that have aggregate weight greater than or equal to 0.5. Coefficients of the selected set from VSD is estimated using OLS. We use the default parameters in \verb+glmvsd+ (e.g. use the Lasso, MCP and SCAD to get the candidate models) except the weight which we use ARM instead. This is because using the default BIC to calculate the weight provides very poor results in some simulation settings.
%------------------------------------------------
		\subsubsection{Performance measures}
			For the datasets for which we know the true sets of covariates (i.e. simulated data and the modified real dataset), we compare the variable selection performance among different methods by the F-measure, the number of false positives (FP), number of false negatives (FN) and the total number of variable selection error (FP+FN). The F-measure is the harmonic mean of precision and recall:
				$$F = \frac{2}{\frac{1}{precision}+\frac{1}{recall}} = \frac{2}{\frac{TP+FP}{TP}+\frac{TP+FN}{TP}} = \frac{2TP}{2TP+FN+FP}$$ 
			Note that comparing the above numbers individually can be misleading. For example, using only FN favors models that select a large number of covariates and using only FP favors models that select fewer number of covariates. Although the F-measure takes both precision and recall into account, assigning same weight to precision and recall is arbitrary. Nevertheless, we use the F-measure as our major measure when we compare the variable selection performance between different methods. \cite{powers2011evaluation} provide a detailed comparison of different evaluation methods. \par
			Although our main focus is variable selection performance, we also compute the prediction mean square errors (MSE) on test set data and the coefficient estimation error ($l_1$ and $l_2$) for CSUV and the comparing methods.\par
%------------------------------------------------
		\subsubsection{Synthetic data \label{syn}} 
			Set $\bm{Y} = \bm{\tilde{X}}\bm{\beta} + \bm{\epsilon}$, $\epsilon_i \overset{i.i.d}{\sim} \mathcal{N}(0, \sigma^2)$. We generate observations with $100$ realizations of $\bm{X}$ using the model specifications below. We then normalize $\bm{X}$ to get $\bm{\tilde{X}}$ so that all covariates have mean $0$ and variance $1$. Except from Model 1, the number of observation is $n = 100$, the number of predictors $p = \{100, 300\}$, the number of true covariates $s = \{5, 10\}$ and $\sigma^2 = 1$. 
			\begin{itemize}
				\item \textbf{(Model 1) modified example 1 from the original Lasso paper \citep{tibshirani1996regression}:} $\bm{\beta} = \{3, 1.5, 0 ,0, 2, 0, 0, 0\}$, $p = 8$ and $n = 50$. Predictors $\bm{X}$ follow $\mathcal{N}(0, \Sigma)$, where $\Sigma_{k,m} = 0.5^{|k-m|}$ and $\sigma = \{1, 3, 6\}$. In the Lasso paper $n = 20$ but here we use $n = 50$ so that there are enough observations for subsampled fit. We include a more challenging SNR with $\sigma = 6$ ($\sigma = 3$ in the Lasso paper).
				\item \textbf{(Model 2) Toeplitz structure:} predictors $\bm{X}$ follow $\mathcal{N}(0, \Sigma)$, where $\Sigma$ is in Toeplitz structure with $\Sigma_{k,m} = \rho^{|k-m|}$ with $\rho = \{0, 0.9\}$.
				\item \textbf{(Model 3) block structure:} predictors $\bm{X}$ follow $\mathcal{N}(0, \Sigma)$, where $\Sigma$ is in block structure with $\Sigma_{k,m} = 1$ for $k = m$. For $k \ne m$, $\Sigma_{k,m} = 0$ except $mod_{10}(m) = mod_{10}(k)$ which $\Sigma_{k,m} = \{0.5, 0.9\}$.
				\item \textbf{(Model 4) factor model:} latent covariates $\phi_j, j = 1,...,J$ are i.i.d. and follow $\mathcal{N}(0, 1)$. Each covariate is generated by $X_k = \sum_{j = 1}^J f_{k,j}\phi_j +\eta_k$, where $f_{k,j}$, $\eta_k$ are i.i.d. and follow $\mathcal{N}(0, 1)$. The number of factor $J= \{2, 10\}$.
				\item \textbf{(Model 5) modified example from \cite{zhang2015cross}:} $\beta_j = 6/j$ for the true covariates $j=1,...,s$ and $\beta_j = 0$ otherwise. Predictors $\bm{X}$ follow $\mathcal{N}(0, \Sigma)$, where $\Sigma_{k,m} = \rho^{|k-m|}, \rho = \{0.5,-0.5\}$. The difference between \cite{zhang2015cross} and the model 5 here is that we use the same $n$ and $p$ as model 2-4.
			\end{itemize}
            For models 2-4, $\lfloor \frac{s}{2}\rfloor$ of the coefficient of the true $s$ are chosen randomly from $U(0.5, 1.5)$ and $\lceil \frac{s}{2}\rceil$ of them are chosen uniformly from $U(-1.5,-0.5)$. The true $\beta$s are chosen randomly among the predictors, and once the $\beta$s are set, the same set of $\beta$s are used for all realizations. 
%------------------------------------------------
		\subsubsection{Real datasets}
			\begin{myexample}[Boston housing data, \citeauthor{harrison1978hedonic}, \citeyear{harrison1978hedonic}]
				{\normalfont The dataset consists of the median value of owner-occupied homes as response and $p=13$ covariates (crime rate, proportion of residential land, etc). Number of observations is $n = 506$. The dataset is publicly available in R with the \verb+MASS+ package. For each simulation, half of the observations are used as the training data and the other half are used as the test set.}
			 \end{myexample}				
			\begin{myexample}[Modified riboflavin data] \label{ribo}
			    {\normalfont Here we re-examine the riboflavin dataset introduced in Example \ref{ribo1}. In order to assess the variable selection performance, we randomly permute all but 10 of the 4088 covariates in the riboflavin dataset across all the observations. The same permutation is used for all permuted covariates to keep the original dependence structure among them. The set of 10 unpermuted covariates is chosen randomly among the 200 covariates with the highest marginal correlation with the response.}
	 		 \end{myexample}
			 The modification for the riboflavin dataset ensures that the permuted covariates cannot be the true covariates in this modified dataset. In the simulation results, we refer the 10 unpermuted covariates as the ``true'' covariates, although in reality they may not be the true covariates. \par
			 For the Boston data, we repeat the process for $m = 100$ times with random cuttings of the training and the test data. For the riboflavin data, we repeat the process for $m = 100$ with random selection of the 10 unpermuted covariates to stabilize the results. 
%------------------------------------------------		
	\subsection{Simulation results} 
		The simulation results are summarized in Table \ref{1_1}-\ref{7_1} in the Appendix \ref{app:table}. Below we discuss the simulation results in detail. 
%------------------------------------------------
		\subsubsection{Verification of claims made in Section \ref{method}}
			In Section \ref{method}, we claim that:
			\begin{itemize}
				\item CSUV-m is designed for variable selection whereas CSUV-s is designed for better prediction.
				\item Performance of CSUV should be similar as long as $q$ is small (e.g. $q = 0$ or $q=5$).
				\item Including more (diverse) methods should improve the performance of CSUV.
				\item Including some similar methods should not worsen the performance of CSUV by much.
			\end{itemize}
			The simulation results support the claims above:
			\begin{itemize}
				\item CSUV-m vs CSUV-s: In general CSUV-m has better variable selection performance in terms of the F-measure. CSUV-s usually has a better prediction performance, and it also has a more stable (not too far off from the best method when CSUV is not performing particularly well) prediction performance in terms of MSE than CSUV-m. This may because CSUV-s selects a larger set of covariates than CSUV-m.
				\item $q = 0$ vs $q = 5$: the performance of CSUV-m when $q = 0$ and $q = 5$ is quite similar in terms of the number of covariates selected, and the prediction and variable selection performance, although $q = 0$ performs slightly better than $q = 5$.
				\item MCP only vs three different methods: here we only consider $q = 0$ as by using $q = 0$ we do not need to worry about the difference in terms of the number of fitted models selected (with $q = 5$ for example, the number of fitted models from three variable selection methods are around three times of the number of fitted models from a single method). In our simulation, CSUV using MCP only in general has worse performance than CSUV using three different constituent methods. In some other cases like the model 3 with parameter setting 7 and 8, both the prediction and variable selection performance of CSUV using MCP only is much worse than CSUV using three different constituent methods.
				\item Including some similar methods: here again we only consider $q = 0$. The results of CSUV using three different constituent methods (Lasso, MCP and SCAD) and five different methods (Lasso, Elastic Net, relaxed Lasso, MCP and SCAD, for which the Lasso, Elastic Net and relaxed Lasso are relatively similar) are very similar.				\end{itemize}
%------------------------------------------------
		\subsubsection{Comparing the performance between CSUV and some existing final model selection procedures}\label{compare:cv}
			In the majority of settings, CSUV-m has a better variable selection performance than the eBIC, delete-$n/2$ cross-validation and VSD-minimizing method in terms of the total number of variable selection error and the F-measure, and a better coefficient estimation performance in terms of the $l_1$ loss. For example, out of the 36 simulation settings that we know the true set of covariates (i.e. the simulated data and the modified riboflavin dataset), CSUV-m has a higher F-measure on 33 of the settings when comparing with the delete-$n/2$ cross-validation and 32 of the settings when comparing with eBIC. CSUV-m also has higher F-measure than VSD-minimizing method on 23 settings. CSUV-m usually selects the smallest set of covariates when comparing with eBIC or delete-$n/2$ cross-validation and the individual variable selection methods. In some cases like model 4 parameter setting 6, it selects a much smaller set of covariates than the truth. While this worsens the prediction performance of CSUV-m and we may view it as a limitation of CSUV-m, it may well due to the limitation of variable selection as a whole: Other methods which select much larger sets of covariates usually include a few more true covariates but inevitably they also include many more false covariates. They may perform better than CSUV-m in terms of prediction, but CSUV-m in general outperforms them in terms of variable selection. \par
			The performance of CSUV-s, on the other hands, is much more difficult to draw conclusions on. CSUV-s is better than delete-$n/2$ cross-validation in terms of variable selection. When comparing with eBIC, while it performs better than eBIC in one measure in some settings, it performs worse than eBIC in some other settings with the same measure. \par
			One encouraging result about CSUV is that in many simulation settings like model 2, CSUV-m outperforms not only the final model selections procedures but it also outperforms \textit{all} individual constituent methods in terms of the F-measure and the total number of variable selection error. In some simulation settings, CSUV performs better than the best individual variable selection method in terms of both prediction and variable selection measured by F-measure. For example in model 2, there are quite a few parameter settings (e.g. parameter setting 2) that the MSE of CSUV is lower and the F-measure is higher than all individual variable selection methods. \par
			For the variable selection performance on the real data, both versions of CSUV perform very well on the riboflavin data example. CSUV-s has the best performance in terms of F-measure and the total number of variable selection error.
%------------------------------------------------
	\subsection{Analysis of the selection by CSUV}				
		\subsubsection{Reasons for selected set to be small for CSUV-m}
			The number of covariates selected by CSUV-m is often small when comparing with other methods and the true size. Investigation into the collection of fitted models $\mathcal{M}$ shows that for many simulation settings, the fitted models in $\mathcal{M}$ can be very different in terms of variable selection. Sometimes all fitted models in $\mathcal{M}$ select different sets of covariates. When the selection decision is so different among $\mathcal{M}$, it is very likely that only a few covariates will have $\tau_j \ge 1/2$. This causes the number of covariates chosen by CSUV-m to be small. Whether a small selected set is desirable depends on the purpose of variable selection. Selecting small(er) number of covariates by this selection rule may cause omission of some true covariates and possibly exclusion of some false covariates that are helpful for prediction. This may result in poor prediction in some situations. On the other hand, the set of covariates selected by CSUV-m often includes fewer false positives than other variable selection methods, as only covariates that are selected by the majority of the subsampled fits are included in CSUV-m. 
%------------------------------------------------
\section{Conclusion} \label{conclusion}
	Many variable selection methods are available. However, there is no clear guideline on how to select which method to use with the data at hand, or how we can trust the set of covariates selected by a method. In practice, cross-validation and information criteria may be used to select the final models: \cite{zhang2015cross} advocate to use the delete-$n/2$ cross-validation and \cite{chen2008extended} extend the use of BIC to high-dimensional data (eBIC). \par
	In this paper we suggest a competitive alternative to these two procedures. We also provide a graphical illustration of the selection uncertainties. CSUV does not attempt to select the best method or to find the optimal regularization parameter. Instead we aggregate the fitted results from different variable selection methods via subsampling, and use a graphical tool to illustrate the uncertainties in selection and estimation. CSUV is very general and can be used with different variable selection methods. The simulation results show that CSUV in general outperforms the delete-$n/2$ cross-validation and eBIC in terms of variable selection. We also show that the graphical tool of CSUV has the capability to aid analysts in variable selection. \par
	
\bigskip
\begin{center}
    {\large\bf SUPPLEMENTARY MATERIAL}
\end{center}

\begin{description}
    \item[Appendix:] Proofs for Proposition \ref{propo:jaccard} and \ref{propo:sign}, heat maps to illustrate selection disagreements and detailed simulation results. (.pdf file)
    %\item[R-code for simulation results:] R code to generate the simulation results shown in this paper. (GNU zipped tar file)
\end{description}

\bibliographystyle{chicago}
\bibliography{csuv}

\appendix
	\section{Appendix}
		\subsection{Proof of Proposition \ref{propo:jaccard}} \label{prop:prove}
			\begin{proof}
				\begin{align*}
					&\operatorname*{arg\ min}_M \frac{1}{|\mathcal{M}|} \sum_{\{k|M_k\in \mathcal{M}\}} Hamming(M_k, M) \\
					\iff & \operatorname*{arg\ min}_M \frac{1}{|\mathcal{M}|} \sum_{\{k|M_k \in \mathcal{M}\}} \sum_{j=1}^p (\mathbbm{1}_{\beta_j^k= 0} \mathbbm{1}_{j\in M}+ \mathbbm{1}_{\
					beta_j^k \ne 0} \mathbbm{1}_{j\notin M}) \\
					\iff & \operatorname*{arg\ min}_M \frac{1}{|\mathcal{M}|} \sum_{j=1}^p \sum_{\{k|M_k \in \mathcal{M}\}} [(1-\mathbbm{1}_{\beta_j^k\ne 0}) \mathbbm{1}_{j\in M}+ \mathbbm{1}_{\beta_j^k\ne 0}(1- \mathbbm{1}_{j\in M})] \\
					\iff & \operatorname*{arg\ min}_M \frac{1}{|\mathcal{M}|} \sum_{j=1}^p \sum_{\{k|M_k \in \mathcal{M}\}} [\mathbbm{1}_{j\in M}(1-2\mathbbm{1}_{\beta_j^k\ne 0})] \\
					\iff & \operatorname*{arg\ min}_M \sum_{j=1}^P \mathbbm{1}_{j \in M} (1-2 \frac{ \sum_{\{k|M_k \in \mathcal{M}\}} \mathbbm{1}_{\beta_j^k\ne 0}}{|\mathcal{M}|}) \\
					\iff & \operatorname*{arg\ min}_M \sum_{j=1}^P \mathbbm{1}_{j \in M} (1-2 \tau_j)
				\end{align*}
			\end{proof}
		\newpage
		\subsection{Proof of Proposition \ref{propo:sign}} \label{prop:prove:sign}
			\begin{proof}
				Let the number of fitted models for covariance $j$ to be estimated with positive sign to be $n_j^+$ and those estimated with negative sign to be $n_j^-$. 
				\begin{align*}
					&\operatorname*{arg\ min}_M \frac{1}{|\mathcal{M}|} \sum_{\{k|M_k\in \mathcal{M}\}} dist(M_k, M) \\
					\iff & \frac{1}{|\mathcal{M}|} \operatorname*{arg\ min}_{\{s_j^M|j=1,...,p\}} \sum_{\{k|M_k \in \mathcal{M}\}}|s_j^M-sign(\beta_j^k)| \\
					\iff & \frac{1}{|\mathcal{M}|} \operatorname*{arg\ min}_{\{s_j^M|j=1,...,p\}} \sum_{\{k|M_k \in \mathcal{M}\}} [(|\mathcal{M}|-n_j^+ +n_j^-)\mathbbm{1}_{s_j^M = +1}+(n_j^+ + n_j^-)\mathbbm{1}_{s_j ^M= 0}+(|\mathcal{M}|+n_j^+ -n_j^- )\mathbbm{1}_{s_j^M=-1}]\\
					\iff & \operatorname*{arg\ min}_{\{s_j^M|j=1,...,p\}} \sum_{\{k|M_k \in \mathcal{M}\}} [(1-\tau_j^+ +\tau_j^-)\mathbbm{1}_{s_j^M = +1}+(\tau_j^+ + \tau_j^-)\mathbbm{1}_{s_j^M = 0}+(1-\tau_j^- +\tau_j^+ )\mathbbm{1}_{s_j^M=-1}] \\
				\end{align*}
					which is equivalent to setting
					$$s_j^M = 
						\begin{cases}
							+1 \text{ if } \tau_j^+ \geq 1/2\\
							-1 \text{ if } \tau_j^- \geq 1/2\\
							0 \text { otherwise}\\
						\end{cases}$$
			\end{proof}
		\subsection{Illustration of selection disagreement via Jaccard distance heatmap} \label{app:heatmap}
\begin{figure}[H]
			\begin{center}
 			\includegraphics[width=0.9\linewidth, scale=0.3]{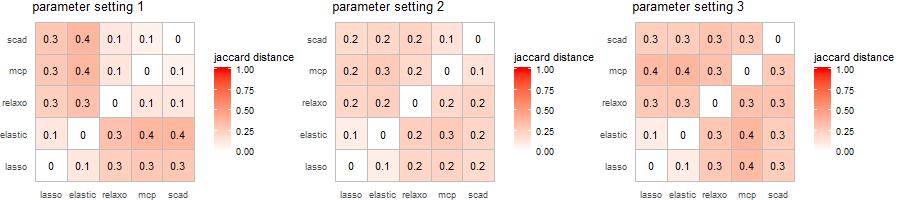}
 			\end{center}
 			\caption{Average Jaccard distance among fitted models from different methods using simulated data from model 1}
 			\label{jaccard:1}
		\end{figure}

\begin{figure}[!h]
			\begin{center}
 			\includegraphics[width=0.9\linewidth, scale=0.3]{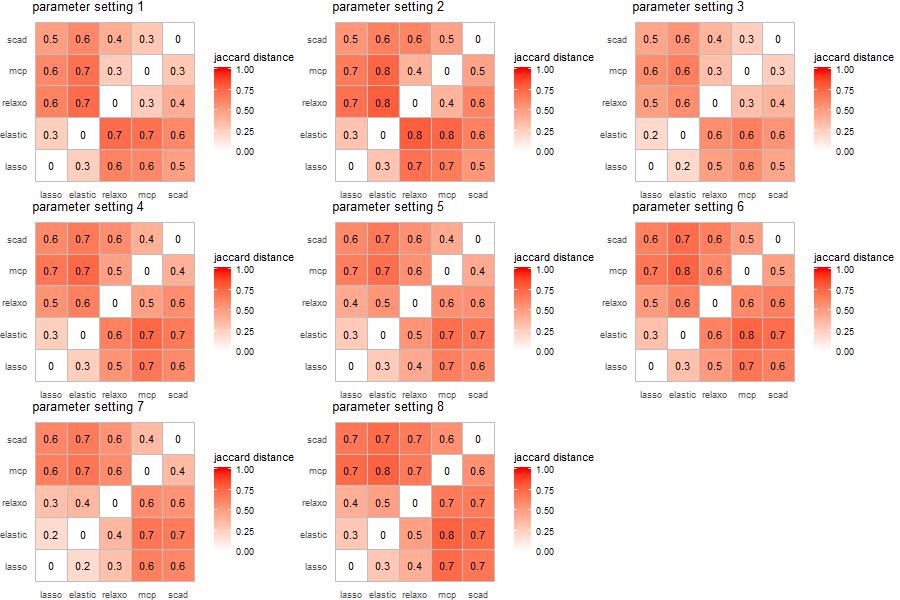}
 			\end{center}
 			\caption{Average Jaccard distance among fitted models from different methods using simulated data from model 2}
 			\label{jaccard:2}
		\end{figure}

\begin{figure}[!h]
			\begin{center}
 			\includegraphics[width=0.9\linewidth, scale=0.3]{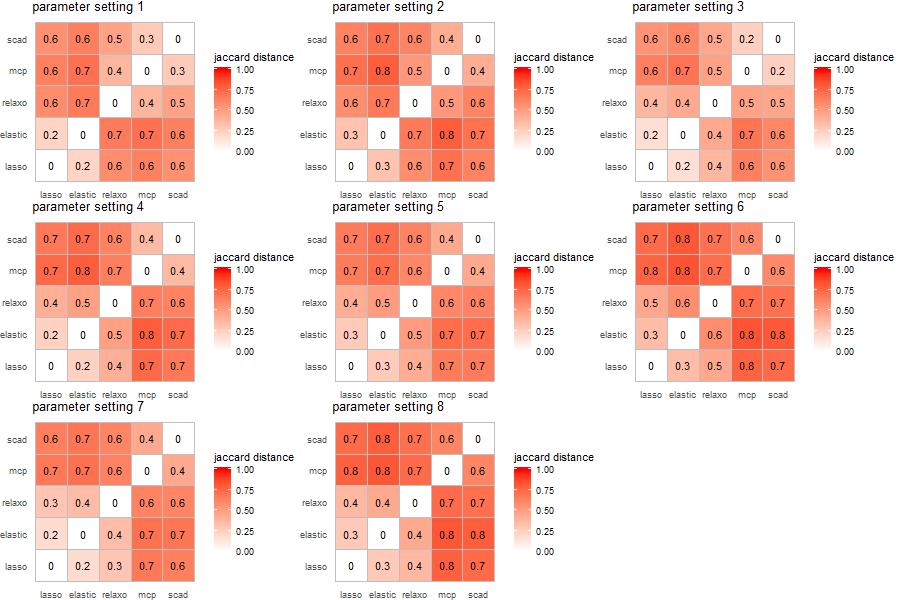}
 			\end{center}
 			\caption{Average Jaccard distance among fitted models from different methods using simulated data from model 3}
 			\label{jaccard:3}
		\end{figure}

\begin{figure}[H]
			\begin{center}
 			\includegraphics[width=0.9\linewidth, scale=0.3]{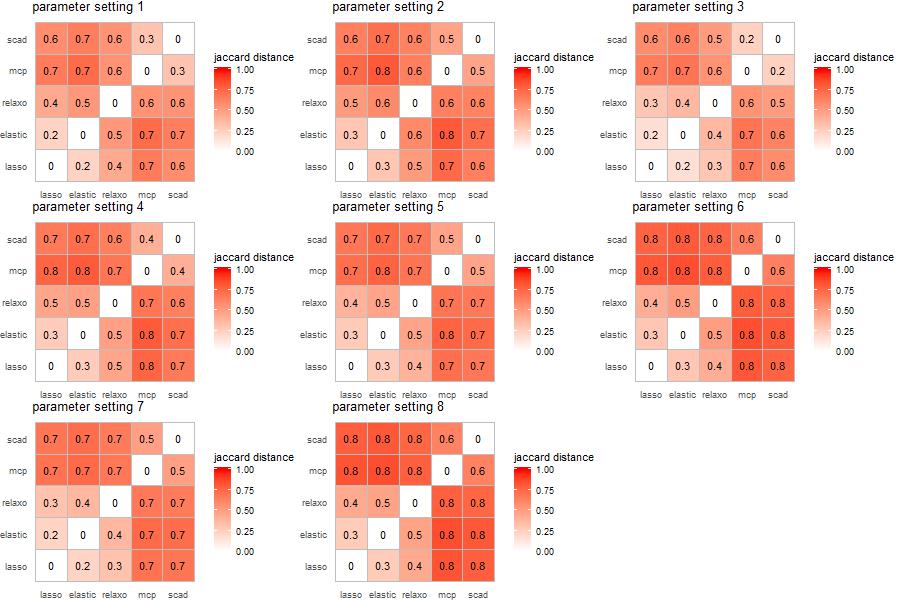}
 			\end{center}
 			\caption{Average Jaccard distance among fitted models from different methods using simulated data from model 4}
 			\label{jaccard:4}
		\end{figure}

\begin{figure}[H]
			\begin{center}
 			\includegraphics[width=0.9\linewidth, scale=0.3]{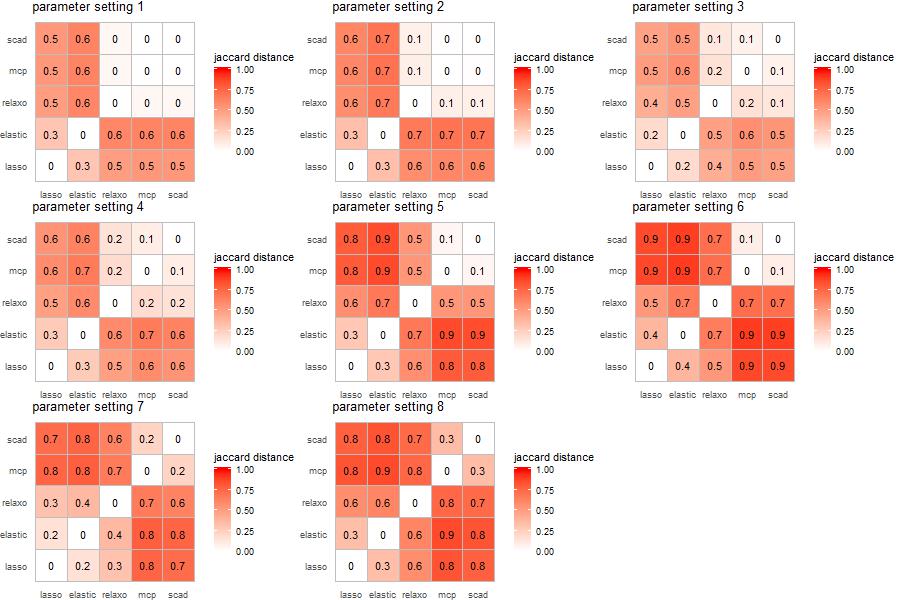}
 			\end{center}
 			\caption{Average Jaccard distance among fitted models from different methods using simulated data from model 5}
 			\label{jaccard:5}
		\end{figure}

			The heat maps in Figure \ref{jaccard:1}-\ref{jaccard:5} show the average Jaccard distance ($dist_{Jaccard}(A, B) = \frac{|A \bigtriangleup B|}{|A \cup B|}$, where $A\bigtriangleup B=(A\cup B)-(A\cap B)$) among selected sets from five different variable selection methods under simulation setting 1-5 (see Section 5 more details). The Jaccard distance shows the relative size of selection disagreement. It ranges from 0 to 1 and larger distance means larger dissimilarity between two selected sets. The Jaccard distance among some selection methods can be quite large, showing that relatively large number of covariates are selected by one method but not by another. Even for methods that are relatively similar like the Lasso and elastic Lasso, the difference among them is still not negligible for most settings.
		\newpage
		\subsection{Tables of the simulation results} \label{app:table}
			% latex table generated in R 3.6.1 by xtable 1.8-4 package
% Mon Jan 13 15:32:14 2020
\begin{table}[ht]
\centering
\begingroup\small
\scalebox{0.75}{
\begin{tabular}{|l|l|l|l|ll|lll|l|}
  \hline
parameters & methods & f & FP+FN & FP & FN & pred.err & l1.diff & l2.diff & size \\ 
  \hline
setting 1 & lasso & 0.76 (0.01) & 2.11 (0.14) & 2.11 & 0 & 1.13 (0.02) & 0.75 (0.04) & 0.39 (0.02) & 5.11 (0.14) \\ 
   & elastic net & 0.71 (0.01) & 2.67 (0.14) & 2.67 & 0 & 1.15 (0.03) & 0.85 (0.04) & 0.42 (0.02) & 5.67 (0.14) \\ 
  rho = 0.5 & relaxed lasso & 0.91 (0.01) & 0.7 (0.1) & 0.7 & 0 & 1.11 (0.03) & 0.6 (0.04) & 0.34 (0.02) & 3.7 (0.1) \\ 
  s = 3 & mcp & 0.94 (0.01) & 0.47 (0.1) & 0.47 & 0 & 1.1 (0.02) & 0.53 (0.04) & 0.32 (0.02) & 3.47 (0.1) \\ 
  p = 8 & scad & 0.94 (0.01) & 0.53 (0.11) & 0.53 & 0 & 1.09 (0.02) & 0.51 (0.04) & 0.31 (0.02) & 3.53 (0.11) \\ 
   & vsd & 0.98 (0) & 0.11 (0.03) & 0.11 & 0 & 1.08 (0.02) & 0.43 (0.02) & 0.28 (0.01) & 3.11 (0.03) \\ 
   & bic & 0.96 (0.01) & 0.35 (0.08) & 0.35 & 0 & 1.09 (0.02) & 0.5 (0.03) & 0.31 (0.02) & 3.35 (0.08) \\ 
   & \textbf{ebic} & \textbf{0.96} (0.01) & 0.35 (0.08) & 0.35 & 0 & \textbf{1.09} (0.02) & \textbf{0.5} (0.03) & \textbf{0.31} (0.02) & 3.35 (0.08) \\ 
   & \textbf{cv} & \underline{0.94} (0.01) & \underline{0.48} (0.11) & \underline{0.48} & 0 & \textbf{1.09} (0.03) & 0.51 (0.04) & \textbf{0.31} (0.02) & 3.48 (0.11) \\ 
   & \textbf{csuv.m.0} & \textbf{0.96} (0.01) & \textbf{0.27} (0.06) & \textbf{0.27} & 0 & \textbf{1.09} (0.02) & \textbf{0.5} (0.03) & \textbf{0.31} (0.02) & 3.27 (0.06) \\ 
   & \textbf{csuv.s.0} & \underline{0.94} (0.01) & 0.42 (0.06) & 0.42 & 0 & \underline{1.1} (0.02) & \underline{0.55} (0.03) & \underline{0.33} (0.02) & 3.42 (0.06) \\ 
   & csuv.m.5 & 0.96 (0.01) & 0.27 (0.06) & 0.27 & 0 & 1.09 (0.02) & 0.5 (0.03) & 0.31 (0.02) & 3.27 (0.06) \\ 
   & csuv.s.5 & 0.94 (0.01) & 0.42 (0.06) & 0.42 & 0 & 1.1 (0.02) & 0.55 (0.03) & 0.33 (0.02) & 3.42 (0.06) \\ 
   & csuv.m.0.all & 0.95 (0.01) & 0.38 (0.07) & 0.38 & 0 & 1.1 (0.02) & 0.53 (0.03) & 0.33 (0.02) & 3.38 (0.07) \\ 
   & csuv.s.0.all & 0.92 (0.01) & 0.56 (0.06) & 0.56 & 0 & 1.11 (0.02) & 0.59 (0.03) & 0.35 (0.02) & 3.56 (0.06) \\ 
   & csuv.m.0.mcp & 0.97 (0.01) & 0.21 (0.06) & 0.21 & 0 & 1.09 (0.02) & 0.47 (0.03) & 0.29 (0.02) & 3.21 (0.06) \\ 
   & csuv.s.0.mcp & 0.97 (0.01) & 0.2 (0.05) & 0.2 & 0 & 1.08 (0.02) & 0.46 (0.03) & 0.29 (0.02) & 3.2 (0.05) \\ 
   \hline
setting 2 & lasso & 0.76 (0.01) & 2.13 (0.14) & 2.11 & 0.02 & 10.12 (0.21) & 2.26 (0.11) & 1.17 (0.05) & 5.09 (0.14) \\ 
   & elastic net & 0.73 (0.01) & 2.49 (0.14) & 2.48 & 0.01 & 10.25 (0.23) & 2.39 (0.11) & 1.2 (0.04) & 5.47 (0.14) \\ 
  rho = 0.5 & relaxed lasso & 0.87 (0.01) & 1.01 (0.12) & 0.96 & 0.05 & 10.11 (0.23) & 1.99 (0.12) & 1.12 (0.05) & 3.91 (0.12) \\ 
  s = 3 & mcp & 0.84 (0.01) & 1.29 (0.13) & 1.15 & 0.14 & 10.38 (0.23) & 2.3 (0.14) & 1.28 (0.06) & 4.01 (0.15) \\ 
  p = 8 & scad & 0.82 (0.01) & 1.53 (0.13) & 1.48 & 0.05 & 10.31 (0.23) & 2.26 (0.13) & 1.26 (0.06) & 4.43 (0.14) \\ 
   & vsd & 0.87 (0.01) & 0.65 (0.07) & 0.08 & 0.57 & 10.79 (0.27) & 2.18 (0.13) & 1.42 (0.08) & 2.51 (0.06) \\ 
   & bic & 0.86 (0.01) & 1.04 (0.11) & 0.92 & 0.12 & 10.24 (0.22) & 2.11 (0.12) & 1.22 (0.06) & 3.8 (0.12) \\ 
   & \textbf{ebic} & 0.86 (0.01) & 1.06 (0.11) & 0.94 & \underline{0.12} & \underline{10.22} (0.22) & 2.11 (0.12) & \underline{1.22} (0.06) & 3.82 (0.13) \\ 
   & \textbf{cv} & \underline{0.76} (0.01) & \underline{2.12} (0.14) & \underline{2.09} & \textbf{0.03} & 10.13 (0.22) & \underline{2.28} (0.11) & 1.18 (0.05) & 5.06 (0.14) \\ 
   & \textbf{csuv.m.0} & \textbf{0.91} (0.01) & \textbf{0.65} (0.09) & \textbf{0.56} & 0.09 & 10.09 (0.22) & \textbf{1.83} (0.11) & \textbf{1.09} (0.06) & 3.47 (0.08) \\ 
   & \textbf{csuv.s.0} & 0.9 (0.01) & 0.72 (0.08) & 0.64 & 0.08 & \textbf{10.08} (0.21) & 1.88 (0.1) & 1.11 (0.05) & 3.56 (0.08) \\ 
   & csuv.m.5 & 0.91 (0.01) & 0.65 (0.09) & 0.56 & 0.09 & 10.09 (0.22) & 1.83 (0.11) & 1.09 (0.06) & 3.47 (0.08) \\ 
   & csuv.s.5 & 0.9 (0.01) & 0.72 (0.08) & 0.64 & 0.08 & 10.08 (0.21) & 1.88 (0.1) & 1.11 (0.05) & 3.56 (0.08) \\ 
   & csuv.m.0.all & 0.91 (0.01) & 0.63 (0.09) & 0.57 & 0.06 & 10.09 (0.22) & 1.83 (0.11) & 1.09 (0.06) & 3.51 (0.08) \\ 
   & csuv.s.0.all & 0.9 (0.01) & 0.73 (0.08) & 0.67 & 0.06 & 10 (0.21) & 1.86 (0.1) & 1.1 (0.05) & 3.61 (0.07) \\ 
   & csuv.m.0.mcp & 0.9 (0.01) & 0.74 (0.1) & 0.57 & 0.17 & 10.25 (0.23) & 1.99 (0.13) & 1.18 (0.07) & 3.4 (0.11) \\ 
   & csuv.s.0.mcp & 0.9 (0.01) & 0.69 (0.09) & 0.55 & 0.14 & 10.15 (0.22) & 1.9 (0.12) & 1.14 (0.06) & 3.41 (0.08) \\ 
   \hline
setting 3 & lasso & 0.73 (0.02) & 2.16 (0.14) & 1.86 & 0.3 & 40.51 (0.85) & 4.33 (0.21) & 2.28 (0.09) & 4.56 (0.16) \\ 
   & elastic net & 0.71 (0.01) & 2.39 (0.14) & 2.14 & 0.25 & 40.66 (0.9) & 4.37 (0.2) & 2.25 (0.08) & 4.89 (0.16) \\ 
  rho = 0.5 & relaxed lasso & 0.73 (0.02) & 1.78 (0.12) & 1.13 & 0.65 & 41.41 (0.91) & 4.52 (0.21) & 2.51 (0.1) & 3.48 (0.16) \\ 
  s = 3 & mcp & 0.66 (0.02) & 2.26 (0.13) & 1.41 & 0.85 & 42.76 (0.98) & 5.54 (0.23) & 2.97 (0.1) & 3.56 (0.2) \\ 
  p = 8 & scad & 0.7 (0.02) & 2.22 (0.14) & 1.67 & 0.55 & 42.47 (0.92) & 5.51 (0.22) & 2.94 (0.1) & 4.12 (0.17) \\ 
   & vsd & 0.54 (0.01) & 2.24 (0.06) & 0.01 & 2.23 & 48.66 (1.21) & 5.94 (0.14) & 3.53 (0.08) & 0.78 (0.06) \\ 
   & bic & 0.69 (0.02) & 1.9 (0.12) & 1.03 & 0.87 & 42.24 (0.95) & 4.87 (0.21) & 2.72 (0.1) & 3.16 (0.16) \\ 
   & \textbf{ebic} & \underline{0.69} (0.02) & 1.9 (0.12) & 1.03 & 0.87 & 42.24 (0.95) & \underline{4.87} (0.21) & \underline{2.72} (0.1) & 3.16 (0.16) \\ 
   & \textbf{cv} & \textbf{0.73} (0.02) & \underline{2.16} (0.14) & \underline{1.86} & \textbf{0.3} & \textbf{40.51} (0.85) & \textbf{4.33} (0.21) & \textbf{2.28} (0.09) & 4.56 (0.16) \\ 
   & \textbf{csuv.m.0} & \textbf{0.73} (0.02) & \textbf{1.45} (0.09) & \textbf{0.43} & \underline{1.02} & \underline{42.38} (0.91) & 4.47 (0.21) & 2.69 (0.1) & 2.41 (0.11) \\ 
   & \textbf{csuv.s.0} & \textbf{0.73} (0.02) & 1.53 (0.11) & 0.63 & 0.9 & 42.16 (0.9) & 4.46 (0.22) & 2.6 (0.1) & 2.73 (0.1) \\ 
   & csuv.m.5 & 0.73 (0.02) & 1.45 (0.09) & 0.43 & 1.02 & 42.38 (0.91) & 4.47 (0.21) & 2.69 (0.1) & 2.41 (0.11) \\ 
   & csuv.s.5 & 0.73 (0.02) & 1.53 (0.11) & 0.63 & 0.9 & 42.16 (0.9) & 4.46 (0.22) & 2.6 (0.1) & 2.73 (0.1) \\ 
   & csuv.m.0.all & 0.74 (0.02) & 1.41 (0.1) & 0.44 & 0.97 & 42.15 (0.89) & 4.39 (0.21) & 2.65 (0.1) & 2.47 (0.11) \\ 
   & csuv.s.0.all & 0.74 (0.02) & 1.51 (0.1) & 0.65 & 0.86 & 41.99 (0.87) & 4.4 (0.2) & 2.6 (0.1) & 2.79 (0.1) \\ 
   & csuv.m.0.mcp & 0.69 (0.02) & 1.59 (0.09) & 0.36 & 1.23 & 43.2 (0.96) & 4.77 (0.21) & 2.85 (0.1) & 2.13 (0.11) \\ 
   & csuv.s.0.mcp & 0.7 (0.02) & 1.6 (0.09) & 0.45 & 1.15 & 42.74 (0.95) & 4.77 (0.22) & 2.8 (0.1) & 2.3 (0.1) \\ 
   \hline
\multicolumn{10}{l}{}\\
\end{tabular}
}
\endgroup
\caption{Model 1: performance of CSUV and methods it compares with. Variable selection performance in terms of F-measure (f), total error (FP+FN), false positives (FP) and false negatives (FN), prediction error in terms of mse (pred.err) and estimation error in terms of l1 and l2 distance (l1.diff and l2.diff) and are shown. The numbers are based on 100 simulations. The last 8 rows are the performance of CSUV with different parameters (e.g. csuv.m.0.mcp corresponds to CSUV with MCP as constituent method and r = 0). A bold number represents the best result among delete-n/2 cross validation, eBIC and CSUV using Lasso, MCP and SCAD while a underlined number represents the worst among them. Standard errors are shown inside the parentheses.} 
\label{1_1}
\end{table}

			% latex table generated in R 3.6.1 by xtable 1.8-4 package
% Mon Jan 13 15:32:14 2020
\begin{table}[ht]
\centering
\begingroup\small
\scalebox{0.75}{
\begin{tabular}{|l|l|l|l|ll|lll|l|}
  \hline
parameters & methods & f & FP+FN & FP & FN & pred.err & l1.diff & l2.diff & size \\ 
  \hline
setting 1 & lasso & 0.43 (0.01) & 15.7 (0.89) & 15.7 & 0 & 1.28 (0.02) & 1.76 (0.07) & 0.53 (0.01) & 20.7 (0.89) \\ 
   & elastic net & 0.35 (0.01) & 21.2 (0.91) & 21.2 & 0 & 1.36 (0.02) & 2.22 (0.07) & 0.6 (0.01) & 26.2 (0.91) \\ 
  rho = 0 & relaxed lasso & 0.88 (0.02) & 2.1 (0.47) & 2.05 & 0.05 & 1.14 (0.02) & 0.81 (0.07) & 0.35 (0.02) & 7 (0.47) \\ 
  s = 5 & mcp & 0.85 (0.02) & 2.3 (0.32) & 2.3 & 0 & 1.1 (0.02) & 0.68 (0.04) & 0.3 (0.01) & 7.3 (0.32) \\ 
  p = 100 & scad & 0.66 (0.02) & 6.16 (0.4) & 6.16 & 0 & 1.09 (0.02) & 0.75 (0.03) & 0.3 (0.01) & 11.16 (0.4) \\ 
   & vsd & 0.99 (0) & 0.09 (0.03) & 0.02 & 0.07 & 1.07 (0.02) & 0.44 (0.02) & 0.24 (0.01) & 4.95 (0.03) \\ 
   & bic & 0.77 (0.02) & 3.6 (0.33) & 3.6 & 0 & 1.09 (0.02) & 0.7 (0.03) & 0.3 (0.01) & 8.6 (0.33) \\ 
   & \textbf{ebic} & 0.85 (0.02) & 2.31 (0.31) & 2.31 & \textbf{0} & 1.09 (0.02) & 0.67 (0.04) & 0.3 (0.01) & 7.31 (0.31) \\ 
   & \textbf{cv} & \underline{0.76} (0.02) & \underline{4.28} (0.46) & \underline{4.28} & \textbf{0} & 1.09 (0.02) & \underline{0.7} (0.03) & 0.29 (0.01) & 9.28 (0.46) \\ 
   & \textbf{csuv.m.0} & \textbf{0.98} (0) & \textbf{0.2} (0.04) & \textbf{0.18} & \underline{0.02} & \textbf{1.06} (0.02) & \textbf{0.47} (0.02) & \textbf{0.25} (0.01) & 5.16 (0.04) \\ 
   & \textbf{csuv.s.0} & 0.93 (0.01) & 0.81 (0.09) & 0.79 & \underline{0.02} & \underline{1.1} (0.02) & 0.62 (0.03) & \underline{0.32} (0.01) & 5.77 (0.09) \\ 
   & csuv.m.5 & 0.99 (0) & 0.15 (0.04) & 0.12 & 0.03 & 1.06 (0.02) & 0.45 (0.02) & 0.25 (0.01) & 5.09 (0.04) \\ 
   & csuv.s.5 & 0.86 (0.01) & 1.63 (0.1) & 1.62 & 0.01 & 1.14 (0.02) & 0.81 (0.03) & 0.38 (0.01) & 6.61 (0.1) \\ 
   & csuv.m.0.all & 0.98 (0) & 0.21 (0.04) & 0.19 & 0.02 & 1.06 (0.02) & 0.47 (0.02) & 0.25 (0.01) & 5.17 (0.04) \\ 
   & csuv.s.0.all & 0.92 (0.01) & 0.92 (0.09) & 0.9 & 0.02 & 1.11 (0.02) & 0.65 (0.03) & 0.33 (0.01) & 5.88 (0.09) \\ 
   & csuv.m.0.mcp & 0.99 (0) & 0.09 (0.03) & 0.04 & 0.05 & 1.06 (0.02) & 0.43 (0.02) & 0.24 (0.01) & 4.99 (0.03) \\ 
   & csuv.s.0.mcp & 0.97 (0) & 0.29 (0.05) & 0.28 & 0.01 & 1.07 (0.02) & 0.49 (0.02) & 0.26 (0.01) & 5.27 (0.05) \\ 
   \hline
setting 2 & lasso & 0.33 (0.01) & 24.65 (1.43) & 24.64 & 0.01 & 1.43 (0.03) & 2.41 (0.1) & 0.65 (0.02) & 29.63 (1.44) \\ 
   & elastic net & 0.26 (0.01) & 32.75 (1.44) & 32.74 & 0.01 & 1.59 (0.03) & 3.05 (0.1) & 0.75 (0.02) & 37.73 (1.44) \\ 
  rho = 0 & relaxed lasso & 0.78 (0.02) & 4.25 (0.81) & 4.12 & 0.13 & 1.28 (0.03) & 1.3 (0.09) & 0.49 (0.02) & 8.99 (0.81) \\ 
  s = 5 & mcp & 0.75 (0.02) & 4.24 (0.39) & 4.23 & 0.01 & 1.13 (0.02) & 0.87 (0.04) & 0.36 (0.01) & 9.22 (0.39) \\ 
  p = 300 & scad & 0.5 (0.02) & 12.21 (0.72) & 12.2 & 0.01 & 1.14 (0.02) & 1.1 (0.04) & 0.37 (0.01) & 17.19 (0.72) \\ 
   & vsd & 0.94 (0.01) & 0.5 (0.09) & 0 & 0.5 & 1.28 (0.05) & 0.76 (0.06) & 0.43 (0.03) & 4.5 (0.09) \\ 
   & bic & 0.55 (0.02) & 12.38 (1.43) & 12.37 & 0.01 & 1.19 (0.02) & 1.32 (0.11) & 0.41 (0.02) & 17.36 (1.43) \\ 
   & \textbf{ebic} & 0.74 (0.02) & 5.84 (1.22) & 5.83 & \textbf{0.01} & 1.15 (0.02) & 0.98 (0.09) & 0.37 (0.01) & 10.82 (1.22) \\ 
   & \textbf{cv} & \underline{0.61} (0.02) & \underline{8.92} (0.75) & \underline{8.91} & \textbf{0.01} & 1.13 (0.02) & \underline{1.01} (0.05) & 0.36 (0.01) & 13.9 (0.75) \\ 
   & \textbf{csuv.m.0} & \textbf{0.97} (0.01) & \textbf{0.28} (0.06) & \textbf{0.13} & \underline{0.15} & \textbf{1.11} (0.02) & \textbf{0.58} (0.03) & \textbf{0.31} (0.02) & 4.98 (0.06) \\ 
   & \textbf{csuv.s.0} & 0.86 (0.01) & 1.71 (0.12) & 1.65 & 0.06 & \underline{1.19} (0.02) & 0.95 (0.04) & \underline{0.44} (0.01) & 6.59 (0.13) \\ 
   & csuv.m.5 & 0.97 (0.01) & 0.3 (0.05) & 0.11 & 0.19 & 1.12 (0.02) & 0.6 (0.04) & 0.33 (0.02) & 4.92 (0.05) \\ 
   & csuv.s.5 & 0.78 (0.01) & 3.02 (0.15) & 3 & 0.02 & 1.25 (0.02) & 1.24 (0.04) & 0.51 (0.01) & 7.98 (0.15) \\ 
   & csuv.m.0.all & 0.97 (0.01) & 0.31 (0.06) & 0.15 & 0.16 & 1.12 (0.02) & 0.59 (0.04) & 0.32 (0.02) & 4.99 (0.05) \\ 
   & csuv.s.0.all & 0.85 (0.01) & 1.87 (0.14) & 1.81 & 0.06 & 1.19 (0.02) & 1 (0.04) & 0.46 (0.01) & 6.75 (0.14) \\ 
   & csuv.m.0.mcp & 0.97 (0.01) & 0.24 (0.05) & 0.01 & 0.23 & 1.13 (0.03) & 0.58 (0.03) & 0.33 (0.02) & 4.78 (0.05) \\ 
   & csuv.s.0.mcp & 0.92 (0.01) & 0.83 (0.07) & 0.72 & 0.11 & 1.15 (0.02) & 0.74 (0.03) & 0.38 (0.01) & 5.61 (0.09) \\ 
   \hline
setting 3 & lasso & 0.46 (0.01) & 24.88 (0.93) & 24.88 & 0 & 1.55 (0.03) & 3.23 (0.1) & 0.73 (0.01) & 34.88 (0.93) \\ 
   & elastic net & 0.42 (0.01) & 29.6 (0.89) & 29.6 & 0 & 1.65 (0.03) & 3.74 (0.1) & 0.79 (0.01) & 39.6 (0.89) \\ 
  rho = 0 & relaxed lasso & 0.8 (0.01) & 5.53 (0.46) & 5.49 & 0.04 & 1.36 (0.03) & 1.95 (0.08) & 0.58 (0.02) & 15.45 (0.46) \\ 
  s = 10 & mcp & 0.88 (0.01) & 3.1 (0.31) & 3.1 & 0 & 1.19 (0.02) & 1.23 (0.05) & 0.42 (0.01) & 13.1 (0.31) \\ 
  p = 100 & scad & 0.7 (0.01) & 8.98 (0.41) & 8.98 & 0 & 1.2 (0.02) & 1.4 (0.05) & 0.43 (0.01) & 18.98 (0.41) \\ 
   & vsd & 0.98 (0.01) & 0.4 (0.09) & 0.1 & 0.3 & 1.25 (0.05) & 1.06 (0.06) & 0.43 (0.03) & 9.8 (0.08) \\ 
   & bic & 0.83 (0.01) & 5.28 (0.72) & 5.28 & 0 & 1.22 (0.03) & 1.41 (0.1) & 0.44 (0.02) & 15.28 (0.72) \\ 
   & \textbf{ebic} & 0.87 (0.01) & 3.27 (0.33) & 3.27 & \textbf{0} & \textbf{1.19} (0.02) & 1.24 (0.05) & \textbf{0.42} (0.01) & 13.27 (0.33) \\ 
   & \textbf{cv} & \underline{0.63} (0.02) & \underline{14.87} (1.16) & \underline{14.87} & \textbf{0} & \underline{1.38} (0.03) & \underline{2.25} (0.14) & \underline{0.58} (0.02) & 24.87 (1.16) \\ 
   & \textbf{csuv.m.0} & \textbf{0.97} (0) & \textbf{0.49} (0.08) & \textbf{0.27} & \underline{0.22} & 1.23 (0.04) & \textbf{1.07} (0.05) & 0.43 (0.02) & 10.05 (0.09) \\ 
   & \textbf{csuv.s.0} & 0.87 (0.01) & 3.19 (0.16) & 3.16 & 0.03 & 1.31 (0.03) & 1.64 (0.05) & 0.54 (0.01) & 13.13 (0.16) \\ 
   & csuv.m.5 & 0.98 (0) & 0.46 (0.08) & 0.25 & 0.21 & 1.22 (0.04) & 1.06 (0.05) & 0.42 (0.02) & 10.04 (0.08) \\ 
   & csuv.s.5 & 0.83 (0) & 4.29 (0.15) & 4.27 & 0.02 & 1.35 (0.03) & 1.86 (0.05) & 0.58 (0.01) & 14.25 (0.15) \\ 
   & csuv.m.0.all & 0.98 (0) & 0.47 (0.07) & 0.3 & 0.17 & 1.21 (0.03) & 1.05 (0.04) & 0.42 (0.02) & 10.13 (0.08) \\ 
   & csuv.s.0.all & 0.86 (0.01) & 3.47 (0.16) & 3.44 & 0.03 & 1.32 (0.03) & 1.7 (0.05) & 0.55 (0.01) & 13.41 (0.17) \\ 
   & csuv.m.0.mcp & 0.98 (0) & 0.36 (0.07) & 0.03 & 0.33 & 1.25 (0.04) & 1.05 (0.05) & 0.43 (0.02) & 9.7 (0.07) \\ 
   & csuv.s.0.mcp & 0.94 (0) & 1.26 (0.08) & 1.19 & 0.07 & 1.24 (0.04) & 1.25 (0.04) & 0.47 (0.02) & 11.12 (0.08) \\ 
   \hline
\multicolumn{10}{l}{}\\
\end{tabular}
}
\endgroup
\caption{Model 2: performance of CSUV and methods it compares with. Variable selection performance in terms of F-measure (f), total error (FP+FN), false positives (FP) and false negatives (FN), prediction error in terms of mse (pred.err) and estimation error in terms of l1 and l2 distance (l1.diff and l2.diff) and are shown. The numbers are based on 100 simulations. The last 8 rows are the performance of CSUV with different parameters (e.g. csuv.m.0.mcp corresponds to CSUV with MCP as constituent method and r = 0). A bold number represents the best result among delete-n/2 cross validation, eBIC and CSUV using Lasso, MCP and SCAD while a underlined number represents the worst among them. Standard errors are shown inside the parentheses.} 
\label{2_1}
\end{table}

			% latex table generated in R 3.6.1 by xtable 1.8-4 package
% Mon Jan 13 15:32:14 2020
\begin{table}[ht]
\centering
\begingroup\small
\scalebox{0.75}{
\begin{tabular}{|l|l|l|l|ll|lll|l|}
  \hline
parameters & methods & f & FP+FN & FP & FN & pred.err & l1.diff & l2.diff & size \\ 
  \hline
setting 4 & lasso & 0.38 (0.01) & 36.64 (1.56) & 36.6 & 0.04 & 1.88 (0.05) & 4.49 (0.14) & 0.94 (0.02) & 46.56 (1.56) \\ 
   & elastic net & 0.31 (0.01) & 47.38 (1.54) & 47.34 & 0.04 & 2.15 (0.05) & 5.62 (0.14) & 1.08 (0.02) & 57.3 (1.55) \\ 
  rho = 0 & relaxed lasso & 0.66 (0.01) & 11.84 (0.9) & 11.61 & 0.23 & 1.7 (0.05) & 3.2 (0.13) & 0.81 (0.02) & 21.38 (0.93) \\ 
  s = 10 & mcp & 0.81 (0.01) & 5.19 (0.35) & 5.16 & 0.03 & 1.24 (0.03) & 1.46 (0.06) & 0.48 (0.02) & 15.13 (0.35) \\ 
  p = 300 & scad & 0.57 (0.01) & 15.88 (0.53) & 15.86 & 0.02 & 1.23 (0.03) & 1.74 (0.06) & 0.48 (0.02) & 25.84 (0.53) \\ 
   & vsd & 0.88 (0.02) & 2.08 (0.28) & 0.03 & 2.05 & 2.36 (0.24) & 2.33 (0.22) & 0.9 (0.08) & 7.98 (0.29) \\ 
   & bic & 0.6 (0.02) & 19.21 (2) & 19.19 & 0.02 & 1.39 (0.05) & 2.44 (0.2) & 0.58 (0.03) & 29.17 (2.01) \\ 
   & \textbf{ebic} & 0.78 (0.02) & 8.05 (1.42) & 8.02 & \textbf{0.03} & \textbf{1.28} (0.03) & \textbf{1.75} (0.15) & \textbf{0.51} (0.02) & 17.99 (1.42) \\ 
   & \textbf{cv} & \underline{0.41} (0.01) & \underline{33.78} (1.67) & \underline{33.74} & 0.04 & \underline{1.8} (0.05) & \underline{4.18} (0.16) & \underline{0.88} (0.02) & 43.7 (1.67) \\ 
   & \textbf{csuv.m.0} & \textbf{0.91} (0.01) & \textbf{1.51} (0.15) & \textbf{0.1} & \underline{1.41} & 1.77 (0.08) & 1.84 (0.1) & 0.8 (0.04) & 8.69 (0.13) \\ 
   & \textbf{csuv.s.0} & 0.76 (0.01) & 6.53 (0.25) & 6.29 & 0.24 & 1.5 (0.04) & 2.35 (0.07) & 0.7 (0.02) & 16.05 (0.27) \\ 
   & csuv.m.5 & 0.91 (0.01) & 1.59 (0.15) & 0.08 & 1.51 & 1.82 (0.08) & 1.89 (0.11) & 0.82 (0.04) & 8.57 (0.14) \\ 
   & csuv.s.5 & 0.72 (0.01) & 7.75 (0.23) & 7.51 & 0.24 & 1.55 (0.04) & 2.57 (0.07) & 0.73 (0.02) & 17.27 (0.26) \\ 
   & csuv.m.0.all & 0.92 (0.01) & 1.48 (0.15) & 0.14 & 1.34 & 1.75 (0.08) & 1.81 (0.11) & 0.79 (0.04) & 8.8 (0.12) \\ 
   & csuv.s.0.all & 0.74 (0.01) & 7.26 (0.27) & 7.03 & 0.23 & 1.53 (0.04) & 2.47 (0.07) & 0.72 (0.02) & 16.8 (0.29) \\ 
   & csuv.m.0.mcp & 0.78 (0.01) & 3.37 (0.18) & 0.01 & 3.36 & 2.86 (0.13) & 3.21 (0.14) & 1.31 (0.04) & 6.65 (0.18) \\ 
   & csuv.s.0.mcp & 0.91 (0.01) & 1.65 (0.14) & 0.45 & 1.2 & 1.73 (0.09) & 1.84 (0.11) & 0.77 (0.04) & 9.25 (0.14) \\ 
   \hline
setting 5 & lasso & 0.39 (0.01) & 16.4 (0.67) & 16.19 & 0.21 & 1.28 (0.02) & 3.51 (0.11) & 1.04 (0.03) & 20.98 (0.68) \\ 
   & elastic net & 0.33 (0.01) & 21.14 (0.69) & 20.99 & 0.15 & 1.31 (0.02) & 4.21 (0.11) & 1.15 (0.02) & 25.84 (0.69) \\ 
  rho = 0.9 & relaxed lasso & 0.49 (0.01) & 10.58 (0.65) & 10 & 0.58 & 1.31 (0.02) & 3.31 (0.13) & 1.07 (0.03) & 14.42 (0.7) \\ 
  s = 5 & mcp & 0.67 (0.02) & 4.26 (0.31) & 3.17 & 1.09 & 1.29 (0.03) & 2.48 (0.17) & 1.09 (0.06) & 7.08 (0.24) \\ 
  p = 100 & scad & 0.65 (0.02) & 4.91 (0.28) & 4.06 & 0.85 & 1.24 (0.03) & 2.01 (0.14) & 0.91 (0.06) & 8.21 (0.22) \\ 
   & vsd & 0.76 (0.02) & 2.1 (0.16) & 0.5 & 1.6 & 1.38 (0.03) & 1.93 (0.13) & 1.01 (0.05) & 3.9 (0.08) \\ 
   & bic & 0.69 (0.02) & 4.15 (0.32) & 3.26 & 0.89 & 1.26 (0.03) & 2.22 (0.17) & 0.96 (0.06) & 7.37 (0.24) \\ 
   & \textbf{ebic} & 0.71 (0.02) & 3.67 (0.27) & 2.77 & 0.9 & \textbf{1.24} (0.03) & 2.07 (0.15) & 0.94 (0.06) & 6.87 (0.2) \\ 
   & \textbf{cv} & \underline{0.47} (0.02) & \underline{13.01} (0.75) & \underline{12.64} & \textbf{0.37} & 1.27 (0.02) & \underline{3.14} (0.14) & \underline{1.01} (0.04) & 17.27 (0.79) \\ 
   & \textbf{csuv.m.0} & \textbf{0.77} (0.02) & \textbf{2.25} (0.17) & \textbf{0.97} & \underline{1.28} & \underline{1.33} (0.03) & \textbf{1.9} (0.11) & \textbf{0.93} (0.05) & 4.69 (0.11) \\ 
   & \textbf{csuv.s.0} & 0.68 (0.01) & 4.21 (0.18) & 3.5 & 0.71 & 1.26 (0.02) & 2.31 (0.1) & 0.95 (0.04) & 7.79 (0.18) \\ 
   & csuv.m.5 & 0.77 (0.02) & 2.16 (0.16) & 0.8 & 1.36 & 1.35 (0.03) & 1.88 (0.11) & 0.95 (0.04) & 4.44 (0.11) \\ 
   & csuv.s.5 & 0.66 (0.01) & 4.45 (0.17) & 3.77 & 0.68 & 1.26 (0.02) & 2.37 (0.1) & 0.95 (0.03) & 8.09 (0.17) \\ 
   & csuv.m.0.all & 0.76 (0.02) & 2.43 (0.17) & 1.23 & 1.2 & 1.31 (0.03) & 1.96 (0.11) & 0.94 (0.04) & 5.03 (0.14) \\ 
   & csuv.s.0.all & 0.65 (0.01) & 4.94 (0.21) & 4.28 & 0.66 & 1.26 (0.03) & 2.48 (0.1) & 0.97 (0.03) & 8.62 (0.22) \\ 
   & csuv.m.0.mcp & 0.69 (0.02) & 2.53 (0.14) & 0.36 & 2.17 & 1.87 (0.07) & 2.37 (0.12) & 1.24 (0.05) & 3.19 (0.09) \\ 
   & csuv.s.0.mcp & 0.72 (0.02) & 2.88 (0.18) & 1.55 & 1.33 & 1.34 (0.03) & 2.24 (0.14) & 1.03 (0.05) & 5.22 (0.08) \\ 
   \hline
setting 6 & lasso & 0.3 (0.01) & 24.79 (1.08) & 24.6 & 0.19 & 1.42 (0.03) & 3.81 (0.15) & 1.04 (0.03) & 29.41 (1.09) \\ 
   & elastic net & 0.23 (0.01) & 34.73 (1.22) & 34.6 & 0.13 & 1.51 (0.03) & 4.91 (0.14) & 1.21 (0.02) & 39.47 (1.21) \\ 
  rho = 0.9 & relaxed lasso & 0.51 (0.02) & 10.84 (0.88) & 10.37 & 0.47 & 1.35 (0.03) & 3.13 (0.15) & 0.99 (0.03) & 14.9 (0.9) \\ 
  s = 5 & mcp & 0.59 (0.02) & 5.52 (0.33) & 4.24 & 1.28 & 1.33 (0.03) & 2.66 (0.16) & 1.16 (0.06) & 7.96 (0.24) \\ 
  p = 300 & scad & 0.57 (0.02) & 7.54 (0.42) & 6.94 & 0.6 & 1.21 (0.02) & 1.78 (0.12) & 0.79 (0.05) & 11.34 (0.36) \\ 
   & vsd & 0.75 (0.02) & 2.47 (0.2) & 0.58 & 1.89 & 1.87 (0.1) & 2.28 (0.16) & 1.12 (0.07) & 3.69 (0.19) \\ 
   & bic & 0.59 (0.02) & 8.1 (1.1) & 7.33 & 0.77 & 1.27 (0.03) & 2.3 (0.19) & 0.92 (0.05) & 11.56 (1.12) \\ 
   & \textbf{ebic} & 0.63 (0.02) & 5.69 (0.73) & 4.65 & 1.04 & 1.29 (0.03) & 2.32 (0.17) & \underline{1} (0.06) & 8.61 (0.72) \\ 
   & \textbf{cv} & \underline{0.44} (0.02) & \underline{16.75} (1.32) & \underline{16.32} & 0.43 & 1.33 (0.03) & \underline{2.89} (0.19) & 0.93 (0.04) & 20.89 (1.34) \\ 
   & \textbf{csuv.m.0} & \textbf{0.79} (0.02) & \textbf{2.01} (0.16) & \textbf{0.91} & \underline{1.1} & \underline{1.4} (0.04) & \textbf{1.81} (0.12) & 0.92 (0.05) & 4.81 (0.1) \\ 
   & \textbf{csuv.s.0} & 0.64 (0.01) & 5.53 (0.24) & 5.19 & \textbf{0.34} & \textbf{1.24} (0.02) & 2.31 (0.1) & \textbf{0.87} (0.03) & 9.85 (0.21) \\ 
   & csuv.m.5 & 0.8 (0.02) & 1.87 (0.15) & 0.76 & 1.11 & 1.43 (0.04) & 1.75 (0.11) & 0.9 (0.05) & 4.65 (0.1) \\ 
   & csuv.s.5 & 0.61 (0.01) & 6.14 (0.19) & 5.84 & 0.3 & 1.25 (0.02) & 2.43 (0.09) & 0.88 (0.03) & 10.54 (0.16) \\ 
   & csuv.m.0.all & 0.8 (0.02) & 2 (0.16) & 1 & 1 & 1.38 (0.04) & 1.74 (0.11) & 0.87 (0.05) & 5 (0.11) \\ 
   & csuv.s.0.all & 0.61 (0.01) & 6.31 (0.27) & 5.98 & 0.33 & 1.26 (0.03) & 2.49 (0.11) & 0.89 (0.03) & 10.65 (0.25) \\ 
   & csuv.m.0.mcp & 0.68 (0.02) & 2.56 (0.15) & 0.39 & 2.17 & 2.11 (0.07) & 2.46 (0.12) & 1.25 (0.05) & 3.22 (0.11) \\ 
   & csuv.s.0.mcp & 0.73 (0.01) & 2.98 (0.17) & 2.05 & 0.93 & 1.3 (0.03) & 2.13 (0.13) & 0.97 (0.05) & 6.12 (0.07) \\ 
   \hline
\multicolumn{10}{l}{}\\
\end{tabular}
}
\endgroup
\caption{Model 2: performance of CSUV and methods it compares with (continue). Variable selection performance in terms of F-measure (f), total error (FP+FN), false positives (FP) and false negatives (FN), prediction error in terms of mse (pred.err) and estimation error in terms of l1 and l2 distance (l1.diff and l2.diff) and are shown. The numbers are based on 100 simulations. The last 8 rows are the performance of CSUV with different parameters (e.g. csuv.m.0.mcp corresponds to CSUV with MCP as constituent method and r = 0). A bold number represents the best result among delete-n/2 cross validation, eBIC and CSUV using Lasso, MCP and SCAD while a underlined number represents the worst among them. Standard errors are shown inside the parentheses.} 
\label{2_2}
\end{table}

			% latex table generated in R 3.6.1 by xtable 1.8-4 package
% Mon Jan 13 15:32:14 2020
\begin{table}[ht]
\centering
\begingroup\small
\scalebox{0.75}{
\begin{tabular}{|l|l|l|l|ll|lll|l|}
  \hline
parameters & methods & f & FP+FN & FP & FN & pred.err & l1.diff & l2.diff & size \\ 
  \hline
setting 7 & lasso & 0.47 (0.01) & 22.4 (0.7) & 21.74 & 0.66 & 1.49 (0.03) & 6.3 (0.17) & 1.49 (0.03) & 31.08 (0.73) \\ 
   & elastic net & 0.42 (0.01) & 27.11 (0.73) & 26.68 & 0.43 & 1.54 (0.03) & 7.11 (0.17) & 1.57 (0.03) & 36.25 (0.73) \\ 
  rho = 0.9 & relaxed lasso & 0.54 (0.01) & 17.19 (0.83) & 16.23 & 0.96 & 1.53 (0.03) & 6.31 (0.22) & 1.55 (0.03) & 25.27 (0.91) \\ 
  s = 10 & mcp & 0.64 (0.01) & 7.8 (0.32) & 4.56 & 3.24 & 1.66 (0.05) & 6.15 (0.23) & 1.98 (0.06) & 11.32 (0.27) \\ 
  p = 100 & scad & 0.66 (0.01) & 8.03 (0.35) & 5.54 & 2.49 & 1.48 (0.04) & 4.97 (0.21) & 1.71 (0.06) & 13.05 (0.3) \\ 
   & vsd & 0.7 (0.02) & 5.13 (0.24) & 1.36 & 3.77 & 1.86 (0.08) & 5.3 (0.22) & 1.89 (0.06) & 7.59 (0.15) \\ 
   & bic & 0.67 (0.01) & 7.44 (0.33) & 4.84 & 2.6 & 1.55 (0.04) & 5.35 (0.22) & 1.78 (0.05) & 12.24 (0.29) \\ 
   & \textbf{ebic} & 0.68 (0.01) & 6.73 (0.29) & 3.94 & \underline{2.79} & 1.55 (0.04) & 5.21 (0.21) & \underline{1.77} (0.06) & 11.15 (0.23) \\ 
   & \textbf{cv} & \underline{0.47} (0.01) & \underline{22.4} (0.7) & \underline{21.74} & \textbf{0.66} & 1.49 (0.03) & \underline{6.3} (0.17) & 1.49 (0.03) & 31.08 (0.73) \\ 
   & \textbf{csuv.m.0} & \textbf{0.76} (0.01) & \textbf{4.57} (0.2) & \textbf{1.84} & 2.73 & \underline{1.66} (0.04) & \textbf{4.35} (0.14) & 1.53 (0.04) & 9.11 (0.15) \\ 
   & \textbf{csuv.s.0} & 0.67 (0.01) & 8.65 (0.24) & 7.19 & 1.46 & \textbf{1.45} (0.03) & 4.87 (0.13) & \textbf{1.44} (0.03) & 15.73 (0.23) \\ 
   & csuv.m.5 & 0.76 (0.01) & 4.54 (0.19) & 1.74 & 2.8 & 1.69 (0.04) & 4.37 (0.14) & 1.54 (0.04) & 8.94 (0.15) \\ 
   & csuv.s.5 & 0.66 (0.01) & 9 (0.23) & 7.54 & 1.46 & 1.45 (0.03) & 4.96 (0.12) & 1.45 (0.03) & 16.08 (0.22) \\ 
   & csuv.m.0.all & 0.76 (0.01) & 4.76 (0.22) & 2.31 & 2.45 & 1.6 (0.04) & 4.25 (0.15) & 1.46 (0.04) & 9.86 (0.16) \\ 
   & csuv.s.0.all & 0.64 (0.01) & 9.97 (0.25) & 8.62 & 1.35 & 1.45 (0.03) & 5.14 (0.11) & 1.46 (0.03) & 17.27 (0.27) \\ 
   & csuv.m.0.mcp & 0.59 (0.01) & 5.95 (0.18) & 0.47 & 5.48 & 3.22 (0.11) & 6.16 (0.18) & 2.19 (0.05) & 4.99 (0.12) \\ 
   & csuv.s.0.mcp & 0.72 (0.01) & 5.48 (0.21) & 2.39 & 3.09 & 1.74 (0.04) & 5.04 (0.17) & 1.74 (0.04) & 9.3 (0.09) \\ 
   \hline
setting 8 & lasso & 0.39 (0.01) & 32.18 (1.08) & 31.63 & 0.55 & 1.72 (0.04) & 6.9 (0.19) & 1.5 (0.03) & 41.08 (1.06) \\ 
   & elastic net & 0.32 (0.01) & 41.56 (0.99) & 41.13 & 0.43 & 1.83 (0.04) & 8.36 (0.16) & 1.68 (0.03) & 50.7 (0.97) \\ 
  rho = 0.9 & relaxed lasso & 0.51 (0.01) & 18.39 (0.75) & 17.43 & 0.96 & 1.64 (0.04) & 6.25 (0.18) & 1.5 (0.03) & 26.47 (0.79) \\ 
  s = 10 & mcp & 0.53 (0.02) & 11.72 (0.57) & 7.88 & 3.84 & 2.09 (0.07) & 7.45 (0.36) & 2.2 (0.08) & 14.04 (0.3) \\ 
  p = 300 & scad & 0.54 (0.01) & 14.05 (0.51) & 11.82 & 2.23 & 1.71 (0.06) & 5.27 (0.26) & 1.7 (0.07) & 19.59 (0.35) \\ 
   & vsd & 0.6 (0.02) & 6.71 (0.33) & 1.82 & 4.89 & 3.24 (0.23) & 6.5 (0.31) & 2.21 (0.09) & 6.93 (0.25) \\ 
   & bic & 0.55 (0.02) & 13.84 (1.12) & 11.25 & 2.59 & 1.79 (0.06) & 5.95 (0.31) & 1.81 (0.07) & 18.66 (1.15) \\ 
   & \textbf{ebic} & 0.57 (0.02) & 10.93 (0.5) & 7.81 & 3.12 & 1.86 (0.05) & 6.22 (0.3) & \underline{1.94} (0.07) & 14.69 (0.36) \\ 
   & \textbf{cv} & \underline{0.39} (0.01) & \underline{32.18} (1.08) & \underline{31.63} & \textbf{0.55} & 1.72 (0.04) & \underline{6.9} (0.19) & \textbf{1.5} (0.03) & 41.08 (1.06) \\ 
   & \textbf{csuv.m.0} & \textbf{0.69} (0.01) & \textbf{5.8} (0.23) & \textbf{2.27} & \underline{3.53} & \underline{2.37} (0.08) & \textbf{5.28} (0.19) & 1.82 (0.05) & 8.74 (0.16) \\ 
   & \textbf{csuv.s.0} & 0.55 (0.01) & 14.1 (0.32) & 12.73 & 1.37 & \textbf{1.65} (0.04) & 5.87 (0.16) & 1.52 (0.03) & 21.36 (0.34) \\ 
   & csuv.m.5 & 0.69 (0.01) & 5.68 (0.22) & 1.96 & 3.72 & 2.57 (0.1) & 5.25 (0.17) & 1.83 (0.05) & 8.24 (0.16) \\ 
   & csuv.s.5 & 0.55 (0.01) & 14.24 (0.33) & 12.87 & 1.37 & 1.64 (0.04) & 5.92 (0.16) & 1.54 (0.04) & 21.5 (0.34) \\ 
   & csuv.m.0.all & 0.69 (0.01) & 5.99 (0.24) & 2.68 & 3.31 & 2.26 (0.08) & 5.17 (0.18) & 1.76 (0.05) & 9.37 (0.17) \\ 
   & csuv.s.0.all & 0.53 (0.01) & 15.93 (0.35) & 14.69 & 1.24 & 1.64 (0.04) & 6.12 (0.16) & 1.53 (0.03) & 23.45 (0.36) \\ 
   & csuv.m.0.mcp & 0.38 (0.01) & 7.95 (0.16) & 0.32 & 7.63 & 7.18 (0.21) & 7.43 (0.16) & 2.53 (0.04) & 2.69 (0.13) \\ 
   & csuv.s.0.mcp & 0.59 (0.01) & 8.07 (0.26) & 3.9 & 4.17 & 2.53 (0.08) & 6.56 (0.21) & 2.08 (0.05) & 9.73 (0.11) \\ 
   \hline
\multicolumn{10}{l}{}\\
\end{tabular}
}
\endgroup
\caption{Model 2: performance of CSUV and methods it compares with (continue). Variable selection performance in terms of F-measure (f), total error (FP+FN), false positives (FP) and false negatives (FN), prediction error in terms of mse (pred.err) and estimation error in terms of l1 and l2 distance (l1.diff and l2.diff) and are shown. The numbers are based on 100 simulations. The last 8 rows are the performance of CSUV with different parameters (e.g. csuv.m.0.mcp corresponds to CSUV with MCP as constituent method and r = 0). A bold number represents the best result among delete-n/2 cross validation, eBIC and CSUV using Lasso, MCP and SCAD while a underlined number represents the worst among them. Standard errors are shown inside the parentheses.} 
\label{2_3}
\end{table}

			% latex table generated in R 3.6.1 by xtable 1.8-4 package
% Mon Jan 13 15:32:14 2020
\begin{table}[ht]
\centering
\begingroup\small
\scalebox{0.75}{
\begin{tabular}{|l|l|l|l|ll|lll|l|}
  \hline
parameters & methods & f & FP+FN & FP & FN & pred.err & l1.diff & l2.diff & size \\ 
  \hline
setting 1 & lasso & 0.38 (0.01) & 17.85 (0.75) & 17.84 & 0.01 & 1.31 (0.02) & 2.35 (0.07) & 0.68 (0.01) & 22.83 (0.75) \\ 
   & elastic net & 0.32 (0.01) & 23.17 (0.81) & 23.17 & 0 & 1.38 (0.02) & 2.91 (0.08) & 0.77 (0.01) & 28.17 (0.81) \\ 
  block.cor = 0.5 & relaxed lasso & 0.62 (0.02) & 7.16 (0.63) & 6.79 & 0.37 & 1.3 (0.03) & 1.76 (0.07) & 0.63 (0.02) & 11.42 (0.69) \\ 
  s = 5 & mcp & 0.85 (0.01) & 2.15 (0.26) & 2.07 & 0.08 & 1.12 (0.03) & 0.77 (0.07) & 0.34 (0.02) & 6.99 (0.26) \\ 
  p = 100 & scad & 0.63 (0.01) & 6.61 (0.37) & 6.58 & 0.03 & 1.13 (0.03) & 0.9 (0.06) & 0.36 (0.02) & 11.55 (0.38) \\ 
   & vsd & 0.92 (0.01) & 0.7 (0.08) & 0.05 & 0.65 & 1.21 (0.03) & 0.87 (0.06) & 0.5 (0.03) & 4.4 (0.08) \\ 
   & bic & 0.81 (0.01) & 2.79 (0.3) & 2.73 & 0.06 & 1.13 (0.03) & 0.82 (0.08) & 0.35 (0.02) & 7.67 (0.31) \\ 
   & \textbf{ebic} & 0.85 (0.01) & 2.15 (0.26) & 2.07 & 0.08 & \textbf{1.12} (0.03) & \textbf{0.77} (0.07) & \textbf{0.34} (0.02) & 6.99 (0.26) \\ 
   & \textbf{cv} & \underline{0.69} (0.02) & \underline{5.62} (0.52) & \underline{5.56} & \textbf{0.06} & 1.13 (0.03) & 0.9 (0.07) & 0.36 (0.02) & 10.5 (0.53) \\ 
   & \textbf{csuv.m.0} & \textbf{0.92} (0.01) & \textbf{0.76} (0.08) & \textbf{0.23} & \underline{0.53} & 1.2 (0.03) & 0.87 (0.06) & 0.48 (0.03) & 4.7 (0.09) \\ 
   & \textbf{csuv.s.0} & 0.85 (0.01) & 1.72 (0.13) & 1.42 & 0.3 & \underline{1.21} (0.03) & \underline{1.06} (0.05) & \underline{0.51} (0.02) & 6.12 (0.15) \\ 
   & csuv.m.5 & 0.92 (0.01) & 0.75 (0.08) & 0.17 & 0.58 & 1.2 (0.03) & 0.88 (0.05) & 0.5 (0.03) & 4.59 (0.08) \\ 
   & csuv.s.5 & 0.81 (0.01) & 2.35 (0.12) & 2.08 & 0.27 & 1.22 (0.03) & 1.17 (0.05) & 0.53 (0.02) & 6.81 (0.16) \\ 
   & csuv.m.0.all & 0.91 (0.01) & 0.83 (0.09) & 0.31 & 0.52 & 1.2 (0.03) & 0.89 (0.06) & 0.49 (0.03) & 4.79 (0.1) \\ 
   & csuv.s.0.all & 0.83 (0.01) & 1.98 (0.14) & 1.7 & 0.28 & 1.21 (0.03) & 1.1 (0.05) & 0.51 (0.02) & 6.42 (0.16) \\ 
   & csuv.m.0.mcp & 0.92 (0.01) & 0.72 (0.08) & 0.07 & 0.65 & 1.21 (0.03) & 0.89 (0.05) & 0.51 (0.03) & 4.42 (0.07) \\ 
   & csuv.s.0.mcp & 0.9 (0.01) & 0.99 (0.08) & 0.5 & 0.49 & 1.2 (0.03) & 0.92 (0.05) & 0.5 (0.03) & 5.01 (0.12) \\ 
   \hline
setting 2 & lasso & 0.29 (0.01) & 26.37 (1.1) & 26.32 & 0.05 & 1.49 (0.03) & 3.19 (0.1) & 0.85 (0.02) & 31.27 (1.1) \\ 
   & elastic net & 0.23 (0.01) & 35.16 (1.2) & 35.1 & 0.06 & 1.62 (0.03) & 3.94 (0.11) & 0.98 (0.02) & 40.04 (1.2) \\ 
  block.cor = 0.5 & relaxed lasso & 0.49 (0.01) & 12.49 (0.9) & 12.23 & 0.26 & 1.48 (0.03) & 2.68 (0.12) & 0.8 (0.03) & 16.97 (0.94) \\ 
  s = 5 & mcp & 0.79 (0.01) & 2.99 (0.25) & 2.9 & 0.09 & 1.15 (0.02) & 0.93 (0.06) & 0.41 (0.02) & 7.81 (0.24) \\ 
  p = 300 & scad & 0.54 (0.01) & 9.86 (0.51) & 9.83 & 0.03 & 1.15 (0.02) & 1.13 (0.05) & 0.41 (0.02) & 14.8 (0.5) \\ 
   & vsd & 0.9 (0.01) & 1.06 (0.14) & 0.05 & 1.01 & 1.56 (0.09) & 1.26 (0.12) & 0.67 (0.06) & 4.04 (0.14) \\ 
   & bic & 0.65 (0.02) & 8.15 (1.17) & 8.08 & 0.07 & 1.18 (0.03) & 1.3 (0.12) & 0.45 (0.02) & 13.01 (1.18) \\ 
   & \textbf{ebic} & 0.79 (0.02) & 3.75 (0.85) & 3.66 & 0.09 & \textbf{1.16} (0.02) & 1.01 (0.1) & \textbf{0.42} (0.02) & 8.57 (0.85) \\ 
   & \textbf{cv} & \underline{0.5} (0.02) & \underline{14.67} (1.34) & \underline{14.61} & \textbf{0.06} & 1.28 (0.03) & \underline{1.85} (0.15) & 0.57 (0.03) & 19.55 (1.34) \\ 
   & \textbf{csuv.m.0} & \textbf{0.92} (0.01) & \textbf{0.72} (0.1) & \textbf{0.18} & \underline{0.54} & \underline{1.3} (0.05) & \textbf{0.96} (0.08) & 0.52 (0.04) & 4.64 (0.07) \\ 
   & \textbf{csuv.s.0} & 0.73 (0.01) & 3.8 (0.21) & 3.67 & 0.13 & 1.28 (0.03) & 1.55 (0.07) & \underline{0.6} (0.02) & 8.54 (0.2) \\ 
   & csuv.m.5 & 0.92 (0.01) & 0.71 (0.1) & 0.16 & 0.55 & 1.3 (0.05) & 0.97 (0.08) & 0.53 (0.04) & 4.61 (0.08) \\ 
   & csuv.s.5 & 0.67 (0.01) & 5.08 (0.17) & 4.96 & 0.12 & 1.32 (0.02) & 1.8 (0.06) & 0.65 (0.02) & 9.84 (0.18) \\ 
   & csuv.m.0.all & 0.92 (0.01) & 0.72 (0.11) & 0.18 & 0.54 & 1.29 (0.05) & 0.96 (0.08) & 0.52 (0.04) & 4.64 (0.07) \\ 
   & csuv.s.0.all & 0.72 (0.01) & 4.13 (0.23) & 4.01 & 0.12 & 1.28 (0.02) & 1.6 (0.07) & 0.6 (0.02) & 8.89 (0.22) \\ 
   & csuv.m.0.mcp & 0.9 (0.01) & 0.84 (0.09) & 0.04 & 0.8 & 1.4 (0.05) & 1.09 (0.07) & 0.62 (0.04) & 4.24 (0.09) \\ 
   & csuv.s.0.mcp & 0.87 (0.01) & 1.38 (0.1) & 1.12 & 0.26 & 1.23 (0.04) & 1.04 (0.07) & 0.5 (0.03) & 5.86 (0.08) \\ 
   \hline
setting 3 & lasso & 0.48 (0.01) & 22.55 (0.82) & 22.53 & 0.02 & 1.52 (0.03) & 3.77 (0.1) & 0.88 (0.01) & 32.51 (0.83) \\ 
   & elastic net & 0.44 (0.01) & 26.43 (0.83) & 26.41 & 0.02 & 1.58 (0.03) & 4.22 (0.1) & 0.94 (0.01) & 36.39 (0.83) \\ 
  block.cor = 0.5 & relaxed lasso & 0.66 (0.01) & 11.12 (0.69) & 10.68 & 0.44 & 1.49 (0.02) & 3.17 (0.1) & 0.87 (0.02) & 20.24 (0.75) \\ 
  s = 10 & mcp & 0.89 (0.01) & 2.62 (0.21) & 2.44 & 0.18 & 1.22 (0.02) & 1.5 (0.07) & 0.54 (0.02) & 12.26 (0.23) \\ 
  p = 100 & scad & 0.76 (0.01) & 6.64 (0.3) & 6.5 & 0.14 & 1.21 (0.02) & 1.55 (0.06) & 0.52 (0.02) & 16.36 (0.3) \\ 
   & vsd & 0.94 (0.01) & 1.17 (0.11) & 0.15 & 1.02 & 1.39 (0.03) & 1.77 (0.08) & 0.73 (0.03) & 9.13 (0.1) \\ 
   & bic & 0.87 (0.01) & 3.28 (0.27) & 3.12 & 0.16 & 1.22 (0.02) & 1.52 (0.07) & 0.54 (0.02) & 12.96 (0.27) \\ 
   & \textbf{ebic} & 0.89 (0.01) & 2.63 (0.21) & 2.45 & 0.18 & \textbf{1.22} (0.02) & \textbf{1.5} (0.07) & \textbf{0.54} (0.02) & 12.27 (0.23) \\ 
   & \textbf{cv} & \underline{0.5} (0.01) & \underline{21.91} (0.89) & \underline{21.86} & \textbf{0.05} & \underline{1.52} (0.03) & \underline{3.72} (0.11) & \underline{0.88} (0.02) & 31.81 (0.9) \\ 
   & \textbf{csuv.m.0} & \textbf{0.93} (0.01) & \textbf{1.39} (0.11) & \textbf{0.54} & \underline{0.85} & 1.36 (0.03) & 1.8 (0.07) & 0.73 (0.03) & 9.69 (0.11) \\ 
   & \textbf{csuv.s.0} & 0.82 (0.01) & 4.47 (0.18) & 4.11 & 0.36 & 1.36 (0.03) & 2.25 (0.07) & 0.73 (0.02) & 13.75 (0.19) \\ 
   & csuv.m.5 & 0.93 (0.01) & 1.42 (0.12) & 0.47 & 0.95 & 1.39 (0.03) & 1.86 (0.07) & 0.76 (0.03) & 9.52 (0.1) \\ 
   & csuv.s.5 & 0.8 (0.01) & 4.86 (0.15) & 4.54 & 0.32 & 1.37 (0.03) & 2.29 (0.07) & 0.73 (0.02) & 14.22 (0.16) \\ 
   & csuv.m.0.all & 0.93 (0.01) & 1.46 (0.12) & 0.66 & 0.8 & 1.35 (0.03) & 1.81 (0.07) & 0.73 (0.03) & 9.86 (0.11) \\ 
   & csuv.s.0.all & 0.8 (0.01) & 5.02 (0.19) & 4.69 & 0.33 & 1.37 (0.03) & 2.32 (0.07) & 0.74 (0.02) & 14.36 (0.2) \\ 
   & csuv.m.0.mcp & 0.89 (0.01) & 1.87 (0.12) & 0.05 & 1.82 & 1.63 (0.04) & 2.34 (0.09) & 0.97 (0.03) & 8.23 (0.11) \\ 
   & csuv.s.0.mcp & 0.93 (0.01) & 1.43 (0.1) & 0.59 & 0.84 & 1.35 (0.03) & 1.8 (0.07) & 0.73 (0.03) & 9.75 (0.1) \\ 
   \hline
\multicolumn{10}{l}{}\\
\end{tabular}
}
\endgroup
\caption{Model 3: performance of CSUV and methods it compares with. Variable selection performance in terms of F-measure (f), total error (FP+FN), false positives (FP) and false negatives (FN), prediction error in terms of mse (pred.err) and estimation error in terms of l1 and l2 distance (l1.diff and l2.diff) and are shown. The numbers are based on 100 simulations. The last 8 rows are the performance of CSUV with different parameters (e.g. csuv.m.0.mcp corresponds to CSUV with MCP as constituent method and r = 0). A bold number represents the best result among delete-n/2 cross validation, eBIC and CSUV using Lasso, MCP and SCAD while a underlined number represents the worst among them. Standard errors are shown inside the parentheses.} 
\label{3_1}
\end{table}

			% latex table generated in R 3.6.1 by xtable 1.8-4 package
% Mon Jan 13 15:32:14 2020
\begin{table}[ht]
\centering
\begingroup\small
\scalebox{0.75}{
\begin{tabular}{|l|l|l|l|ll|lll|l|}
  \hline
parameters & methods & f & FP+FN & FP & FN & pred.err & l1.diff & l2.diff & size \\ 
  \hline
setting 4 & lasso & 0.37 (0.01) & 35.15 (1.15) & 35.03 & 0.12 & 1.77 (0.04) & 5.29 (0.13) & 1.11 (0.02) & 44.91 (1.15) \\ 
   & elastic net & 0.31 (0.01) & 45.8 (1.39) & 45.65 & 0.15 & 1.97 (0.05) & 6.59 (0.17) & 1.27 (0.02) & 55.5 (1.39) \\ 
  block.cor = 0.5 & relaxed lasso & 0.51 (0.01) & 20.03 (0.83) & 19.75 & 0.28 & 1.73 (0.04) & 4.62 (0.15) & 1.04 (0.03) & 29.47 (0.86) \\ 
  s = 10 & mcp & 0.84 (0.01) & 3.71 (0.27) & 3.31 & 0.4 & 1.39 (0.06) & 1.96 (0.14) & 0.67 (0.04) & 12.91 (0.21) \\ 
  p = 300 & scad & 0.68 (0.01) & 9.88 (0.44) & 9.75 & 0.13 & 1.3 (0.04) & 1.96 (0.11) & 0.62 (0.03) & 19.62 (0.41) \\ 
   & vsd & 0.91 (0.01) & 1.83 (0.2) & 0.21 & 1.62 & 1.88 (0.17) & 2.31 (0.16) & 0.92 (0.06) & 8.59 (0.18) \\ 
   & bic & 0.73 (0.02) & 10.55 (1.62) & 10.37 & 0.18 & 1.37 (0.04) & 2.35 (0.19) & 0.68 (0.04) & 20.19 (1.62) \\ 
   & \textbf{ebic} & \textbf{0.85} (0.01) & 4.28 (0.81) & 3.97 & 0.31 & \textbf{1.33} (0.04) & \textbf{1.9} (0.13) & \textbf{0.64} (0.03) & 13.66 (0.81) \\ 
   & \textbf{cv} & \underline{0.37} (0.01) & \underline{35.15} (1.15) & \underline{35.03} & \textbf{0.12} & 1.77 (0.04) & \underline{5.29} (0.13) & \underline{1.11} (0.02) & 44.91 (1.15) \\ 
   & \textbf{csuv.m.0} & \textbf{0.85} (0.01) & \textbf{2.77} (0.17) & \textbf{0.88} & \underline{1.89} & \underline{1.92} (0.07) & 2.73 (0.12) & 1.08 (0.04) & 8.99 (0.13) \\ 
   & \textbf{csuv.s.0} & 0.61 (0.01) & 12.58 (0.35) & 12.06 & 0.52 & 1.64 (0.04) & 3.82 (0.11) & 1 (0.02) & 21.54 (0.37) \\ 
   & csuv.m.5 & 0.85 (0.01) & 2.78 (0.17) & 0.73 & 2.05 & 2.01 (0.08) & 2.79 (0.13) & 1.11 (0.04) & 8.68 (0.13) \\ 
   & csuv.s.5 & 0.6 (0.01) & 13 (0.35) & 12.45 & 0.55 & 1.66 (0.04) & 3.89 (0.12) & 1 (0.03) & 21.9 (0.37) \\ 
   & csuv.m.0.all & 0.85 (0.01) & 2.85 (0.18) & 1.12 & 1.73 & 1.87 (0.07) & 2.65 (0.12) & 1.04 (0.04) & 9.39 (0.14) \\ 
   & csuv.s.0.all & 0.58 (0.01) & 14.46 (0.41) & 13.95 & 0.51 & 1.68 (0.04) & 4.11 (0.12) & 1.02 (0.03) & 23.44 (0.44) \\ 
   & csuv.m.0.mcp & 0.6 (0.02) & 5.53 (0.16) & 0.04 & 5.49 & 4.62 (0.2) & 5.23 (0.13) & 1.9 (0.04) & 4.55 (0.16) \\ 
   & csuv.s.0.mcp & 0.82 (0.01) & 3.47 (0.21) & 1.37 & 2.1 & 2.07 (0.08) & 3.14 (0.15) & 1.18 (0.04) & 9.27 (0.07) \\ 
   \hline
setting 5 & lasso & 0.38 (0.01) & 14.34 (0.77) & 13.22 & 1.12 & 1.27 (0.02) & 4.03 (0.14) & 1.28 (0.02) & 17.1 (0.85) \\ 
   & elastic net & 0.31 (0.01) & 19.43 (0.84) & 18.45 & 0.98 & 1.3 (0.02) & 4.77 (0.15) & 1.38 (0.02) & 22.47 (0.91) \\ 
  block.cor = 0.9 & relaxed lasso & 0.48 (0.01) & 8.15 (0.5) & 6.46 & 1.69 & 1.26 (0.02) & 3.42 (0.11) & 1.26 (0.03) & 9.77 (0.6) \\ 
  s = 5 & mcp & 0.58 (0.02) & 4.88 (0.22) & 3.03 & 1.85 & 1.28 (0.02) & 2.99 (0.15) & 1.34 (0.05) & 6.18 (0.26) \\ 
  p = 100 & scad & 0.6 (0.02) & 4.7 (0.27) & 3.04 & 1.66 & 1.25 (0.02) & 2.7 (0.15) & 1.26 (0.05) & 6.38 (0.32) \\ 
   & vsd & 0.65 (0.02) & 2.86 (0.14) & 0.46 & 2.4 & 1.25 (0.03) & 2.41 (0.13) & 1.25 (0.04) & 3.06 (0.03) \\ 
   & bic & 0.61 (0.02) & 4.35 (0.21) & 2.7 & 1.65 & 1.26 (0.03) & 2.74 (0.15) & 1.26 (0.05) & 6.05 (0.29) \\ 
   & \textbf{ebic} & 0.62 (0.02) & 4.13 (0.2) & 2.39 & 1.74 & \underline{1.26} (0.03) & 2.66 (0.14) & 1.24 (0.05) & 5.65 (0.28) \\ 
   & \textbf{cv} & \underline{0.47} (0.02) & \underline{10.44} (0.76) & \underline{9.07} & \textbf{1.37} & \underline{1.26} (0.02) & \underline{3.53} (0.16) & \underline{1.28} (0.03) & 12.7 (0.86) \\ 
   & \textbf{csuv.m.0} & \textbf{0.66} (0.01) & \textbf{2.99} (0.12) & \textbf{0.85} & \underline{2.14} & 1.25 (0.03) & \textbf{2.41} (0.09) & \textbf{1.19} (0.03) & 3.71 (0.09) \\ 
   & \textbf{csuv.s.0} & 0.58 (0.01) & 4.72 (0.16) & 2.89 & 1.83 & \textbf{1.22} (0.02) & 2.95 (0.09) & 1.22 (0.03) & 6.06 (0.14) \\ 
   & csuv.m.5 & 0.66 (0.01) & 2.93 (0.12) & 0.74 & 2.19 & 1.25 (0.03) & 2.37 (0.09) & 1.19 (0.03) & 3.55 (0.09) \\ 
   & csuv.s.5 & 0.57 (0.01) & 4.88 (0.14) & 3.03 & 1.85 & 1.23 (0.02) & 2.98 (0.09) & 1.23 (0.03) & 6.18 (0.13) \\ 
   & csuv.m.0.all & 0.63 (0.01) & 3.42 (0.16) & 1.28 & 2.14 & 1.25 (0.03) & 2.6 (0.1) & 1.22 (0.03) & 4.14 (0.11) \\ 
   & csuv.s.0.all & 0.53 (0.01) & 5.71 (0.17) & 3.91 & 1.8 & 1.23 (0.02) & 3.16 (0.09) & 1.25 (0.03) & 7.11 (0.16) \\ 
   & csuv.m.0.mcp & 0.53 (0.02) & 3.51 (0.12) & 0.18 & 3.33 & 2.42 (0.1) & 3.1 (0.11) & 1.56 (0.04) & 1.85 (0.09) \\ 
   & csuv.s.0.mcp & 0.61 (0.02) & 3.45 (0.18) & 0.99 & 2.46 & 1.39 (0.04) & 2.91 (0.15) & 1.39 (0.05) & 3.53 (0.08) \\ 
   \hline
setting 6 & lasso & 0.28 (0.01) & 21.2 (1.05) & 19.8 & 1.4 & 1.4 (0.02) & 5.37 (0.15) & 1.6 (0.02) & 23.4 (1.1) \\ 
   & elastic net & 0.21 (0.01) & 30.54 (1.35) & 29.32 & 1.22 & 1.41 (0.02) & 6.18 (0.2) & 1.66 (0.02) & 33.1 (1.41) \\ 
  block.cor = 0.9 & relaxed lasso & 0.34 (0.01) & 14.1 (0.77) & 12.28 & 1.82 & 1.41 (0.03) & 5.18 (0.17) & 1.65 (0.03) & 15.46 (0.83) \\ 
  s = 5 & mcp & 0.42 (0.02) & 7.21 (0.26) & 4.52 & 2.69 & 1.45 (0.03) & 4.8 (0.16) & 1.85 (0.04) & 6.83 (0.25) \\ 
  p = 300 & scad & 0.4 (0.02) & 8.83 (0.35) & 6.53 & 2.3 & 1.42 (0.03) & 4.42 (0.19) & 1.74 (0.05) & 9.23 (0.32) \\ 
   & vsd & 0.45 (0.02) & 4.8 (0.14) & 1.14 & 3.66 & 2 (0.06) & 4.4 (0.11) & 1.94 (0.03) & 2.48 (0.12) \\ 
   & bic & 0.42 (0.02) & 9.21 (0.91) & 6.9 & 2.31 & 1.44 (0.03) & 4.76 (0.21) & 1.77 (0.05) & 9.59 (0.95) \\ 
   & \textbf{ebic} & 0.44 (0.02) & 6.91 (0.27) & 4.39 & 2.52 & 1.43 (0.03) & 4.59 (0.17) & \underline{1.79} (0.05) & 6.87 (0.26) \\ 
   & \textbf{cv} & \underline{0.29} (0.01) & \underline{20.18} (1.01) & \underline{18.7} & \textbf{1.48} & 1.39 (0.02) & \underline{5.31} (0.15) & \textbf{1.62} (0.02) & 22.22 (1.06) \\ 
   & \textbf{csuv.m.0} & \textbf{0.51} (0.02) & \textbf{3.97} (0.12) & \textbf{0.81} & \underline{3.16} & \underline{2.06} (0.06) & \textbf{3.78} (0.1) & 1.72 (0.03) & 2.65 (0.12) \\ 
   & \textbf{csuv.s.0} & 0.44 (0.01) & 7.26 (0.19) & 5.08 & 2.18 & \textbf{1.38} (0.03) & 4.45 (0.11) & 1.65 (0.03) & 7.9 (0.13) \\ 
   & csuv.m.5 & 0.51 (0.02) & 3.86 (0.12) & 0.65 & 3.21 & 2.1 (0.06) & 3.71 (0.1) & 1.71 (0.03) & 2.44 (0.11) \\ 
   & csuv.s.5 & 0.44 (0.01) & 7.46 (0.19) & 5.32 & 2.14 & 1.38 (0.03) & 4.45 (0.1) & 1.64 (0.03) & 8.18 (0.13) \\ 
   & csuv.m.0.all & 0.51 (0.02) & 4.02 (0.13) & 1.02 & 3 & 1.94 (0.06) & 3.8 (0.1) & 1.7 (0.03) & 3.02 (0.13) \\ 
   & csuv.s.0.all & 0.42 (0.01) & 8.34 (0.2) & 6.34 & 2 & 1.37 (0.03) & 4.51 (0.1) & 1.62 (0.03) & 9.34 (0.17) \\ 
   & csuv.m.0.mcp & 0.42 (0.01) & 4.26 (0.09) & 0.14 & 4.12 & 2.89 (0.06) & 3.93 (0.07) & 1.84 (0.02) & 1.02 (0.09) \\ 
   & csuv.s.0.mcp & 0.47 (0.02) & 5.21 (0.17) & 2.35 & 2.86 & 1.65 (0.05) & 4.39 (0.11) & 1.78 (0.03) & 4.49 (0.08) \\ 
   \hline
\multicolumn{10}{l}{}\\
\end{tabular}
}
\endgroup
\caption{Model 3: performance of CSUV and methods it compares with (continue). Variable selection performance in terms of F-measure (f), total error (FP+FN), false positives (FP) and false negatives (FN), prediction error in terms of mse (pred.err) and estimation error in terms of l1 and l2 distance (l1.diff and l2.diff) and are shown. The numbers are based on 100 simulations. The last 8 rows are the performance of CSUV with different parameters (e.g. csuv.m.0.mcp corresponds to CSUV with MCP as constituent method and r = 0). A bold number represents the best result among delete-n/2 cross validation, eBIC and CSUV using Lasso, MCP and SCAD while a underlined number represents the worst among them. Standard errors are shown inside the parentheses.} 
\label{3_2}
\end{table}

			% latex table generated in R 3.6.1 by xtable 1.8-4 package
% Mon Jan 13 15:32:14 2020
\begin{table}[ht]
\centering
\begingroup\small
\scalebox{0.75}{
\begin{tabular}{|l|l|l|l|ll|lll|l|}
  \hline
parameters & methods & f & FP+FN & FP & FN & pred.err & l1.diff & l2.diff & size \\ 
  \hline
setting 7 & lasso & 0.5 (0.01) & 18.54 (0.66) & 17.19 & 1.35 & 1.42 (0.02) & 6.55 (0.16) & 1.65 (0.03) & 25.84 (0.69) \\ 
   & elastic net & 0.44 (0.01) & 22.85 (0.63) & 21.67 & 1.18 & 1.43 (0.02) & 7.18 (0.15) & 1.7 (0.02) & 30.49 (0.67) \\ 
  block.cor = 0.9 & relaxed lasso & 0.56 (0.01) & 13.74 (0.64) & 11.91 & 1.83 & 1.42 (0.03) & 6.37 (0.18) & 1.68 (0.03) & 20.08 (0.72) \\ 
  s = 10 & mcp & 0.53 (0.01) & 8.51 (0.29) & 3.35 & 5.16 & 1.73 (0.04) & 7.19 (0.25) & 2.34 (0.06) & 8.19 (0.13) \\ 
  p = 100 & scad & 0.59 (0.01) & 7.76 (0.28) & 3.23 & 4.53 & 1.61 (0.04) & 6.23 (0.21) & 2.17 (0.06) & 8.7 (0.16) \\ 
   & vsd & 0.6 (0.01) & 6.52 (0.23) & 1.53 & 4.99 & 1.8 (0.06) & 6.56 (0.22) & 2.3 (0.06) & 6.54 (0.1) \\ 
   & bic & 0.6 (0.01) & 7.5 (0.28) & 3.04 & 4.46 & 1.59 (0.04) & 6.11 (0.21) & 2.11 (0.06) & 8.58 (0.21) \\ 
   & \textbf{ebic} & 0.6 (0.01) & 7.27 (0.27) & 2.77 & \underline{4.5} & 1.6 (0.04) & 6.12 (0.22) & \underline{2.12} (0.06) & 8.27 (0.14) \\ 
   & \textbf{cv} & \underline{0.5} (0.01) & \underline{18.54} (0.66) & \underline{17.19} & \textbf{1.35} & 1.42 (0.02) & \underline{6.55} (0.16) & \textbf{1.65} (0.03) & 25.84 (0.69) \\ 
   & \textbf{csuv.m.0} & \textbf{0.69} (0.01) & \textbf{5.88} (0.23) & \textbf{2.47} & 3.41 & \underline{1.66} (0.05) & \textbf{5.24} (0.16) & 1.77 (0.04) & 9.06 (0.18) \\ 
   & \textbf{csuv.s.0} & 0.62 (0.01) & 9.49 (0.27) & 7.22 & 2.27 & \textbf{1.38} (0.02) & 5.69 (0.14) & \textbf{1.65} (0.03) & 14.95 (0.22) \\ 
   & csuv.m.5 & 0.68 (0.01) & 5.92 (0.23) & 2.38 & 3.54 & 1.71 (0.06) & 5.34 (0.16) & 1.8 (0.04) & 8.84 (0.18) \\ 
   & csuv.s.5 & 0.62 (0.01) & 9.76 (0.26) & 7.54 & 2.22 & 1.38 (0.02) & 5.67 (0.13) & 1.64 (0.03) & 15.32 (0.21) \\ 
   & csuv.m.0.all & 0.7 (0.01) & 5.99 (0.25) & 2.86 & 3.13 & 1.58 (0.04) & 5.21 (0.17) & 1.73 (0.04) & 9.73 (0.18) \\ 
   & csuv.s.0.all & 0.6 (0.01) & 10.58 (0.28) & 8.54 & 2.04 & 1.38 (0.02) & 5.79 (0.13) & 1.64 (0.03) & 16.5 (0.25) \\ 
   & csuv.m.0.mcp & 0.4 (0.01) & 7.64 (0.13) & 0.31 & 7.33 & 5.38 (0.28) & 7.79 (0.14) & 2.71 (0.04) & 2.98 (0.12) \\ 
   & csuv.s.0.mcp & 0.6 (0.01) & 7.01 (0.21) & 2.25 & 4.76 & 1.98 (0.06) & 6.57 (0.19) & 2.22 (0.05) & 7.49 (0.08) \\ 
   \hline
setting 8 & lasso & 0.33 (0.01) & 34.07 (0.92) & 32.27 & 1.8 & 1.59 (0.03) & 10.03 (0.23) & 2.1 (0.03) & 40.47 (0.9) \\ 
   & elastic net & 0.29 (0.01) & 42.66 (0.87) & 41.11 & 1.55 & 1.62 (0.03) & 10.86 (0.21) & 2.15 (0.03) & 49.56 (0.84) \\ 
  block.cor = 0.9 & relaxed lasso & 0.35 (0.01) & 28.37 (0.78) & 25.88 & 2.49 & 1.63 (0.03) & 10.14 (0.22) & 2.19 (0.04) & 33.39 (0.88) \\ 
  s = 10 & mcp & 0.38 (0.02) & 12.94 (0.35) & 6.81 & 6.13 & 2.2 (0.06) & 11.21 (0.31) & 3.09 (0.06) & 10.68 (0.18) \\ 
  p = 300 & scad & 0.43 (0.02) & 12.2 (0.39) & 6.8 & 5.4 & 2.01 (0.05) & 9.94 (0.32) & 2.89 (0.07) & 11.4 (0.21) \\ 
   & vsd & 0.44 (0.02) & 10 (0.31) & 3.78 & 6.22 & 2.73 (0.17) & 9.8 (0.29) & 2.97 (0.06) & 7.56 (0.14) \\ 
   & bic & 0.42 (0.02) & 14 (1) & 8.7 & 5.3 & 2.05 (0.05) & 10.53 (0.34) & 2.92 (0.07) & 13.4 (1.09) \\ 
   & \textbf{ebic} & 0.44 (0.02) & 11.7 (0.35) & 6.29 & 5.41 & 2.03 (0.05) & \underline{10.05} (0.31) & \underline{2.9} (0.06) & 10.88 (0.18) \\ 
   & \textbf{cv} & \underline{0.33} (0.01) & \underline{34.07} (0.92) & \underline{32.27} & \textbf{1.8} & \textbf{1.59} (0.03) & 10.03 (0.23) & \textbf{2.1} (0.03) & 40.47 (0.9) \\ 
   & \textbf{csuv.m.0} & \textbf{0.52} (0.01) & \textbf{7.84} (0.22) & \textbf{2.15} & \underline{5.69} & \underline{3.96} (0.16) & \textbf{7.74} (0.21) & 2.52 (0.05) & 6.46 (0.16) \\ 
   & \textbf{csuv.s.0} & 0.42 (0.01) & 20.14 (0.32) & 17.25 & 2.89 & 1.65 (0.03) & 9.4 (0.2) & 2.21 (0.04) & 24.36 (0.25) \\ 
   & csuv.m.5 & 0.52 (0.01) & 7.69 (0.21) & 1.87 & 5.82 & 4.29 (0.16) & 7.61 (0.2) & 2.5 (0.05) & 6.05 (0.15) \\ 
   & csuv.s.5 & 0.41 (0.01) & 20.32 (0.32) & 17.36 & 2.96 & 1.67 (0.03) & 9.45 (0.2) & 2.21 (0.04) & 24.4 (0.23) \\ 
   & csuv.m.0.all & 0.53 (0.01) & 7.91 (0.21) & 2.46 & 5.45 & 3.55 (0.15) & 7.77 (0.19) & 2.53 (0.05) & 7.01 (0.16) \\ 
   & csuv.s.0.all & 0.41 (0.01) & 21.06 (0.34) & 18.27 & 2.79 & 1.65 (0.03) & 9.47 (0.2) & 2.19 (0.04) & 25.48 (0.26) \\ 
   & csuv.m.0.mcp & 0.23 (0.01) & 9.67 (0.09) & 0.22 & 9.45 & 10.76 (0.22) & 8.96 (0.1) & 2.97 (0.03) & 0.77 (0.09) \\ 
   & csuv.s.0.mcp & 0.36 (0.02) & 11.97 (0.28) & 5.33 & 6.64 & 3.41 (0.12) & 10.68 (0.25) & 3.08 (0.05) & 8.69 (0.05) \\ 
   \hline
\multicolumn{10}{l}{}\\
\end{tabular}
}
\endgroup
\caption{Model 3: performance of CSUV and methods it compares with (continue). Variable selection performance in terms of F-measure (f), total error (FP+FN), false positives (FP) and false negatives (FN), prediction error in terms of mse (pred.err) and estimation error in terms of l1 and l2 distance (l1.diff and l2.diff) and are shown. The numbers are based on 100 simulations. The last 8 rows are the performance of CSUV with different parameters (e.g. csuv.m.0.mcp corresponds to CSUV with MCP as constituent method and r = 0). A bold number represents the best result among delete-n/2 cross validation, eBIC and CSUV using Lasso, MCP and SCAD while a underlined number represents the worst among them. Standard errors are shown inside the parentheses.} 
\label{3_3}
\end{table}

			% latex table generated in R 3.6.1 by xtable 1.8-4 package
% Mon Jan 13 15:32:14 2020
\begin{table}[ht]
\centering
\begingroup\small
\scalebox{0.75}{
\begin{tabular}{|l|l|l|l|ll|lll|l|}
  \hline
parameters & methods & f & FP+FN & FP & FN & pred.err & l1.diff & l2.diff & size \\ 
  \hline
setting 1 & lasso & 0.37 (0.01) & 18.19 (0.89) & 17.98 & 0.21 & 1.85 (0.09) & 3.02 (0.13) & 0.9 (0.03) & 22.77 (0.89) \\ 
   & elastic net & 0.32 (0.01) & 23.2 (0.97) & 23.01 & 0.19 & 1.98 (0.08) & 3.57 (0.13) & 0.98 (0.03) & 27.82 (0.98) \\ 
  num.factor = 2 & relaxed lasso & 0.5 (0.02) & 11.16 (0.77) & 10.68 & 0.48 & 1.89 (0.08) & 2.77 (0.12) & 0.91 (0.03) & 15.2 (0.81) \\ 
  s = 5 & mcp & 0.75 (0.02) & 3.26 (0.28) & 2.57 & 0.69 & 1.71 (0.09) & 1.67 (0.11) & 0.76 (0.04) & 6.88 (0.26) \\ 
  p = 100 & scad & 0.65 (0.02) & 5.65 (0.37) & 5.17 & 0.48 & 1.67 (0.09) & 1.62 (0.1) & 0.73 (0.04) & 9.69 (0.37) \\ 
   & vsd & 0.82 (0.02) & 1.71 (0.15) & 0.19 & 1.52 & 2.22 (0.15) & 1.9 (0.13) & 0.98 (0.06) & 3.67 (0.13) \\ 
   & bic & 0.72 (0.02) & 4.01 (0.32) & 3.41 & 0.6 & 1.69 (0.1) & 1.67 (0.11) & 0.74 (0.04) & 7.81 (0.31) \\ 
   & \textbf{ebic} & 0.75 (0.02) & 3.18 (0.28) & 2.52 & 0.66 & \textbf{1.7} (0.1) & \textbf{1.62} (0.11) & \textbf{0.74} (0.04) & 6.86 (0.27) \\ 
   & \textbf{cv} & \underline{0.47} (0.02) & \underline{14.49} (1.1) & \underline{14.18} & \textbf{0.31} & 1.8 (0.09) & \underline{2.66} (0.15) & \underline{0.85} (0.03) & 18.87 (1.12) \\ 
   & \textbf{csuv.m.0} & \textbf{0.81} (0.02) & \textbf{1.81} (0.15) & \textbf{0.82} & \underline{0.99} & \underline{1.95} (0.13) & 1.73 (0.11) & \underline{0.85} (0.05) & 4.83 (0.13) \\ 
   & \textbf{csuv.s.0} & 0.69 (0.01) & 4.03 (0.23) & 3.34 & 0.69 & 1.84 (0.09) & 2.05 (0.1) & \underline{0.85} (0.04) & 7.65 (0.23) \\ 
   & csuv.m.5 & 0.82 (0.02) & 1.7 (0.14) & 0.67 & 1.03 & 1.95 (0.13) & 1.71 (0.11) & 0.85 (0.05) & 4.64 (0.12) \\ 
   & csuv.s.5 & 0.67 (0.01) & 4.43 (0.21) & 3.73 & 0.7 & 1.84 (0.09) & 2.12 (0.1) & 0.86 (0.04) & 8.03 (0.21) \\ 
   & csuv.m.0.all & 0.81 (0.02) & 1.93 (0.18) & 0.98 & 0.95 & 1.87 (0.11) & 1.71 (0.11) & 0.83 (0.05) & 5.03 (0.15) \\ 
   & csuv.s.0.all & 0.66 (0.01) & 4.67 (0.24) & 4 & 0.67 & 1.84 (0.09) & 2.14 (0.1) & 0.86 (0.04) & 8.33 (0.26) \\ 
   & csuv.m.0.mcp & 0.78 (0.02) & 1.73 (0.13) & 0.15 & 1.58 & 2.32 (0.14) & 2.03 (0.11) & 1.04 (0.05) & 3.57 (0.1) \\ 
   & csuv.s.0.mcp & 0.78 (0.02) & 2.19 (0.17) & 1.08 & 1.11 & 2.03 (0.11) & 1.91 (0.11) & 0.92 (0.05) & 4.97 (0.1) \\ 
   \hline
setting 2 & lasso & 0.3 (0.01) & 26.02 (1.59) & 25.54 & 0.48 & 2.19 (0.08) & 3.84 (0.15) & 1.09 (0.03) & 30.06 (1.64) \\ 
   & elastic net & 0.24 (0.01) & 32.73 (1.45) & 32.32 & 0.41 & 2.41 (0.09) & 4.46 (0.14) & 1.19 (0.03) & 36.91 (1.49) \\ 
  num.factor = 2 & relaxed lasso & 0.43 (0.01) & 14.49 (1.11) & 13.72 & 0.77 & 2.17 (0.09) & 3.44 (0.13) & 1.07 (0.03) & 17.95 (1.18) \\ 
  s = 5 & mcp & 0.68 (0.02) & 4.14 (0.28) & 3.19 & 0.95 & 2.02 (0.12) & 1.94 (0.13) & 0.89 (0.05) & 7.24 (0.26) \\ 
  p = 300 & scad & 0.53 (0.01) & 8.71 (0.45) & 8.12 & 0.59 & 1.83 (0.1) & 1.87 (0.11) & 0.8 (0.04) & 12.53 (0.45) \\ 
   & vsd & 0.68 (0.02) & 2.52 (0.15) & 0.19 & 2.33 & 3.16 (0.22) & 2.61 (0.14) & 1.32 (0.06) & 2.86 (0.13) \\ 
   & bic & 0.57 (0.02) & 10.8 (1.67) & 10.12 & 0.68 & 1.91 (0.11) & 2.35 (0.2) & 0.86 (0.05) & 14.44 (1.7) \\ 
   & \textbf{ebic} & 0.66 (0.02) & 5.83 (1.15) & 4.94 & 0.89 & \textbf{2.03} (0.13) & \textbf{2.08} (0.16) & \textbf{0.88} (0.05) & 9.05 (1.17) \\ 
   & \textbf{cv} & \underline{0.4} (0.02) & \underline{20.34} (1.67) & \underline{19.83} & \textbf{0.51} & 2.12 (0.09) & \underline{3.26} (0.17) & 1.01 (0.04) & 24.32 (1.7) \\ 
   & \textbf{csuv.m.0} & \textbf{0.75} (0.02) & \textbf{2.09} (0.13) & \textbf{0.52} & \underline{1.57} & \underline{2.28} (0.12) & 2.11 (0.1) & \underline{1.07} (0.05) & 3.95 (0.14) \\ 
   & \textbf{csuv.s.0} & 0.62 (0.01) & 5.37 (0.27) & 4.44 & 0.93 & 2.05 (0.09) & 2.49 (0.1) & 0.99 (0.04) & 8.51 (0.32) \\ 
   & csuv.m.5 & 0.74 (0.02) & 2.13 (0.14) & 0.44 & 1.69 & 2.42 (0.13) & 2.18 (0.11) & 1.11 (0.05) & 3.75 (0.13) \\ 
   & csuv.s.5 & 0.58 (0.01) & 6.04 (0.25) & 5.06 & 0.98 & 2.11 (0.09) & 2.63 (0.09) & 1.02 (0.04) & 9.08 (0.31) \\ 
   & csuv.m.0.all & 0.75 (0.02) & 2.15 (0.13) & 0.64 & 1.51 & 2.22 (0.11) & 2.1 (0.1) & 1.05 (0.05) & 4.13 (0.15) \\ 
   & csuv.s.0.all & 0.59 (0.01) & 6.2 (0.32) & 5.31 & 0.89 & 2.07 (0.09) & 2.58 (0.1) & 1 (0.04) & 9.42 (0.37) \\ 
   & csuv.m.0.mcp & 0.66 (0.02) & 2.47 (0.12) & 0.04 & 2.43 & 2.92 (0.15) & 2.61 (0.1) & 1.33 (0.04) & 2.61 (0.12) \\ 
   & csuv.s.0.mcp & 0.71 (0.02) & 2.63 (0.15) & 1.03 & 1.6 & 2.57 (0.15) & 2.41 (0.12) & 1.17 (0.05) & 4.43 (0.13) \\ 
   \hline
setting 3 & lasso & 0.45 (0.01) & 25.15 (0.93) & 24.76 & 0.39 & 2.6 (0.11) & 5.16 (0.15) & 1.23 (0.03) & 34.37 (0.95) \\ 
   & elastic net & 0.41 (0.01) & 29.57 (0.97) & 29.24 & 0.33 & 2.73 (0.11) & 5.84 (0.15) & 1.3 (0.02) & 38.91 (0.99) \\ 
  num.factor = 2 & relaxed lasso & 0.56 (0.01) & 16.19 (0.9) & 15.26 & 0.93 & 2.74 (0.12) & 4.82 (0.14) & 1.25 (0.03) & 24.33 (1) \\ 
  s = 10 & mcp & 0.79 (0.01) & 4.32 (0.24) & 2.68 & 1.64 & 2.62 (0.16) & 3.08 (0.16) & 1.14 (0.05) & 11.04 (0.22) \\ 
  p = 100 & scad & 0.74 (0.01) & 6.33 (0.27) & 5.11 & 1.22 & 2.49 (0.14) & 3.02 (0.13) & 1.11 (0.05) & 13.89 (0.29) \\ 
   & vsd & 0.71 (0.02) & 4.26 (0.24) & 0.19 & 4.07 & 4.24 (0.29) & 4.36 (0.19) & 1.67 (0.06) & 6.12 (0.24) \\ 
   & bic & 0.79 (0.01) & 4.77 (0.26) & 3.47 & 1.3 & 2.47 (0.14) & 2.96 (0.15) & 1.09 (0.05) & 12.17 (0.25) \\ 
   & \textbf{ebic} & \textbf{0.8} (0.01) & 4.32 (0.23) & 2.85 & 1.47 & \textbf{2.56} (0.14) & \textbf{2.98} (0.15) & \textbf{1.11} (0.05) & 11.38 (0.25) \\ 
   & \textbf{cv} & \underline{0.47} (0.01) & \underline{24} (1.03) & \underline{23.48} & \textbf{0.52} & 2.67 (0.11) & \underline{5.1} (0.15) & 1.24 (0.03) & 32.96 (1.09) \\ 
   & \textbf{csuv.m.0} & 0.79 (0.01) & \textbf{3.71} (0.16) & \textbf{1.05} & \underline{2.66} & \underline{3.29} (0.17) & 3.66 (0.14) & \underline{1.4} (0.04) & 8.39 (0.2) \\ 
   & \textbf{csuv.s.0} & 0.71 (0.01) & 7.01 (0.29) & 5.28 & 1.73 & 2.82 (0.14) & 3.85 (0.12) & 1.27 (0.04) & 13.55 (0.34) \\ 
   & csuv.m.5 & 0.79 (0.01) & 3.73 (0.17) & 0.94 & 2.79 & 3.39 (0.18) & 3.71 (0.14) & 1.42 (0.05) & 8.15 (0.19) \\ 
   & csuv.s.5 & 0.7 (0.01) & 7.14 (0.27) & 5.45 & 1.69 & 2.87 (0.14) & 3.86 (0.12) & 1.26 (0.04) & 13.76 (0.32) \\ 
   & csuv.m.0.all & 0.8 (0.01) & 3.74 (0.17) & 1.25 & 2.49 & 3.19 (0.18) & 3.58 (0.13) & 1.37 (0.04) & 8.76 (0.21) \\ 
   & csuv.s.0.all & 0.69 (0.01) & 7.88 (0.33) & 6.3 & 1.58 & 2.84 (0.14) & 3.96 (0.13) & 1.26 (0.04) & 14.72 (0.37) \\ 
   & csuv.m.0.mcp & 0.67 (0.01) & 4.84 (0.15) & 0.11 & 4.73 & 5.18 (0.32) & 5.18 (0.16) & 1.95 (0.05) & 5.38 (0.16) \\ 
   & csuv.s.0.mcp & 0.75 (0.01) & 4.26 (0.19) & 0.93 & 3.33 & 4.23 (0.29) & 4.28 (0.17) & 1.61 (0.05) & 7.6 (0.17) \\ 
   \hline
\multicolumn{10}{l}{}\\
\end{tabular}
}
\endgroup
\caption{Model 4: performance of CSUV and methods it compares with. Variable selection performance in terms of F-measure (f), total error (FP+FN), false positives (FP) and false negatives (FN), prediction error in terms of mse (pred.err) and estimation error in terms of l1 and l2 distance (l1.diff and l2.diff) and are shown. The numbers are based on 100 simulations. The last 8 rows are the performance of CSUV with different parameters (e.g. csuv.m.0.mcp corresponds to CSUV with MCP as constituent method and r = 0). A bold number represents the best result among delete-n/2 cross validation, eBIC and CSUV using Lasso, MCP and SCAD while a underlined number represents the worst among them. Standard errors are shown inside the parentheses.} 
\label{4_1}
\end{table}

			% latex table generated in R 3.6.1 by xtable 1.8-4 package
% Mon Jan 13 15:32:14 2020
\begin{table}[ht]
\centering
\begingroup\small
\scalebox{0.75}{
\begin{tabular}{|l|l|l|l|ll|lll|l|}
  \hline
parameters & methods & f & FP+FN & FP & FN & pred.err & l1.diff & l2.diff & size \\ 
  \hline
setting 4 & lasso & 0.32 (0.01) & 40.26 (1.61) & 39.05 & 1.21 & 3.71 (0.18) & 7.52 (0.2) & 1.64 (0.03) & 47.84 (1.67) \\ 
   & elastic net & 0.28 (0.01) & 47.74 (1.62) & 46.54 & 1.2 & 4.11 (0.19) & 8.45 (0.19) & 1.74 (0.03) & 55.34 (1.68) \\ 
  num.factor = 2 & relaxed lasso & 0.43 (0.01) & 23.75 (1.41) & 21.6 & 2.15 & 3.87 (0.22) & 6.88 (0.21) & 1.67 (0.04) & 29.45 (1.59) \\ 
  s = 10 & mcp & 0.67 (0.01) & 7.02 (0.35) & 4.06 & 2.96 & 4 (0.26) & 4.64 (0.2) & 1.63 (0.06) & 11.1 (0.36) \\ 
  p = 300 & scad & 0.58 (0.01) & 11.98 (0.47) & 9.91 & 2.07 & 3.48 (0.22) & 4.36 (0.18) & 1.49 (0.06) & 17.84 (0.49) \\ 
   & vsd & 0.5 (0.02) & 7.11 (0.24) & 0.26 & 6.85 & 6.99 (0.38) & 6.89 (0.23) & 2.41 (0.07) & 3.41 (0.23) \\ 
   & bic & 0.54 (0.02) & 22.46 (2.35) & 20.64 & 1.82 & 3.43 (0.2) & 5.61 (0.3) & 1.52 (0.05) & 28.82 (2.45) \\ 
   & \textbf{ebic} & \textbf{0.66} (0.02) & 9.52 (1.27) & 7 & 2.52 & \textbf{3.63} (0.22) & \textbf{4.53} (0.21) & \textbf{1.53} (0.05) & 14.48 (1.35) \\ 
   & \textbf{cv} & \underline{0.36} (0.01) & \underline{36.39} (1.7) & \underline{35.02} & \textbf{1.37} & 3.64 (0.17) & \underline{7.12} (0.21) & 1.63 (0.03) & 43.65 (1.81) \\ 
   & \textbf{csuv.m.0} & 0.61 (0.02) & \textbf{5.79} (0.17) & \textbf{0.63} & \underline{5.16} & \underline{5.54} (0.31) & 5.6 (0.16) & \underline{2.03} (0.05) & 5.47 (0.21) \\ 
   & \textbf{csuv.s.0} & 0.56 (0.01) & 10.06 (0.34) & 6.51 & 3.55 & 4.4 (0.22) & 5.77 (0.17) & 1.79 (0.05) & 12.96 (0.47) \\ 
   & csuv.m.5 & 0.6 (0.02) & 5.89 (0.17) & 0.55 & 5.34 & 5.65 (0.31) & 5.74 (0.16) & 2.07 (0.05) & 5.21 (0.2) \\ 
   & csuv.s.5 & 0.55 (0.01) & 10.51 (0.33) & 6.95 & 3.56 & 4.49 (0.23) & 5.91 (0.16) & 1.81 (0.05) & 13.39 (0.46) \\ 
   & csuv.m.0.all & 0.62 (0.01) & 5.86 (0.18) & 0.85 & 5.01 & 5.38 (0.3) & 5.6 (0.16) & 2.02 (0.05) & 5.84 (0.22) \\ 
   & csuv.s.0.all & 0.55 (0.01) & 11.23 (0.39) & 7.89 & 3.34 & 4.32 (0.21) & 5.88 (0.16) & 1.77 (0.05) & 14.55 (0.51) \\ 
   & csuv.m.0.mcp & 0.46 (0.02) & 6.98 (0.16) & 0.05 & 6.93 & 7.18 (0.43) & 6.98 (0.17) & 2.46 (0.05) & 3.12 (0.16) \\ 
   & csuv.s.0.mcp & 0.58 (0.02) & 6.33 (0.2) & 0.9 & 5.43 & 5.84 (0.34) & 6.06 (0.18) & 2.16 (0.05) & 5.47 (0.18) \\ 
   \hline
setting 5 & lasso & 0.39 (0.01) & 15.08 (0.68) & 14.51 & 0.57 & 2.5 (0.09) & 4 (0.16) & 1.22 (0.03) & 18.94 (0.71) \\ 
   & elastic net & 0.33 (0.01) & 20.63 (0.96) & 20.2 & 0.43 & 2.76 (0.1) & 4.84 (0.21) & 1.32 (0.03) & 24.77 (0.97) \\ 
  num.factor = 10 & relaxed lasso & 0.42 (0.01) & 12.22 (0.73) & 11.21 & 1.01 & 2.83 (0.13) & 4.3 (0.21) & 1.32 (0.04) & 15.2 (0.8) \\ 
  s = 5 & mcp & 0.55 (0.02) & 5.41 (0.29) & 3.58 & 1.83 & 3.36 (0.21) & 3.69 (0.2) & 1.43 (0.07) & 6.75 (0.21) \\ 
  p = 100 & scad & 0.51 (0.02) & 6.87 (0.33) & 5.25 & 1.62 & 3.59 (0.25) & 3.64 (0.2) & 1.42 (0.07) & 8.63 (0.27) \\ 
   & vsd & 0.57 (0.03) & 3.84 (0.22) & 1.02 & 2.82 & 3.96 (0.24) & 3.65 (0.18) & 1.59 (0.07) & 3.2 (0.13) \\ 
   & bic & 0.57 (0.02) & 5.46 (0.29) & 3.83 & 1.63 & 3.36 (0.21) & 3.66 (0.2) & 1.4 (0.07) & 7.2 (0.23) \\ 
   & \textbf{ebic} & 0.57 (0.02) & 5.24 (0.29) & 3.52 & 1.72 & \underline{3.27} (0.21) & 3.56 (0.2) & \underline{1.38} (0.07) & 6.8 (0.21) \\ 
   & \textbf{cv} & \underline{0.39} (0.01) & \underline{14.96} (0.7) & \underline{14.3} & \textbf{0.66} & \textbf{2.53} (0.1) & \underline{4} (0.16) & \textbf{1.22} (0.04) & 18.64 (0.74) \\ 
   & \textbf{csuv.m.0} & \textbf{0.63} (0.02) & \textbf{3.61} (0.2) & \textbf{1.72} & \underline{1.89} & 2.86 (0.14) & \textbf{3.08} (0.14) & 1.31 (0.05) & 4.83 (0.14) \\ 
   & \textbf{csuv.s.0} & 0.53 (0.01) & 6.52 (0.22) & 5.24 & 1.28 & 2.61 (0.11) & 3.42 (0.13) & 1.25 (0.04) & 8.96 (0.23) \\ 
   & csuv.m.5 & 0.62 (0.02) & 3.48 (0.19) & 1.48 & 2 & 2.99 (0.15) & 3.12 (0.15) & 1.35 (0.05) & 4.48 (0.13) \\ 
   & csuv.s.5 & 0.52 (0.01) & 6.67 (0.21) & 5.34 & 1.33 & 2.63 (0.11) & 3.43 (0.13) & 1.26 (0.04) & 9.01 (0.21) \\ 
   & csuv.m.0.all & 0.62 (0.02) & 3.77 (0.21) & 1.93 & 1.84 & 2.82 (0.13) & 3.11 (0.14) & 1.31 (0.05) & 5.09 (0.15) \\ 
   & csuv.s.0.all & 0.52 (0.01) & 7.29 (0.25) & 6.11 & 1.18 & 2.61 (0.12) & 3.5 (0.13) & 1.25 (0.04) & 9.93 (0.27) \\ 
   & csuv.m.0.mcp & 0.55 (0.02) & 3.45 (0.16) & 0.48 & 2.97 & 3.78 (0.21) & 3.43 (0.14) & 1.59 (0.05) & 2.51 (0.09) \\ 
   & csuv.s.0.mcp & 0.56 (0.02) & 4.49 (0.21) & 2.46 & 2.03 & 3.14 (0.15) & 3.49 (0.16) & 1.42 (0.05) & 5.43 (0.09) \\ 
   \hline
setting 6 & lasso & 0.28 (0.01) & 22.46 (1.13) & 21.37 & 1.09 & 3.19 (0.15) & 5.23 (0.2) & 1.46 (0.03) & 25.28 (1.15) \\ 
   & elastic net & 0.23 (0.01) & 30.03 (1.38) & 29.08 & 0.95 & 3.48 (0.15) & 5.98 (0.21) & 1.55 (0.03) & 33.13 (1.39) \\ 
  num.factor = 10 & relaxed lasso & 0.32 (0.01) & 15.45 (0.87) & 13.71 & 1.74 & 3.55 (0.19) & 5.2 (0.21) & 1.56 (0.04) & 16.97 (0.92) \\ 
  s = 5 & mcp & 0.42 (0.02) & 7.03 (0.29) & 4.27 & 2.76 & 4.64 (0.27) & 4.86 (0.21) & 1.79 (0.06) & 6.51 (0.22) \\ 
  p = 300 & scad & 0.37 (0.02) & 9.32 (0.3) & 7.05 & 2.27 & 4.29 (0.29) & 4.58 (0.22) & 1.71 (0.07) & 9.78 (0.25) \\ 
   & vsd & 0.48 (0.02) & 4.78 (0.18) & 1.06 & 3.72 & 5.31 (0.3) & 4.62 (0.16) & 1.98 (0.06) & 2.34 (0.12) \\ 
   & bic & 0.41 (0.02) & 10.11 (1.19) & 7.7 & 2.41 & 4.57 (0.31) & 5.2 (0.28) & 1.77 (0.07) & 10.29 (1.24) \\ 
   & \textbf{ebic} & 0.42 (0.02) & 7.11 (0.31) & 4.46 & 2.65 & \underline{4.69} (0.32) & 4.83 (0.23) & \underline{1.77} (0.07) & 6.81 (0.24) \\ 
   & \textbf{cv} & \underline{0.29} (0.01) & \underline{21.97} (1.16) & \underline{20.83} & \textbf{1.14} & \textbf{3.24} (0.16) & \underline{5.19} (0.21) & \textbf{1.47} (0.03) & 24.69 (1.19) \\ 
   & \textbf{csuv.m.0} & \textbf{0.5} (0.02) & \textbf{4.27} (0.16) & \textbf{1.35} & \underline{2.92} & 3.96 (0.2) & \textbf{3.85} (0.12) & 1.66 (0.04) & 3.43 (0.12) \\ 
   & \textbf{csuv.s.0} & 0.39 (0.01) & 9.17 (0.28) & 7.05 & 2.12 & 3.61 (0.18) & 4.51 (0.14) & 1.56 (0.04) & 9.93 (0.29) \\ 
   & csuv.m.5 & 0.49 (0.02) & 4.23 (0.15) & 1.25 & 2.98 & 4.03 (0.2) & 3.86 (0.12) & 1.68 (0.04) & 3.27 (0.13) \\ 
   & csuv.s.5 & 0.39 (0.01) & 9.12 (0.28) & 6.97 & 2.15 & 3.68 (0.18) & 4.57 (0.14) & 1.58 (0.04) & 9.82 (0.28) \\ 
   & csuv.m.0.all & 0.5 (0.02) & 4.42 (0.18) & 1.67 & 2.75 & 3.82 (0.2) & 3.81 (0.13) & 1.61 (0.04) & 3.92 (0.14) \\ 
   & csuv.s.0.all & 0.37 (0.01) & 10.27 (0.32) & 8.26 & 2.01 & 3.52 (0.16) & 4.63 (0.14) & 1.55 (0.04) & 11.25 (0.34) \\ 
   & csuv.m.0.mcp & 0.43 (0.01) & 4.38 (0.12) & 0.37 & 4.01 & 4.91 (0.26) & 4.23 (0.12) & 1.93 (0.04) & 1.36 (0.08) \\ 
   & csuv.s.0.mcp & 0.41 (0.02) & 5.95 (0.17) & 2.92 & 3.03 & 4.48 (0.24) & 4.63 (0.13) & 1.8 (0.04) & 4.89 (0.1) \\ 
   \hline
\multicolumn{10}{l}{}\\
\end{tabular}
}
\endgroup
\caption{Model 4: performance of CSUV and methods it compares with (continue). Variable selection performance in terms of F-measure (f), total error (FP+FN), false positives (FP) and false negatives (FN), prediction error in terms of mse (pred.err) and estimation error in terms of l1 and l2 distance (l1.diff and l2.diff) and are shown. The numbers are based on 100 simulations. The last 8 rows are the performance of CSUV with different parameters (e.g. csuv.m.0.mcp corresponds to CSUV with MCP as constituent method and r = 0). A bold number represents the best result among delete-n/2 cross validation, eBIC and CSUV using Lasso, MCP and SCAD while a underlined number represents the worst among them. Standard errors are shown inside the parentheses.} 
\label{4_2}
\end{table}

			% latex table generated in R 3.6.1 by xtable 1.8-4 package
% Mon Jan 13 15:32:14 2020
\begin{table}[ht]
\centering
\begingroup\small
\scalebox{0.75}{
\begin{tabular}{|l|l|l|l|ll|lll|l|}
  \hline
parameters & methods & f & FP+FN & FP & FN & pred.err & l1.diff & l2.diff & size \\ 
  \hline
setting 7 & lasso & 0.44 (0.01) & 21.75 (0.68) & 19.99 & 1.76 & 4.57 (0.2) & 7.63 (0.19) & 1.84 (0.04) & 28.23 (0.77) \\ 
   & elastic net & 0.4 (0.01) & 27.39 (0.79) & 26.07 & 1.32 & 4.83 (0.23) & 8.52 (0.24) & 1.89 (0.04) & 34.75 (0.86) \\ 
  num.factor = 10 & relaxed lasso & 0.46 (0.01) & 19.07 (0.88) & 16.61 & 2.46 & 5.34 (0.26) & 8.24 (0.27) & 2 (0.04) & 24.15 (1.05) \\ 
  s = 10 & mcp & 0.51 (0.02) & 9.26 (0.28) & 4.25 & 5.01 & 7.58 (0.44) & 8.19 (0.24) & 2.47 (0.06) & 9.24 (0.27) \\ 
  p = 100 & scad & 0.51 (0.02) & 9.76 (0.32) & 5.05 & 4.71 & 7.41 (0.43) & 7.93 (0.25) & 2.43 (0.06) & 10.34 (0.22) \\ 
   & vsd & 0.42 (0.02) & 8.28 (0.23) & 1.28 & 7 & 8.89 (0.52) & 8.34 (0.23) & 2.7 (0.06) & 4.28 (0.18) \\ 
   & bic & 0.55 (0.02) & 8.96 (0.39) & 4.57 & 4.39 & 6.89 (0.41) & 7.68 (0.25) & 2.32 (0.06) & 10.18 (0.38) \\ 
   & \textbf{ebic} & 0.55 (0.02) & 8.62 (0.3) & 3.98 & 4.64 & \underline{7} (0.41) & \underline{7.7} (0.25) & \underline{2.36} (0.06) & 9.34 (0.25) \\ 
   & \textbf{cv} & \underline{0.44} (0.01) & \underline{21.75} (0.68) & \underline{19.99} & \textbf{1.76} & \textbf{4.57} (0.2) & 7.63 (0.19) & \textbf{1.84} (0.04) & 28.23 (0.77) \\ 
   & \textbf{csuv.m.0} & \textbf{0.58} (0.01) & \textbf{7.1} (0.2) & \textbf{2.14} & \underline{4.96} & 6.13 (0.36) & \textbf{6.78} (0.17) & 2.19 (0.04) & 7.18 (0.2) \\ 
   & \textbf{csuv.s.0} & 0.53 (0.01) & 10.88 (0.25) & 7.16 & 3.72 & 5.34 (0.24) & 7.08 (0.18) & 2.05 (0.04) & 13.44 (0.32) \\ 
   & csuv.m.5 & 0.57 (0.01) & 7.11 (0.2) & 1.98 & 5.13 & 6.45 (0.38) & 6.84 (0.17) & 2.22 (0.04) & 6.85 (0.21) \\ 
   & csuv.s.5 & 0.53 (0.01) & 10.87 (0.25) & 7.15 & 3.72 & 5.25 (0.22) & 7.05 (0.18) & 2.04 (0.04) & 13.43 (0.32) \\ 
   & csuv.m.0.all & 0.58 (0.01) & 7.41 (0.21) & 2.66 & 4.75 & 5.94 (0.32) & 6.78 (0.18) & 2.16 (0.05) & 7.91 (0.22) \\ 
   & csuv.s.0.all & 0.52 (0.01) & 11.72 (0.24) & 8.26 & 3.46 & 5.27 (0.22) & 7.18 (0.17) & 2.03 (0.04) & 14.8 (0.33) \\ 
   & csuv.m.0.mcp & 0.39 (0.01) & 8.05 (0.16) & 0.53 & 7.52 & 8.89 (0.47) & 8.21 (0.17) & 2.73 (0.05) & 3.01 (0.12) \\ 
   & csuv.s.0.mcp & 0.53 (0.01) & 7.86 (0.23) & 2.4 & 5.46 & 7.07 (0.38) & 7.58 (0.2) & 2.38 (0.05) & 6.94 (0.13) \\ 
   \hline
setting 8 & lasso & 0.31 (0.01) & 32.71 (1.19) & 29.58 & 3.13 & 6.43 (0.26) & 10.07 (0.22) & 2.25 (0.03) & 36.45 (1.33) \\ 
   & elastic net & 0.28 (0.01) & 38.5 (1.24) & 35.73 & 2.77 & 6.7 (0.26) & 10.7 (0.19) & 2.29 (0.03) & 42.96 (1.35) \\ 
  num.factor = 10 & relaxed lasso & 0.32 (0.01) & 25.3 (1.17) & 21.11 & 4.19 & 7.22 (0.31) & 10.41 (0.25) & 2.39 (0.04) & 26.92 (1.43) \\ 
  s = 10 & mcp & 0.35 (0.01) & 12.31 (0.29) & 5.46 & 6.85 & 9.21 (0.39) & 10.25 (0.24) & 2.91 (0.05) & 8.61 (0.23) \\ 
  p = 300 & scad & 0.34 (0.01) & 13.98 (0.35) & 7.49 & 6.49 & 9.88 (0.46) & 9.88 (0.25) & 2.89 (0.06) & 11 (0.31) \\ 
   & vsd & 0.25 (0.01) & 10.75 (0.15) & 1.72 & 9.03 & 12.25 (0.59) & 10.71 (0.17) & 3.36 (0.05) & 2.69 (0.13) \\ 
   & bic & 0.36 (0.02) & 16.89 (1.3) & 11.03 & 5.86 & 9.08 (0.44) & 10.25 (0.28) & 2.76 (0.06) & 15.17 (1.52) \\ 
   & \textbf{ebic} & 0.37 (0.01) & 12.05 (0.31) & 5.52 & 6.53 & \underline{9.34} (0.44) & 9.78 (0.23) & \underline{2.82} (0.05) & 8.99 (0.26) \\ 
   & \textbf{cv} & \underline{0.31} (0.01) & \underline{32.71} (1.19) & \underline{29.58} & \textbf{3.13} & \textbf{6.43} (0.26) & \underline{10.07} (0.22) & \textbf{2.25} (0.03) & 36.45 (1.33) \\ 
   & \textbf{csuv.m.0} & \textbf{0.41} (0.01) & \textbf{9} (0.2) & \textbf{2.15} & \underline{6.85} & 8.16 (0.35) & \textbf{8.41} (0.18) & 2.65 (0.05) & 5.3 (0.2) \\ 
   & \textbf{csuv.s.0} & 0.37 (0.01) & 15.23 (0.26) & 9.75 & 5.48 & 7.55 (0.3) & 9.2 (0.16) & 2.47 (0.04) & 14.27 (0.35) \\ 
   & csuv.m.5 & 0.42 (0.01) & 8.87 (0.2) & 1.88 & 6.99 & 8.32 (0.36) & 8.38 (0.19) & 2.68 (0.05) & 4.89 (0.19) \\ 
   & csuv.s.5 & 0.37 (0.01) & 15.25 (0.26) & 9.77 & 5.48 & 7.41 (0.28) & 9.2 (0.16) & 2.47 (0.04) & 14.29 (0.34) \\ 
   & csuv.m.0.all & 0.42 (0.01) & 9.23 (0.22) & 2.56 & 6.67 & 8.15 (0.34) & 8.46 (0.18) & 2.62 (0.04) & 5.89 (0.2) \\ 
   & csuv.s.0.all & 0.36 (0.01) & 16.64 (0.29) & 11.47 & 5.17 & 7.39 (0.29) & 9.41 (0.16) & 2.45 (0.04) & 16.3 (0.38) \\ 
   & csuv.m.0.mcp & 0.24 (0.01) & 9.66 (0.13) & 0.65 & 9.01 & 11.13 (0.5) & 9.61 (0.16) & 3.17 (0.05) & 1.64 (0.09) \\ 
   & csuv.s.0.mcp & 0.34 (0.01) & 10.65 (0.22) & 3.27 & 7.38 & 9.61 (0.45) & 9.75 (0.19) & 2.92 (0.05) & 5.89 (0.1) \\ 
   \hline
\multicolumn{10}{l}{}\\
\end{tabular}
}
\endgroup
\caption{Model 4: performance of CSUV and methods it compares with (continue). Variable selection performance in terms of F-measure (f), total error (FP+FN), false positives (FP) and false negatives (FN), prediction error in terms of mse (pred.err) and estimation error in terms of l1 and l2 distance (l1.diff and l2.diff) and are shown. The numbers are based on 100 simulations. The last 8 rows are the performance of CSUV with different parameters (e.g. csuv.m.0.mcp corresponds to CSUV with MCP as constituent method and r = 0). A bold number represents the best result among delete-n/2 cross validation, eBIC and CSUV using Lasso, MCP and SCAD while a underlined number represents the worst among them. Standard errors are shown inside the parentheses.} 
\label{4_3}
\end{table}

			% latex table generated in R 3.6.1 by xtable 1.8-4 package
% Mon Jan 13 15:32:14 2020
\begin{table}[ht]
\centering
\begingroup\small
\scalebox{0.75}{
\begin{tabular}{|l|l|l|l|ll|lll|l|}
  \hline
parameters & methods & f & FP+FN & FP & FN & pred.err & l1.diff & l2.diff & size \\ 
  \hline
setting 1 & lasso & 0.58 (0.02) & 9.6 (0.72) & 9.6 & 0 & 1.18 (0.02) & 1.19 (0.05) & 0.42 (0.01) & 14.6 (0.72) \\ 
   & elastic net & 0.49 (0.02) & 12.7 (0.76) & 12.7 & 0 & 1.23 (0.02) & 1.42 (0.06) & 0.46 (0.01) & 17.7 (0.76) \\ 
  rho = 0.5 & relaxed lasso & 0.95 (0.01) & 0.84 (0.32) & 0.84 & 0 & 1.08 (0.02) & 0.67 (0.04) & 0.32 (0.01) & 5.84 (0.32) \\ 
  s = 5 & mcp & 1 (0) & 0 (0) & 0 & 0 & 1.06 (0.02) & 0.6 (0.03) & 0.33 (0.01) & 5 (0) \\ 
  p = 100 & scad & 1 (0) & 0.01 (0.01) & 0.01 & 0 & 1.21 (0.03) & 1.02 (0.05) & 0.57 (0.03) & 5.01 (0.01) \\ 
   & vsd & 1 (0) & 0.03 (0.02) & 0.03 & 0 & 1.04 (0.02) & 0.54 (0.02) & 0.29 (0.01) & 5.03 (0.02) \\ 
   & bic & 0.92 (0.02) & 1.41 (0.33) & 1.41 & 0 & 1.1 (0.02) & 0.73 (0.04) & 0.36 (0.02) & 6.41 (0.33) \\ 
   & \textbf{ebic} & 1 (0) & \underline{0.01} (0.01) & \underline{0.01} & 0 & \underline{1.07} (0.02) & \underline{0.62} (0.03) & \underline{0.33} (0.02) & 5.01 (0.01) \\ 
   & \textbf{cv} & 1 (0) & \textbf{0} (0) & \textbf{0} & 0 & 1.06 (0.02) & 0.6 (0.03) & \underline{0.33} (0.01) & 5 (0) \\ 
   & \textbf{csuv.m.0} & 1 (0) & \textbf{0} (0) & \textbf{0} & 0 & \textbf{1.04} (0.02) & \textbf{0.52} (0.02) & \textbf{0.28} (0.01) & 5 (0) \\ 
   & \textbf{csuv.s.0} & 1 (0) & \textbf{0} (0) & \textbf{0} & 0 & \textbf{1.04} (0.02) & \textbf{0.52} (0.02) & \textbf{0.28} (0.01) & 5 (0) \\ 
   & csuv.m.5 & 1 (0) & 0 (0) & 0 & 0 & 1.04 (0.02) & 0.52 (0.02) & 0.28 (0.01) & 5 (0) \\ 
   & csuv.s.5 & 1 (0) & 0 (0) & 0 & 0 & 1.04 (0.02) & 0.52 (0.02) & 0.28 (0.01) & 5 (0) \\ 
   & csuv.m.0.all & 1 (0) & 0.01 (0.01) & 0.01 & 0 & 1.04 (0.02) & 0.53 (0.02) & 0.28 (0.01) & 5.01 (0.01) \\ 
   & csuv.s.0.all & 0.99 (0) & 0.06 (0.02) & 0.06 & 0 & 1.05 (0.02) & 0.55 (0.02) & 0.29 (0.01) & 5.06 (0.02) \\ 
   & csuv.m.0.mcp & 1 (0) & 0 (0) & 0 & 0 & 1.04 (0.02) & 0.52 (0.02) & 0.28 (0.01) & 5 (0) \\ 
   & csuv.s.0.mcp & 1 (0) & 0 (0) & 0 & 0 & 1.04 (0.02) & 0.52 (0.02) & 0.28 (0.01) & 5 (0) \\ 
   \hline
setting 2 & lasso & 0.49 (0.02) & 14.89 (1.18) & 14.89 & 0 & 1.23 (0.02) & 1.46 (0.07) & 0.45 (0.01) & 19.89 (1.18) \\ 
   & elastic net & 0.39 (0.01) & 18.8 (1.03) & 18.8 & 0 & 1.32 (0.02) & 1.74 (0.06) & 0.52 (0.01) & 23.8 (1.03) \\ 
  rho = 0.5 & relaxed lasso & 0.95 (0.01) & 0.99 (0.4) & 0.99 & 0 & 1.08 (0.02) & 0.65 (0.05) & 0.31 (0.01) & 5.99 (0.4) \\ 
  s = 5 & mcp & 1 (0) & 0.03 (0.02) & 0.03 & 0 & 1.06 (0.02) & 0.57 (0.03) & 0.3 (0.01) & 5.03 (0.02) \\ 
  p = 300 & scad & 0.99 (0) & 0.07 (0.03) & 0.07 & 0 & 1.18 (0.02) & 0.94 (0.05) & 0.52 (0.03) & 5.07 (0.03) \\ 
   & vsd & 1 (0) & 0.01 (0.01) & 0.01 & 0 & 1.05 (0.02) & 0.49 (0.02) & 0.26 (0.01) & 5.01 (0.01) \\ 
   & bic & 0.6 (0.03) & 10.86 (1.11) & 10.86 & 0 & 1.21 (0.02) & 1.28 (0.07) & 0.43 (0.02) & 15.86 (1.11) \\ 
   & \textbf{ebic} & \underline{0.99} (0) & \underline{0.08} (0.03) & \underline{0.08} & 0 & \underline{1.07} (0.02) & \underline{0.61} (0.04) & \underline{0.32} (0.02) & 5.08 (0.03) \\ 
   & \textbf{cv} & \textbf{1} (0) & 0.03 (0.02) & 0.03 & 0 & 1.06 (0.02) & 0.57 (0.03) & 0.3 (0.01) & 5.03 (0.02) \\ 
   & \textbf{csuv.m.0} & \textbf{1} (0) & \textbf{0.01} (0.01) & \textbf{0.01} & 0 & \textbf{1.05} (0.02) & \textbf{0.49} (0.02) & \textbf{0.26} (0.01) & 5.01 (0.01) \\ 
   & \textbf{csuv.s.0} & \textbf{1} (0) & 0.03 (0.02) & 0.03 & 0 & \textbf{1.05} (0.02) & 0.5 (0.02) & \textbf{0.26} (0.01) & 5.03 (0.02) \\ 
   & csuv.m.5 & 1 (0) & 0.01 (0.01) & 0.01 & 0 & 1.05 (0.02) & 0.49 (0.02) & 0.26 (0.01) & 5.01 (0.01) \\ 
   & csuv.s.5 & 1 (0) & 0.01 (0.01) & 0.01 & 0 & 1.05 (0.02) & 0.49 (0.02) & 0.26 (0.01) & 5.01 (0.01) \\ 
   & csuv.m.0.all & 1 (0) & 0.03 (0.02) & 0.03 & 0 & 1.05 (0.02) & 0.5 (0.02) & 0.26 (0.01) & 5.03 (0.02) \\ 
   & csuv.s.0.all & 0.99 (0) & 0.14 (0.03) & 0.14 & 0 & 1.07 (0.02) & 0.55 (0.03) & 0.28 (0.01) & 5.14 (0.03) \\ 
   & csuv.m.0.mcp & 1 (0) & 0 (0) & 0 & 0 & 1.04 (0.02) & 0.49 (0.02) & 0.26 (0.01) & 5 (0) \\ 
   & csuv.s.0.mcp & 1 (0) & 0 (0) & 0 & 0 & 1.04 (0.02) & 0.49 (0.02) & 0.26 (0.01) & 5 (0) \\ 
   \hline
setting 3 & lasso & 0.65 (0.01) & 12.17 (0.65) & 12.17 & 0 & 1.29 (0.02) & 1.97 (0.06) & 0.56 (0.01) & 22.17 (0.65) \\ 
   & elastic net & 0.6 (0.01) & 14.68 (0.71) & 14.68 & 0 & 1.34 (0.02) & 2.2 (0.06) & 0.59 (0.01) & 24.68 (0.71) \\ 
  rho = 0.5 & relaxed lasso & 0.97 (0.01) & 0.89 (0.32) & 0.85 & 0.04 & 1.15 (0.02) & 1.27 (0.05) & 0.46 (0.01) & 10.81 (0.32) \\ 
  s = 10 & mcp & 0.93 (0.01) & 1.23 (0.09) & 0.02 & 1.21 & 1.91 (0.05) & 3.43 (0.1) & 1.34 (0.04) & 8.81 (0.09) \\ 
  p = 100 & scad & 0.95 (0.01) & 0.87 (0.09) & 0.12 & 0.75 & 2.41 (0.06) & 4.35 (0.09) & 1.65 (0.03) & 9.37 (0.09) \\ 
   & vsd & 0.99 (0) & 0.25 (0.05) & 0 & 0.25 & 1.16 (0.02) & 1.29 (0.06) & 0.52 (0.02) & 9.75 (0.05) \\ 
   & bic & 0.83 (0.02) & 5.52 (0.73) & 5.44 & 0.08 & 1.66 (0.05) & 2.72 (0.11) & 0.95 (0.05) & 15.36 (0.74) \\ 
   & \textbf{ebic} & 0.94 (0.01) & 1.24 (0.19) & 0.71 & \underline{0.53} & \underline{1.95} (0.06) & \underline{3.37} (0.12) & \underline{1.29} (0.05) & 10.18 (0.22) \\ 
   & \textbf{cv} & \underline{0.67} (0.01) & \underline{11.01} (0.68) & \underline{10.83} & 0.18 & 1.37 (0.04) & 2.15 (0.09) & 0.65 (0.03) & 20.65 (0.73) \\ 
   & \textbf{csuv.m.0} & \textbf{0.99} (0) & \textbf{0.14} (0.03) & \textbf{0.02} & 0.12 & \textbf{1.13} (0.02) & 1.17 (0.04) & 0.46 (0.02) & 9.9 (0.04) \\ 
   & \textbf{csuv.s.0} & \textbf{0.99} (0) & 0.21 (0.05) & 0.15 & \textbf{0.06} & \textbf{1.13} (0.02) & \textbf{1.15} (0.04) & \textbf{0.44} (0.02) & 10.09 (0.04) \\ 
   & csuv.m.5 & 0.99 (0) & 0.18 (0.04) & 0.01 & 0.17 & 1.15 (0.02) & 1.2 (0.05) & 0.48 (0.02) & 9.84 (0.04) \\ 
   & csuv.s.5 & 0.99 (0) & 0.3 (0.06) & 0.23 & 0.07 & 1.14 (0.02) & 1.18 (0.04) & 0.45 (0.02) & 10.16 (0.05) \\ 
   & csuv.m.0.all & 0.99 (0) & 0.13 (0.03) & 0.04 & 0.09 & 1.13 (0.02) & 1.15 (0.04) & 0.45 (0.02) & 9.95 (0.04) \\ 
   & csuv.s.0.all & 0.98 (0) & 0.46 (0.07) & 0.42 & 0.04 & 1.14 (0.02) & 1.21 (0.04) & 0.46 (0.01) & 10.38 (0.07) \\ 
   & csuv.m.0.mcp & 0.91 (0.01) & 1.62 (0.09) & 0 & 1.62 & 1.74 (0.05) & 2.88 (0.13) & 1.17 (0.05) & 8.38 (0.09) \\ 
   & csuv.s.0.mcp & 0.89 (0) & 1.92 (0.05) & 0 & 1.92 & 1.91 (0.05) & 3.23 (0.09) & 1.32 (0.03) & 8.08 (0.05) \\ 
   \hline
\multicolumn{10}{l}{}\\
\end{tabular}
}
\endgroup
\caption{Model 5: performance of CSUV and methods it compares with. Variable selection performance in terms of F-measure (f), total error (FP+FN), false positives (FP) and false negatives (FN), prediction error in terms of mse (pred.err) and estimation error in terms of l1 and l2 distance (l1.diff and l2.diff) and are shown. The numbers are based on 100 simulations. The last 8 rows are the performance of CSUV with different parameters (e.g. csuv.m.0.mcp corresponds to CSUV with MCP as constituent method and r = 0). A bold number represents the best result among delete-n/2 cross validation, eBIC and CSUV using Lasso, MCP and SCAD while a underlined number represents the worst among them. Standard errors are shown inside the parentheses.} 
\label{5_1}
\end{table}

			% latex table generated in R 3.6.1 by xtable 1.8-4 package
% Mon Jan 13 15:32:14 2020
\begin{table}[ht]
\centering
\begingroup\small
\scalebox{0.75}{
\begin{tabular}{|l|l|l|l|ll|lll|l|}
  \hline
parameters & methods & f & FP+FN & FP & FN & pred.err & l1.diff & l2.diff & size \\ 
  \hline
setting 4 & lasso & 0.55 (0.01) & 19.23 (1.1) & 19.23 & 0 & 1.4 (0.02) & 2.38 (0.07) & 0.62 (0.01) & 29.23 (1.1) \\ 
   & elastic net & 0.48 (0.01) & 24.02 (1.09) & 24.02 & 0 & 1.5 (0.03) & 2.76 (0.07) & 0.67 (0.01) & 34.02 (1.09) \\ 
  rho = 0.5 & relaxed lasso & 0.96 (0.01) & 1.3 (0.44) & 1.23 & 0.07 & 1.19 (0.02) & 1.39 (0.06) & 0.5 (0.02) & 11.16 (0.44) \\ 
  s = 10 & mcp & 0.92 (0) & 1.47 (0.09) & 0.13 & 1.34 & 1.89 (0.04) & 3.39 (0.09) & 1.33 (0.03) & 8.79 (0.08) \\ 
  p = 300 & scad & 0.93 (0.01) & 1.31 (0.11) & 0.51 & 0.8 & 2.41 (0.06) & 4.29 (0.09) & 1.62 (0.03) & 9.71 (0.12) \\ 
   & vsd & 0.96 (0) & 0.76 (0.07) & 0.02 & 0.74 & 1.35 (0.03) & 1.77 (0.08) & 0.74 (0.03) & 9.28 (0.07) \\ 
   & bic & 0.61 (0.02) & 16.37 (1.2) & 16.32 & 0.05 & 1.49 (0.04) & 2.54 (0.09) & 0.72 (0.03) & 26.27 (1.22) \\ 
   & \textbf{ebic} & 0.93 (0.01) & 1.44 (0.17) & 0.76 & \underline{0.68} & \underline{2.07} (0.06) & \underline{3.54} (0.11) & \underline{1.36} (0.04) & 10.08 (0.21) \\ 
   & \textbf{cv} & \underline{0.56} (0.02) & \underline{18.5} (1.16) & \underline{18.38} & \textbf{0.12} & 1.44 (0.03) & 2.49 (0.08) & 0.67 (0.02) & 28.26 (1.2) \\ 
   & \textbf{csuv.m.0} & \textbf{0.99} (0) & \textbf{0.27} (0.05) & \textbf{0.02} & 0.25 & 1.19 (0.02) & 1.3 (0.06) & 0.52 (0.02) & 9.77 (0.05) \\ 
   & \textbf{csuv.s.0} & 0.98 (0) & 0.4 (0.06) & 0.26 & 0.14 & \textbf{1.17} (0.02) & \textbf{1.26} (0.05) & \textbf{0.49} (0.02) & 10.12 (0.06) \\ 
   & csuv.m.5 & 0.98 (0) & 0.33 (0.05) & 0.02 & 0.31 & 1.21 (0.02) & 1.34 (0.06) & 0.54 (0.02) & 9.71 (0.05) \\ 
   & csuv.s.5 & 0.97 (0) & 0.72 (0.07) & 0.64 & 0.08 & 1.17 (0.02) & 1.3 (0.04) & 0.49 (0.02) & 10.56 (0.07) \\ 
   & csuv.m.0.all & 0.99 (0) & 0.24 (0.05) & 0.04 & 0.2 & 1.17 (0.02) & 1.26 (0.05) & 0.5 (0.02) & 9.84 (0.05) \\ 
   & csuv.s.0.all & 0.96 (0) & 0.8 (0.08) & 0.72 & 0.08 & 1.19 (0.02) & 1.32 (0.05) & 0.5 (0.02) & 10.64 (0.07) \\ 
   & csuv.m.0.mcp & 0.88 (0.01) & 2.1 (0.08) & 0 & 2.1 & 2.07 (0.08) & 3.49 (0.13) & 1.4 (0.05) & 7.9 (0.08) \\ 
   & csuv.s.0.mcp & 0.89 (0) & 2.01 (0.05) & 0 & 2.01 & 1.99 (0.05) & 3.36 (0.1) & 1.37 (0.03) & 7.99 (0.05) \\ 
   \hline
setting 5 & lasso & 0.3 (0.01) & 25.34 (0.85) & 25.34 & 0 & 1.46 (0.03) & 3.36 (0.09) & 0.92 (0.02) & 30.34 (0.85) \\ 
   & elastic net & 0.23 (0) & 34.05 (0.8) & 34.05 & 0 & 1.69 (0.04) & 4.51 (0.1) & 1.1 (0.02) & 39.05 (0.8) \\ 
  rho = -0.5 & relaxed lasso & 0.62 (0.01) & 7.04 (0.48) & 7.04 & 0 & 1.27 (0.03) & 1.76 (0.09) & 0.57 (0.02) & 12.04 (0.48) \\ 
  s = 5 & mcp & 0.98 (0.01) & 0.26 (0.06) & 0.26 & 0 & 1.05 (0.02) & 0.55 (0.02) & 0.28 (0.01) & 5.26 (0.06) \\ 
  p = 100 & scad & 0.96 (0.01) & 0.55 (0.11) & 0.55 & 0 & 1.05 (0.02) & 0.55 (0.02) & 0.28 (0.01) & 5.55 (0.11) \\ 
   & vsd & 1 (0) & 0.01 (0.01) & 0.01 & 0 & 1.05 (0.02) & 0.53 (0.02) & 0.28 (0.01) & 5.01 (0.01) \\ 
   & bic & 0.95 (0.01) & 0.6 (0.11) & 0.6 & 0 & 1.05 (0.02) & 0.56 (0.02) & 0.28 (0.01) & 5.6 (0.11) \\ 
   & \textbf{ebic} & \underline{0.96} (0.01) & \underline{0.43} (0.09) & \underline{0.43} & 0 & \textbf{1.05} (0.02) & 0.55 (0.02) & \textbf{0.28} (0.01) & 5.43 (0.09) \\ 
   & \textbf{cv} & 0.97 (0.01) & 0.41 (0.09) & 0.41 & 0 & \textbf{1.05} (0.02) & 0.55 (0.02) & \textbf{0.28} (0.01) & 5.41 (0.09) \\ 
   & \textbf{csuv.m.0} & \textbf{1} (0) & \textbf{0.03} (0.02) & \textbf{0.03} & 0 & \textbf{1.05} (0.02) & \textbf{0.54} (0.02) & \textbf{0.28} (0.01) & 5.03 (0.02) \\ 
   & \textbf{csuv.s.0} & 0.98 (0) & 0.24 (0.05) & 0.24 & 0 & \underline{1.07} (0.02) & \underline{0.61} (0.03) & \underline{0.31} (0.01) & 5.24 (0.05) \\ 
   & csuv.m.5 & 1 (0) & 0.03 (0.02) & 0.03 & 0 & 1.05 (0.02) & 0.54 (0.02) & 0.28 (0.01) & 5.03 (0.02) \\ 
   & csuv.s.5 & 0.94 (0.01) & 0.68 (0.07) & 0.68 & 0 & 1.1 (0.02) & 0.74 (0.03) & 0.36 (0.01) & 5.68 (0.07) \\ 
   & csuv.m.0.all & 1 (0) & 0.03 (0.02) & 0.03 & 0 & 1.05 (0.02) & 0.54 (0.02) & 0.28 (0.01) & 5.03 (0.02) \\ 
   & csuv.s.0.all & 0.98 (0) & 0.27 (0.05) & 0.27 & 0 & 1.07 (0.02) & 0.62 (0.03) & 0.32 (0.01) & 5.27 (0.05) \\ 
   & csuv.m.0.mcp & 1 (0) & 0.02 (0.01) & 0.02 & 0 & 1.05 (0.02) & 0.53 (0.02) & 0.28 (0.01) & 5.02 (0.01) \\ 
   & csuv.s.0.mcp & 1 (0) & 0.03 (0.02) & 0.03 & 0 & 1.05 (0.02) & 0.54 (0.02) & 0.28 (0.01) & 5.03 (0.02) \\ 
   \hline
setting 6 & lasso & 0.21 (0) & 40.76 (1.16) & 40.76 & 0 & 1.86 (0.05) & 5.06 (0.14) & 1.28 (0.03) & 45.76 (1.16) \\ 
   & elastic net & 0.15 (0) & 56.91 (1.19) & 56.9 & 0.01 & 2.72 (0.08) & 7.8 (0.19) & 1.79 (0.04) & 61.89 (1.19) \\ 
  rho = -0.5 & relaxed lasso & 0.41 (0.01) & 17.7 (1.17) & 17.64 & 0.06 & 1.74 (0.06) & 3.86 (0.2) & 0.98 (0.04) & 22.58 (1.18) \\ 
  s = 5 & mcp & 0.96 (0.01) & 0.48 (0.1) & 0.48 & 0 & 1.05 (0.02) & 0.56 (0.03) & 0.28 (0.01) & 5.48 (0.1) \\ 
  p = 300 & scad & 0.92 (0.01) & 1.09 (0.16) & 1.09 & 0 & 1.05 (0.02) & 0.56 (0.02) & 0.28 (0.01) & 6.09 (0.16) \\ 
   & vsd & 1 (0) & 0.04 (0.02) & 0.04 & 0 & 1.05 (0.02) & 0.53 (0.02) & 0.28 (0.01) & 5.04 (0.02) \\ 
   & bic & 0.82 (0.03) & 7.95 (1.89) & 7.95 & 0 & 1.19 (0.05) & 1.31 (0.21) & 0.41 (0.04) & 12.95 (1.89) \\ 
   & \textbf{ebic} & 0.93 (0.01) & 0.81 (0.11) & 0.81 & \textbf{0} & \textbf{1.05} (0.02) & \textbf{0.56} (0.03) & \textbf{0.28} (0.01) & 5.81 (0.11) \\ 
   & \textbf{cv} & \underline{0.92} (0.01) & \underline{1.04} (0.16) & \underline{1.04} & \textbf{0} & \textbf{1.05} (0.02) & \textbf{0.56} (0.02) & \textbf{0.28} (0.01) & 6.04 (0.16) \\ 
   & \textbf{csuv.m.0} & \textbf{1} (0) & \textbf{0.05} (0.02) & \textbf{0.04} & \underline{0.01} & 1.07 (0.02) & \textbf{0.56} (0.04) & 0.3 (0.02) & 5.03 (0.02) \\ 
   & \textbf{csuv.s.0} & 0.96 (0.01) & 0.51 (0.08) & 0.51 & \textbf{0} & \underline{1.1} (0.02) & \underline{0.7} (0.04) & \underline{0.35} (0.01) & 5.51 (0.08) \\ 
   & csuv.m.5 & 0.99 (0) & 0.06 (0.02) & 0.05 & 0.01 & 1.07 (0.02) & 0.57 (0.04) & 0.3 (0.02) & 5.04 (0.02) \\ 
   & csuv.s.5 & 0.86 (0.01) & 1.73 (0.11) & 1.73 & 0 & 1.18 (0.02) & 1.03 (0.04) & 0.46 (0.01) & 6.73 (0.11) \\ 
   & csuv.m.0.all & 1 (0) & 0.05 (0.02) & 0.04 & 0.01 & 1.07 (0.02) & 0.56 (0.04) & 0.3 (0.02) & 5.03 (0.02) \\ 
   & csuv.s.0.all & 0.95 (0.01) & 0.54 (0.08) & 0.54 & 0 & 1.1 (0.02) & 0.71 (0.04) & 0.35 (0.02) & 5.54 (0.08) \\ 
   & csuv.m.0.mcp & 0.99 (0) & 0.05 (0.03) & 0.02 & 0.03 & 1.08 (0.03) & 0.59 (0.05) & 0.31 (0.03) & 4.99 (0.03) \\ 
   & csuv.s.0.mcp & 0.99 (0) & 0.12 (0.04) & 0.12 & 0 & 1.06 (0.02) & 0.56 (0.03) & 0.29 (0.01) & 5.12 (0.04) \\ 
   \hline
\multicolumn{10}{l}{}\\
\end{tabular}
}
\endgroup
\caption{Model 5: performance of CSUV and methods it compares with (continue). Variable selection performance in terms of F-measure (f), total error (FP+FN), false positives (FP) and false negatives (FN), prediction error in terms of mse (pred.err) and estimation error in terms of l1 and l2 distance (l1.diff and l2.diff) and are shown. The numbers are based on 100 simulations. The last 8 rows are the performance of CSUV with different parameters (e.g. csuv.m.0.mcp corresponds to CSUV with MCP as constituent method and r = 0). A bold number represents the best result among delete-n/2 cross validation, eBIC and CSUV using Lasso, MCP and SCAD while a underlined number represents the worst among them. Standard errors are shown inside the parentheses.} 
\label{5_2}
\end{table}

			% latex table generated in R 3.6.1 by xtable 1.8-4 package
% Mon Jan 13 15:32:14 2020
\begin{table}[ht]
\centering
\begingroup\small
\scalebox{0.75}{
\begin{tabular}{|l|l|l|l|ll|lll|l|}
  \hline
parameters & methods & f & FP+FN & FP & FN & pred.err & l1.diff & l2.diff & size \\ 
  \hline
setting 7 & lasso & 0.37 (0) & 34.06 (0.68) & 33.96 & 0.1 & 1.86 (0.04) & 5.83 (0.12) & 1.24 (0.03) & 43.86 (0.7) \\ 
   & elastic net & 0.35 (0) & 38.37 (0.66) & 38.29 & 0.08 & 2.09 (0.05) & 6.83 (0.13) & 1.4 (0.03) & 48.21 (0.68) \\ 
  rho = -0.5 & relaxed lasso & 0.48 (0.01) & 22.32 (1.04) & 21.73 & 0.59 & 1.87 (0.04) & 5.2 (0.14) & 1.18 (0.03) & 31.14 (1.15) \\ 
  s = 10 & mcp & 0.85 (0.01) & 3.04 (0.21) & 1.53 & 1.51 & 1.56 (0.04) & 2.69 (0.12) & 1.06 (0.04) & 10.02 (0.15) \\ 
  p = 100 & scad & 0.76 (0.01) & 5.14 (0.26) & 3.25 & 1.89 & 1.87 (0.05) & 3.81 (0.11) & 1.45 (0.03) & 11.36 (0.23) \\ 
   & vsd & 0.81 (0.01) & 3.01 (0.19) & 0.16 & 2.85 & 1.81 (0.06) & 3.24 (0.14) & 1.3 (0.05) & 7.31 (0.18) \\ 
   & bic & 0.84 (0.01) & 3.58 (0.46) & 2.21 & 1.37 & 1.58 (0.04) & 2.78 (0.13) & 1.08 (0.04) & 10.84 (0.46) \\ 
   & \textbf{ebic} & \textbf{0.85} (0.01) & \textbf{3.09} (0.21) & 1.59 & \textbf{1.5} & \textbf{1.58} (0.04) & \textbf{2.74} (0.12) & \textbf{1.07} (0.05) & 10.09 (0.16) \\ 
   & \textbf{cv} & 0.77 (0.01) & \underline{5.01} (0.27) & \underline{3.09} & 1.92 & 1.83 (0.05) & 3.68 (0.12) & 1.41 (0.04) & 11.17 (0.24) \\ 
   & \textbf{csuv.m.0} & \underline{0.76} (0.01) & 3.92 (0.13) & \textbf{0.22} & \underline{3.7} & \underline{2.08} (0.05) & \underline{3.95} (0.11) & \underline{1.57} (0.04) & 6.52 (0.13) \\ 
   & \textbf{csuv.s.0} & 0.8 (0.01) & 3.83 (0.18) & 1.56 & 2.27 & 1.85 (0.06) & 3.44 (0.13) & 1.3 (0.04) & 9.29 (0.19) \\ 
   & csuv.m.5 & 0.75 (0.01) & 3.99 (0.13) & 0.17 & 3.82 & 2.11 (0.05) & 4.03 (0.11) & 1.6 (0.03) & 6.35 (0.13) \\ 
   & csuv.s.5 & 0.79 (0.01) & 4.15 (0.18) & 1.89 & 2.26 & 1.92 (0.06) & 3.59 (0.13) & 1.33 (0.04) & 9.63 (0.17) \\ 
   & csuv.m.0.all & 0.76 (0.01) & 3.87 (0.13) & 0.25 & 3.62 & 2.06 (0.05) & 3.91 (0.11) & 1.55 (0.03) & 6.63 (0.13) \\ 
   & csuv.s.0.all & 0.8 (0.01) & 3.86 (0.17) & 1.71 & 2.15 & 1.84 (0.06) & 3.4 (0.12) & 1.28 (0.04) & 9.56 (0.18) \\ 
   & csuv.m.0.mcp & 0.72 (0.01) & 4.25 (0.12) & 0.03 & 4.22 & 2.17 (0.05) & 4.22 (0.11) & 1.67 (0.03) & 5.81 (0.12) \\ 
   & csuv.s.0.mcp & 0.79 (0.01) & 3.55 (0.16) & 0.54 & 3.01 & 1.97 (0.06) & 3.67 (0.12) & 1.44 (0.04) & 7.53 (0.14) \\ 
   \hline
setting 8 & lasso & 0.26 (0) & 50.38 (1.39) & 48.77 & 1.61 & 3.4 (0.11) & 10.41 (0.21) & 2.23 (0.05) & 57.16 (1.54) \\ 
   & elastic net & 0.22 (0) & 59.24 (1.43) & 57.32 & 1.92 & 4.64 (0.17) & 13.28 (0.23) & 2.75 (0.06) & 65.4 (1.63) \\ 
  rho = -0.5 & relaxed lasso & 0.35 (0.01) & 26.61 (1.59) & 22.95 & 3.66 & 3.76 (0.14) & 9.65 (0.24) & 2.33 (0.06) & 29.29 (1.96) \\ 
  s = 10 & mcp & 0.76 (0.01) & 5.25 (0.28) & 3.29 & 1.96 & 1.68 (0.05) & 3.11 (0.13) & 1.18 (0.05) & 11.33 (0.25) \\ 
  p = 300 & scad & 0.63 (0.01) & 9.48 (0.37) & 7.35 & 2.13 & 1.97 (0.05) & 4.2 (0.12) & 1.5 (0.04) & 15.22 (0.33) \\ 
   & vsd & 0.67 (0.01) & 4.87 (0.16) & 0.02 & 4.85 & 2.49 (0.07) & 4.98 (0.16) & 1.89 (0.05) & 5.17 (0.16) \\ 
   & bic & 0.61 (0.02) & 20.31 (2.56) & 18.77 & 1.54 & 2.04 (0.09) & 5.1 (0.37) & 1.4 (0.06) & 27.23 (2.63) \\ 
   & \textbf{ebic} & \textbf{0.75} (0.01) & \textbf{5.32} (0.28) & 3.37 & \textbf{1.95} & \textbf{1.69} (0.05) & \textbf{3.12} (0.13) & \textbf{1.18} (0.05) & 11.42 (0.26) \\ 
   & \textbf{cv} & 0.67 (0.01) & \underline{8.23} (0.42) & \underline{6.13} & 2.1 & 1.87 (0.05) & 3.85 (0.13) & 1.4 (0.04) & 14.03 (0.37) \\ 
   & \textbf{csuv.m.0} & \underline{0.59} (0.01) & 5.77 (0.11) & \textbf{0.09} & \underline{5.68} & \underline{2.96} (0.08) & \underline{5.94} (0.14) & \underline{2.2} (0.04) & 4.41 (0.12) \\ 
   & \textbf{csuv.s.0} & 0.64 (0.01) & 6.02 (0.17) & 1.64 & 4.38 & 2.61 (0.09) & 5.37 (0.16) & 1.89 (0.05) & 7.26 (0.23) \\ 
   & csuv.m.5 & 0.59 (0.01) & 5.77 (0.11) & 0.08 & 5.69 & 2.99 (0.09) & 5.95 (0.15) & 2.2 (0.04) & 4.39 (0.12) \\ 
   & csuv.s.5 & 0.64 (0.01) & 6.69 (0.2) & 2.65 & 4.04 & 2.56 (0.07) & 5.41 (0.15) & 1.84 (0.04) & 8.61 (0.22) \\ 
   & csuv.m.0.all & 0.59 (0.01) & 5.77 (0.11) & 0.09 & 5.68 & 2.96 (0.08) & 5.94 (0.14) & 2.2 (0.04) & 4.41 (0.12) \\ 
   & csuv.s.0.all & 0.64 (0.01) & 6.01 (0.17) & 1.66 & 4.35 & 2.61 (0.08) & 5.37 (0.16) & 1.89 (0.05) & 7.31 (0.23) \\ 
   & csuv.m.0.mcp & 0.56 (0.01) & 6.05 (0.1) & 0 & 6.05 & 3.18 (0.1) & 6.32 (0.14) & 2.32 (0.04) & 3.95 (0.1) \\ 
   & csuv.s.0.mcp & 0.63 (0.01) & 5.63 (0.14) & 0.58 & 5.05 & 2.81 (0.09) & 5.61 (0.15) & 2.05 (0.05) & 5.53 (0.17) \\ 
   \hline
\multicolumn{10}{l}{}\\
\end{tabular}
}
\endgroup
\caption{Model 5: performance of CSUV and methods it compares with (continue). Variable selection performance in terms of F-measure (f), total error (FP+FN), false positives (FP) and false negatives (FN), prediction error in terms of mse (pred.err) and estimation error in terms of l1 and l2 distance (l1.diff and l2.diff) and are shown. The numbers are based on 100 simulations. The last 8 rows are the performance of CSUV with different parameters (e.g. csuv.m.0.mcp corresponds to CSUV with MCP as constituent method and r = 0). A bold number represents the best result among delete-n/2 cross validation, eBIC and CSUV using Lasso, MCP and SCAD while a underlined number represents the worst among them. Standard errors are shown inside the parentheses.} 
\label{5_3}
\end{table}

			% latex table generated in R 3.6.1 by xtable 1.8-4 package
% Mon Jan 13 15:32:14 2020
\begin{table}[ht]
\centering
\begingroup\small
\scalebox{0.8}{
\begin{tabular}{|l|l|l|}
  \hline
methods & pred.err & size \\ 
  \hline
lasso & 26.08 (0.39) & 12.39 (0.09) \\ 
  elastic net & 26.12 (0.4) & 12.38 (0.1) \\ 
  relaxed lasso & 26.53 (0.42) & 11.21 (0.14) \\ 
  mcp & 26.16 (0.39) & 11.29 (0.15) \\ 
  scad & 26.1 (0.39) & 11.54 (0.12) \\ 
  vsd & 28.54 (0.42) & 6.03 (0.16) \\ 
  bic & 26.2 (0.4) & 11.05 (0.15) \\ 
  \textbf{ebic} & 26.19 (0.4) & 11.07 (0.15) \\ 
  \textbf{cv} & \textbf{26.08} (0.39) & 12.36 (0.09) \\ 
  \textbf{csuv.m.0} & 26.64 (0.41) & 10.02 (0.17) \\ 
  \textbf{csuv.s.0} & \underline{26.73} (0.41) & 9.9 (0.16) \\ 
  csuv.m.5 & 26.58 (0.41) & 10.07 (0.16) \\ 
  csuv.s.5 & 26.63 (0.41) & 10.07 (0.15) \\ 
  csuv.m.0.all & 26.72 (0.41) & 9.95 (0.17) \\ 
  csuv.s.0.all & 26.76 (0.41) & 9.81 (0.16) \\ 
  csuv.m.0.mcp & 26.33 (0.4) & 10.52 (0.16) \\ 
  csuv.s.0.mcp & 26.39 (0.4) & 10.46 (0.15) \\ 
   \hline
\multicolumn{3}{l}{}\\
\end{tabular}
}
\endgroup
\caption{Boston data: performance of CSUV and methods it compares with. The numbers are based on 100 simulations. The last 8 rows are the performance of CSUV with different parameters (e.g. csuv.m.0.mcp corresponds to CSUV with MCP as constituent method and r = 0). A bold number represents the best result among delete-n/2 cross validation, eBIC and CSUV using Lasso, MCP and SCAD while a underlined number represents the worst among them. Standard errors are shown inside the parentheses.} 
\label{6_1}
\end{table}

			% latex table generated in R 3.6.1 by xtable 1.8-4 package
% Mon Jan 13 15:32:14 2020
\begin{table}[ht]
\centering
\begingroup\small
\scalebox{0.8}{
\begin{tabular}{|l|l|l|ll|l|}
  \hline
methods & f & FP+FN & FP & FN & size \\ 
  \hline
lasso & 0.46 (0.01) & 19.96 (1.45) & 16.88 & 3.08 & 23.8 (1.48) \\ 
  elastic net & 0.45 (0.01) & 22.59 (1.51) & 20.44 & 2.15 & 28.29 (1.53) \\ 
  relaxed lasso & 0.56 (0.02) & 10.15 (0.81) & 5.85 & 4.3 & 11.55 (0.97) \\ 
  mcp & 0.45 (0.01) & 8.97 (0.22) & 2.62 & 6.35 & 6.27 (0.25) \\ 
  scad & 0.48 (0.02) & 13.01 (0.67) & 8.35 & 4.66 & 13.69 (0.56) \\ 
  vsd & NaN (NA) & 10 (0) & 0 & 10 & 0 (0) \\ 
  bic & 0.43 (0.01) & 15.78 (1.47) & 10.5 & 5.28 & 15.22 (1.65) \\ 
  \textbf{ebic} & \underline{0.44} (0.01) & 11.33 (1.18) & 5.09 & 6.24 & 8.85 (1.31) \\ 
  \textbf{cv} & 0.45 (0.02) & \underline{19.3} (1.45) & \underline{15.82} & \textbf{3.48} & 22.34 (1.54) \\ 
  \textbf{csuv.m.0} & 0.45 (0.01) & 7.04 (0.12) & \textbf{0.03} & \underline{7.01} & 3.02 (0.12) \\ 
  \textbf{csuv.s.0} & \textbf{0.65} (0.01) & \textbf{5.49} (0.19) & 0.99 & 4.5 & 6.49 (0.32) \\ 
  csuv.m.5 & 0.43 (0.01) & 7.26 (0.12) & 0.02 & 7.24 & 2.78 (0.12) \\ 
  csuv.s.5 & 0.69 (0.01) & 5.07 (0.19) & 1.08 & 3.99 & 7.09 (0.3) \\ 
  csuv.m.0.all & 0.49 (0.01) & 6.72 (0.13) & 0.02 & 6.7 & 3.32 (0.13) \\ 
  csuv.s.0.all & 0.69 (0.01) & 5.16 (0.22) & 1.3 & 3.86 & 7.44 (0.35) \\ 
  csuv.m.0.mcp & 0.28 (0.01) & 8.61 (0.09) & 0 & 8.61 & 1.39 (0.09) \\ 
  csuv.s.0.mcp & 0.44 (0.01) & 7.19 (0.09) & 0.05 & 7.14 & 2.91 (0.09) \\ 
   \hline
\multicolumn{6}{l}{}\\
\end{tabular}
}
\endgroup
\caption{Riboflavin data with permutation: performance of CSUV and methods it compares with. The numbers are based on 100 simulations. The last 8 rows are the performance of CSUV with different parameters (e.g. csuv.m.0.mcp corresponds to CSUV with MCP as constituent method and r = 0). A bold number represents the best result among delete-n/2 cross validation, eBIC and CSUV using Lasso, MCP and SCAD while a underlined number represents the worst among them. Standard errors are shown inside the parentheses.} 
\label{7_1}
\end{table}

\end{document}